\documentclass[nofootinbib,aps,11pt]{revtex4-1}
\usepackage{graphicx}
\usepackage{hyperref}
\usepackage{cancel}
\usepackage{amssymb}
\usepackage{textcomp}
\usepackage{amsmath}
\usepackage{bm}
\usepackage{times}
\usepackage{epsfig}
\usepackage{color}
\usepackage{multirow}

\begin{document}
\title{\Large New Avenues of Heavy Neutral Lepton at Muon Collider}
\bigskip

\author{Fa-Xin Yang$^1$}
\email{yfx@qfnu.edu.cn}
\author{Feng-Lan Shao$^1$}
\email{shaofl@mail.sdu.edu.cn}
\author{Zhi-Long Han$^2$}
\email{sps\_hanzl@ujn.edu.cn}
\author{Honglei Li$^2$}
\email{sps\_lihl@ujn.edu.cn}

\affiliation{$^1$School of Physics and Physical Engineering, Qufu Normal University, Qufu, Shandong 273165, China	}
\affiliation{$^2$School of Physics and Technology, University of Jinan, Jinan, Shandong 250022, China}

\date{\today}

\begin{abstract}
	With initial state radiation, the multi-TeV muon collider can be regarded as an electroweak boson collider. The dominant production mode of the certain process becomes the vector boson fusion channel, because the corresponding cross section typically increases logarithmically at high energies. This also holds true for new physics beyond the standard model. Within the $U(1)$ gauged extension of seesaw models, the heavy neutral lepton has additional interactions with the new gauge boson $Z'$ and heavy Higgs $H$. In this paper, we investigate the production of heavy neutral lepton $N$ via the new vector boson fusion processes $Z'Z'\to H\to NN$ with and $Z'Z'\to NN$ without heavy Higgs at the multi-TeV muon collider. Different from the canonical vector boson fusion processes $WW/ZZ\to H\to NN$, the new process $Z'Z'\to H\to NN$ is not suppressed by the small mixing angle $\alpha$ between the Higgs bosons. Meanwhile, the pair production process $Z'Z'\to NN$ is also viable even for  heavy Higgs $m_H> \sqrt{s}$. Therefore, these new avenues provide alternative pathways to probe the intrinsic feature of the heavy neutral lepton. We then perform a detailed analysis of the lepton number violation signals via the new vector boson fusion with heavy Higgs $\mu^+\mu^-\to \mu^+\mu^- H \to \mu^+\mu^- NN$ and without heavy Higgs $\mu^+\mu^-\to \mu^+\mu^- NN$, followed by $N\to \mu^\pm jj$, where the two jets from $W$ boson decay are treated as one fat-jet $J$.

\end{abstract}

\maketitle

\section{Introduction}
The observations of neutrino oscillation indicate non-zero neutrino mass  \cite{Super-Kamiokande:1998kpq,SNO:2002tuh,DayaBay:2012fng}, but the origin of such tiny neutrino mass is still unresolved in particle physics. To naturally explain the sub-eV neutrino masses, we can introduce heavy neutral leptons $N$ through the seesaw mechanism \cite{Minkowski:1977sc,Mohapatra:1979ia,Schechter:1980gr,Schechter:1981cv}. In the minimal scenario, the present colliders are sensitive to heavy neutral lepton around the GeV-scale and relatively large mixing parameter between the light and heavy neutrino $V_{\ell N}$ \cite{Abdullahi:2022jlv}. For instance, the current search for the same-sign dilepton signature $pp\to W^{\pm(*)}\to \ell^\pm N\to \ell^\pm \ell^\pm jj$ has excluded $|V_{\ell N}|^2\gtrsim2\times10^{-6}$ with $m_N=20$~GeV \cite{ATLAS:2019kpx}. In the future, the region within $|V_{\ell N}|^2\gtrsim5\times10^{-10}$ and $m_N\lesssim40$ GeV could be tested through the displaced vertex signature \cite{Drewes:2019fou}. However, the natural seesaw induced mixing $V_{\ell N}\sim \sqrt{m_\nu/m_N}$ is too small to be detected at colliders \cite{Abdullahi:2022jlv}.

In order to explain the tentative new physics, the $U(1)_{L_\mu-L_\tau}$ symmetry is extensively studied, such as the muon anomalous magnetic dipole moment  \cite{He:1990pn,Baek:2001kca,Muong-2:2021ojo},  lepton universality anomaly in $b\to s \ell^+\ell^-$ decay \cite{Altmannshofer:2014cfa,Altmannshofer:2016jzy,Han:2019diw,LHCb:2021trn}, dark matter \cite{Baek:2008nz,Foldenauer:2018zrz,Holst:2021lzm,Wang:2025kit,Bernal:2025szh}, neutrino mixing \cite{Ma:2001md,Baek:2015mna,Baek:2015fea,Asai:2017ryy,Asai:2018ocx,Ibe:2025rwk}, and leptogenesis \cite{Borah:2021mri,Eijima:2023yiw,Granelli:2023egb,Wada:2024cbe}. As the new gauge boson $Z'$ does not directly couple to electrons and quarks, the experimental limits from current colliders are relatively loose above the electroweak scale \cite{ATLAS:2023vxg,ATLAS:2024uvu,NA64:2024klw}. Recently, the multi-TeV muon collider is proposed \cite{Delahaye:2019omf,Accettura:2023ked,InternationalMuonCollider:2025sys}, which is quite promising to investigate the $U(1)_{L_\mu-L_\tau}$ symmetry \cite{Huang:2021nkl,Sun:2023ylp,Dasgupta:2023zrh}. Therefore, we consider the $U(1)_{L_\mu-L_\tau}$ gauge extension of the type-I seesaw mechanism in this study.

Depending on the specific interactions, there are various production channels of heavy neutral lepton $N$ at the high energy muon collider. Through the mixing with the light neutrino, the production of the heavy-light neutrino pair $\mu^+\mu^-\to N\nu$ is the dominant process \cite{Chakraborty:2022pcc,Mekala:2023diu,Kwok:2023dck,Li:2023tbx,Wang:2023zhh,Cao:2024rzb,Li:2026kes}, which could be also mediated by the dipole operator $\bar{\nu}\sigma^{\mu \nu} N F_{\mu \nu}$ \cite{Barducci:2024kig,Frigerio:2024jlh,Brdar:2025iua,Vignaroli:2025pwn}. The associated production process $\mu^+\mu^-\to N W^\pm \ell^\mp$ is also induced by the heavy-light neutrino mixing or the dipole operator \cite{Antonov:2023otp,Barducci:2024kig}. With coupling to the new gauge boson $Z'$, the productions of heavy neutral lepton  pair $\mu^+\mu^-\to Z^{\prime *} \to NN$ and $\mu^+\mu^+\to Z'\gamma \to NN\gamma$ are promising at the muon collider \cite{Liu:2021akf,He:2024dwh,Bi:2024pkk}. Meanwhile, the new Higgs-strahlung process $\mu^+\mu^-\to Z'H\to NN + NN$ could lead to the novel same-sign tetralepton signature $4\mu^\pm+4J$ with $N\to \mu^\pm J$ \cite{Zhang:2026kwz}, where the fat-jets $J$ come from the hadronic decay of $W$ bosons. The Yukawa interaction with charged Higgs boson $H^\pm$ would mediate the direct heavy neutral lepton pair production $\mu^+\mu^-\to NN$ and associated production $\mu^+\mu^-\to N H^\pm \mu^\mp$ \cite{Batra:2023ssq,Batra:2023mds}. Such Yukawa coupling also contributes to the production of heavy neutral lepton at the same-sign muon collider via the processes as $\mu^+\mu^+\to H^+H^+\to \mu^+N+\mu^+N$ and $\mu^+\mu^+\to NH^+\mu^+$ \cite{Yan:2026wrb}.

With forward gauge boson radiation, the multi-TeV muon collider is effectively an electroweak gauge boson collider \cite{Costantini:2020stv,Han:2020uid,Ruiz:2021tdt}. Due to the logarithmic increase of the cross section, the vector boson fusion channels become the dominant contribution at the multi-TeV muon collider. In this way, the heavy neutral lepton can be produced through the vector boson fusion processes $W^\pm Z/\gamma\to \ell^\pm N$ by the mixing with the light neutrino \cite{Li:2023lkl,Mikulenko:2023ezx,Dehghani:2025xkd}. On the other hand, production of heavy neutral lepton pair through the vector boson fusion channel as $W^+W^- \to NN$ can be mediated by the standard model Higgs boson $h$ \cite{Das:2025rlt}, the new heavy Higgs boson $H$ \cite{Yang:2025jxc}, and the axion-like particle $a$ \cite{Marcos:2024yfm}. Provided new gauge interaction of the muon, such as the $U(1)_{L_\mu-L_\tau}$ symmetry, the new gauge boson $Z'$ can also be  emitted from the high energy muon \cite{Asadi:2026kpt}, which results in new vector boson fusion processes, such as $Z'Z'\to H$ \cite{Das:2022mmh}.

In this paper, we propose the new avenues of heavy neutral lepton $N$ via the new vector boson fusion processes $Z'Z'\to H\to NN$ with and $Z'Z'\to NN$ without the heavy Higgs at the multi-TeV muon collider. Since the interactions between the new gauge boson $Z'$ and heavy Higgs $H$ are not  suppressed by the small mixing angle $\alpha$ between the Higgs bosons, the process $Z'Z'\to H\to NN$ remains viable in the Higgs decoupling limit $\alpha\to 0$. On the other hand, in the  heavy Higgs scenario with $m_H> \sqrt{s}$, the heavy Higgs resonance channel $Z'Z'\to H\to NN$ is kinematically not allowed, but the pair production of a heavy neutral lepton pair $Z'Z'\to NN$ is still possible through the exchange of a heavy neutral lepton. Due to the Majorana nature of the heavy neutral lepton $N$, further decay $N\to \mu^\pm J$ leads to the lepton number violation signature $\mu^\pm \mu^\pm JJ$ from the heavy neutral lepton pair $NN$, where the fat-jets $J$ come from the hadronic decay of $W$ bosons.

The rest of this paper is organized as follows. In Section \ref{SEC:MD}, we briefly introduce the gauged  $U(1)_{L_\mu-L_\tau}$ extension of the Type-I seesaw model. Decay properties of the heavy Higgs $H$ and heavy neutral lepton $N$ are considered in Section \ref{SEC:DP}.  The lepton number violation signature $\mu^+\mu^- \mu^\pm\mu^\pm JJ$ from the new vector boson fusion processes $Z'Z'\to H\to NN$ with and $Z'Z'\to NN$ without heavy Higgs are studied in Section \ref{SEC:Sig1} and Section \ref{SEC:Sig2}, respectively. Section \ref{SEC:CL} is the conclusion.

\section{The Model}\label{SEC:MD}

In this paper, we focus on the $U(1)_{L_\mu-L_\tau}$ gauge extension of the Type-I seesaw model. Three generations of heavy neutral lepton $(N_e, N_\mu, N_\tau)$ are introduced to generate tiny neutrino masses, which have $U(1)_{L_\mu-L_\tau}$ charge $(0,1,-1)$. The gauged $U(1)_{L_\mu-L\tau}$ symmetry is broken spontaneously by a scalar singlet $S$ with $U(1)_{L_\mu-L_\tau}$ charge $+1$. Neglecting the kinetic mixing term for simplicity \cite{Hapitas:2021ilr}, the relevant kinetic terms of the new fields are
\begin{equation}
	\mathcal{L}\supset-\frac{1}{4}F'_{\mu \nu}F'^{\mu\nu} + |D_\mu S|^2,
\end{equation}
where $F'_{\mu\nu}$ is the field strength of the $U(1)_{L_\mu-L_\tau}$ symmetry, and $D_\mu=\partial_\mu-ig'Z'_\mu$ is the covariant derivative with $g'$ being the $L_\mu-L_\tau$ gauge coupling constant.  After $S$ develops nonzero vacuum expectation value $v_S$, the mass of new gauge boson $Z'$ is derived as \cite{Nomura:2020vnk}
\begin{equation}
	m_{Z'}=g' v_S.
\end{equation}

The new gauge interactions of the relevant fermions are
\begin{equation}
	\mathcal{L}\supset g' (\bar{\mu}\gamma^\mu\mu-\bar{\tau}\gamma^\mu \tau +\bar{\nu}_\mu \gamma^\mu P_L\nu_\mu-\bar{\nu}_\tau \gamma^\mu P_L\nu_\tau+\bar{N}_\mu \gamma^\mu P_R N_\mu-\bar{N}_\tau \gamma^\mu P_R N_\tau)Z'_\mu.
\end{equation}
Currently, there are various constraints on the new gauge boson $Z'$ \cite{Wang:2025kit}. In this paper, we assume $m_{Z'}>100$ GeV. The relevant constraint is from the neutrino trident production at the CCFR experiment \cite{CCFR:1991lpl}, which roughly requires $v_s=m_{Z'}/g'\gtrsim550$ GeV \cite{Altmannshofer:2014pba}.

Under the $L_\mu-L_\tau$ symmetry, the Yukawa interactions and mass terms  are given by  \cite{Asai:2018ocx}
\begin{eqnarray}
	\mathcal{L}&\supset& - y_e \bar{L}_e \tilde{\Phi} N_e - y_\mu \bar{L}_\mu \tilde{\Phi} N_\mu - y_\tau \bar{L}_\tau \tilde{\Phi} N_\tau -\frac{1}{2} M_{ee} \overline{N^c_e}N_e \\ \nonumber
	&&-y_{e\mu} S^* \overline{N^c_e}N_\mu -y_{e\tau} S \overline{N^c_e}N_\tau -M_{\mu\tau}\overline{N^c_\mu}N_\tau+ \text{h.c.},
\end{eqnarray}
where $L_e,L_\mu,L_\tau$ are the lepton doublets, and $\tilde{\Phi}=i\tau_2 \Phi^*$. After the spontaneous symmetry breaking, the Dirac neutrino mass matrix and heavy neutral lepton mass matrix are 
\begin{align}
	M_D=\left(
	\begin{array}{c c c}
		\frac{y_e v_0}{\sqrt{2}} & 0 & 0\\
		0 & \frac{y_\mu v_0}{\sqrt{2}}&0 \\
		0 & 0 & \frac{y_\tau v_0}{\sqrt{2}}
	\end{array}
	\right), \quad
	M_N =
	\left(
	\begin{array}{c c c}
		M_{ee} & \frac{y_{e\mu}v_s}{\sqrt{2}} & \frac{y_{e\tau}v_s}{\sqrt{2}}\\
		\frac{y_{e\mu}v_s}{\sqrt{2}} & 0& M_{\mu\tau} \\
		\frac{y_{e\tau}v_s}{\sqrt{2}} & M_{\mu\tau} & 0
	\end{array}\right).
\end{align}
Light neutrino masses are generated via the type-I seesaw mechanism
\begin{equation}\label{Eq:SS}
	M_\nu\simeq - M_D M_N^{-1} M_D^T.
\end{equation}

The scalar potential  $V(\Phi,S)$ involving the SM Higgs doublet $\Phi$ and scalar singlet $S$ is given by
\begin{align}
	V(\Phi,S)=m_1^2H\Phi^\dagger \Phi+m_2^2 S^\dag S+\lambda_1(\Phi^\dagger \Phi)^2+\lambda_2 (S^\dag S)^2+\lambda_3(\Phi^\dagger \Phi)(S^\dag S),
\end{align}
where $m_1^2<0,m_2^2<0$ is responsible for breaking the symmetry spontaneously. The stability
of the potential is satisfied with the condition
\begin{equation}
	\lambda_1>0,~\lambda_2>0, ~4\lambda_1 \lambda_2 - \lambda_3^2>0.
\end{equation}

After the symmetry breaking, the vacuum expectation values of the scalars are denoted as $\langle\Phi\rangle= v_0/\sqrt{2}$ and $\langle S \rangle= v_s/\sqrt{2}$, where $v_0=246$ GeV.  The mass matrix for the two CP-even Higgs bosons $\phi^0$ and $s^0$ can be found as
\begin{align}
	M^2(\phi^0,s^0)=\begin{pmatrix}
		2\lambda_1v_0^2 & \lambda_3v_0v_s \\ 
		\lambda_3v_0v_s & 2\lambda_2v_s^2 
	\end{pmatrix}.
\end{align}
The mass matrix is diagonalized by an orthogonal transformation
\begin{align}
	\begin{pmatrix}
		h\\
		H
	\end{pmatrix}
	=\begin{pmatrix}
		\cos\alpha & -\sin\alpha \\ 
		\sin\alpha & \cos\alpha 
	\end{pmatrix}
	\begin{pmatrix}
		\phi^0\\
		s^0
	\end{pmatrix},
\end{align}
where the mixing angle $\alpha$ is derived as
\begin{align}
	\tan(2\alpha)=\frac{\lambda_3v_0v_s}{\lambda_2v_s^2-\lambda_1v_0^2}.
\end{align}
Then, the masses for the two physical Higgs states $h,H$ are
\begin{align}
	m_{h}^2=\lambda_1v_0^2+\lambda_2v_s^2-\sqrt{(\lambda_1v_0^2-\lambda_2v_s^2)^2+(\lambda_3v_0v_s)^2},\\
	m_{H}^2=\lambda_1v_0^2+\lambda_2v_s^2+\sqrt{(\lambda_1v_0^2-\lambda_2v_s^2)^2+(\lambda_3v_0v_s)^2},
\end{align}
where $h$ is the SM Higgs, and $H$ is the additional heavy Higgs. To satisfy the current experimental constraints, $\sin\alpha\lesssim0.1$ is required \cite{Lewis:2024yvj}. In this paper, we consider the decoupling limit $\sin\alpha=0$ for simplicity. Then the 
interactions between the heavy Higgs $H$ and new gauge boson $Z'$ can be written as,
\begin{align}\label{Hzp}
	\mathcal{L}= \frac{m_{Z'}^2}{v_s}Z^{\prime \mu}Z'_\mu H+\frac{m_{Z'}^2}{2v_s^2}Z^{\prime \mu}Z'_\mu H^2.
 \end{align}

\section{Decay Properties}\label{SEC:DP}

In this Section, we  consider the decay properties of the heavy Higgs $H$. Since we have set $\sin\alpha=0$ to forbid the mixing between the neutral scalars, the heavy Higgs $H$ will not directly couple to the SM particles. There are still two decay modes of the heavy Higgs. One is $H\to NN$ through the Yukawa interaction, and the other one is $H\to Z'Z'$ through the new gauge interaction.
The corresponding partial decay widths are
\begin{eqnarray}
\Gamma(H\rightarrow NN)&=& \frac{m_N^2 m_H}{4\pi v_s^2} \left(1-\frac{4m_N^2}{m_H^2}\right)^{3/2},\\
\Gamma(H\rightarrow Z'Z')&=&\frac{m_H^3}{32\pi v_s^2} \left(1-\frac{4m_{Z'}^2}{m_H^2}+ \frac{12m_{Z'}^4}{m_H^4}\right)\left(1-\frac{4m_{Z'}^2}{m_H^2}\right)^{1/2}.
\end{eqnarray}

\begin{figure}
	\begin{center}
		\includegraphics[width=0.45\linewidth]{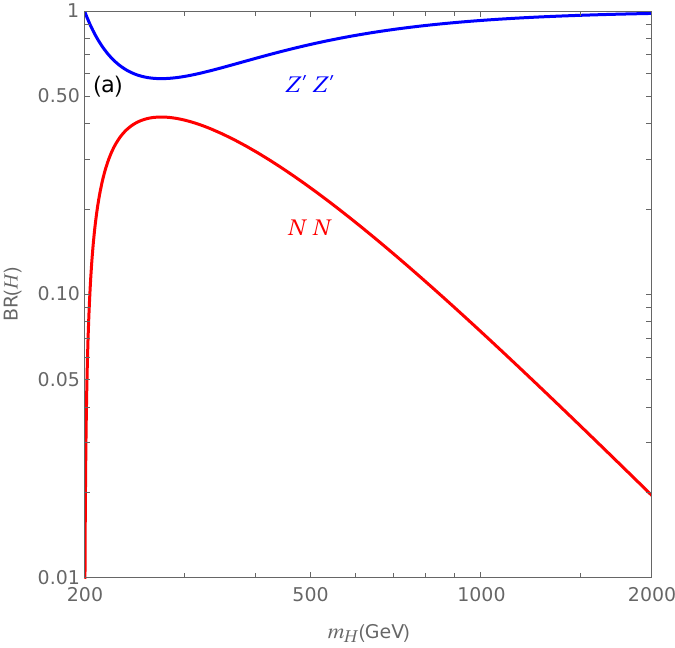}
		\includegraphics[width=0.45\linewidth]{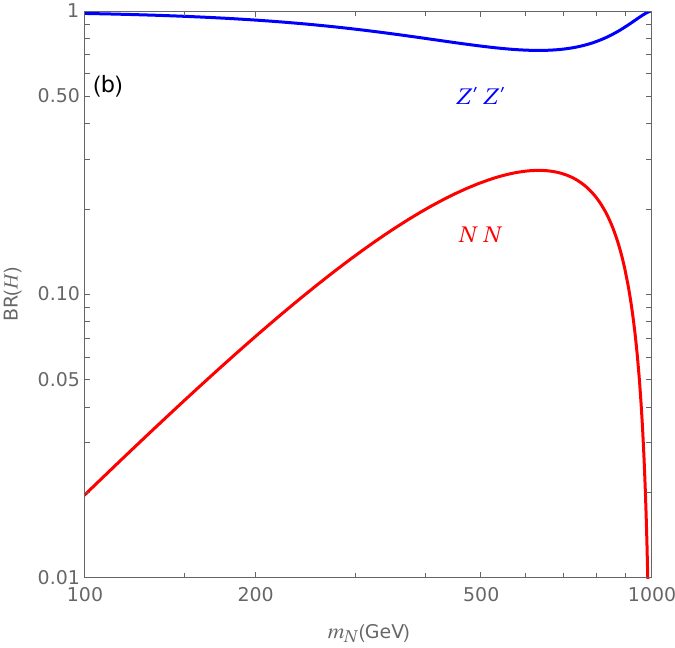}
	\end{center}
	\caption{Branching ratio of the heavy Higgs $H$ as a function of $m_{H}$ in panel (a) and $m_N$ in panel (b). In panel (a), we set $m_N=100$ GeV and $m_{Z'}=100$ GeV. In panel (b), we set $m_{H}=2000$ GeV and $m_{Z'}=100$ GeV.}
	\label{fig1}
\end{figure}

Branching ratios of the heavy Higgs are shown in Figure \ref{fig1}. As we investigate the new vector boson fusion processes, a relatively light gauge boson is assumed. In panel (a) of Figure \ref{fig1}, both $m_N$ and $m_{Z'}$ are fixed as 100 GeV. We report that the heavy Higgs always dominantly decays into a pair of $Z'$. Around $m_H\simeq 270$ GeV, the decay mode of $H\to NN$ achieves the largest branching ratio of $\text{BR}(H\to NN)\approx 0.4$. In the heavy Higgs limit $m_{H}\gg m_N$, the branching ratio of $H\to NN$ quickly decreases, which is because of $\Gamma(H\to Z'Z')$ being proportional to $m_H^3$. This indicates that the new production channel $Z'Z'\to H\to NN$ is unpromising when $m_H\gg m_N$.  In panel (b) of Figure \ref{fig1}, we then fix $m_H=2000$~GeV and $m_{Z'}=100$ GeV. The  branching ratio $\text{BR}(H\to NN)$ reaches the largest value of 0.3 when $m_N\sim650$ GeV. When $m_N$ approaches $m_H/2$, the branching ratio of $H\to NN$ is clearly suppressed by the final state phase space.

Because a too light heavy neutral lepton will lead to a too small branching ratio of $H\to NN$, we assume that the heavy neutral lepton is above 100 GeV. Through mixings with the light neutrinos, the two body decay modes $N\to \ell^\pm W^\mp, \nu_\ell Z, \nu_\ell h$ are the dominant channels. The corresponding partial decay widths are 
\begin{eqnarray}
	\Gamma(N\to \ell^\pm W^\mp) &=& \frac{g^2}{64\pi} |V_{\ell N}|^2 \frac{m_N^3}{m_W^2}\left(1-\frac{m_W^2}{m_N^2}\right)^2\left(1+2\frac{m_W^2}{m_N^2}\right), \\
	\Gamma(N\to \nu_\ell Z) &=& \frac{g^2}{128\pi} |V_{\ell N}|^2 \frac{m_N^3}{m_W^2}\left(1-\frac{m_Z^2}{m_N^2}\right)^2\left(1+2\frac{m_Z^2}{m_N^2}\right),\\
	\Gamma(N\to \nu_\ell h) & =& \frac{g^2}{128\pi} |V_{\ell N}|^2 \frac{m_N^3}{m_W^2}\left(1-\frac{m_h^2}{m_N^2}\right)^2.
\end{eqnarray}

For sufficient large heavy neutral lepton mass, all the final state particles can be treated as massless. Then, the branching ratios simply satisfy
\begin{equation}
	\text{BR}(N\to \ell^\pm W^\mp):\text{BR}(N\to \nu_\ell Z):\text{BR}(\nu_\ell h)\simeq 2:1:1.
\end{equation}
In this paper, we focus on the fully visible mode $N\to \ell^\pm W^\mp \to \ell^\pm jj$, which would induce obvious lepton number violation signatures via the new gauge boson fusion processes.

\section{Signature with Heavy Higgs}\label{SEC:Sig1}

\begin{figure}
	\begin{center}
		\includegraphics[width=0.45\linewidth,height=0.35\linewidth]{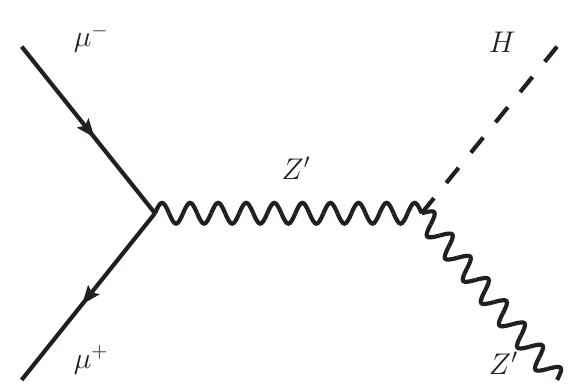}
		\includegraphics[width=0.45\linewidth]{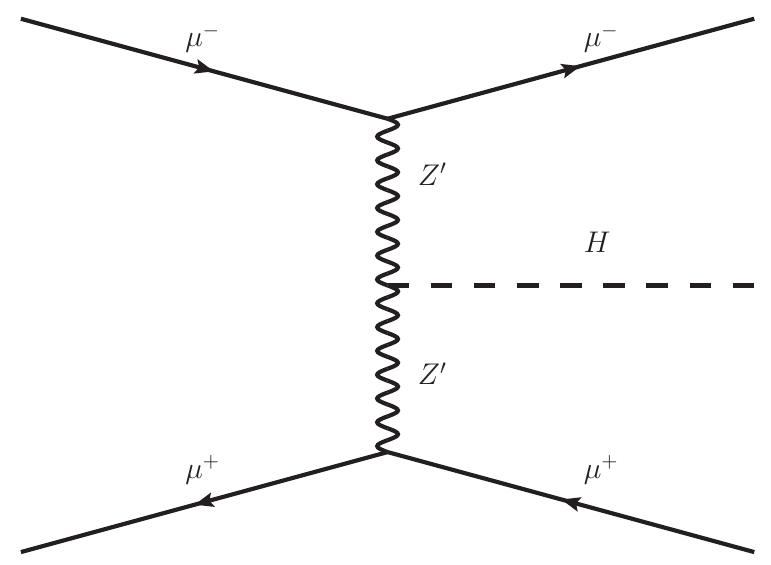}
	\end{center}
	\caption{Feynman diagrams of the heavy Higgs-strahlung $\mu^+\mu^-\to Z'H$ (left) and the $Z'Z'$-fusion $\mu^+\mu^-\rightarrow \mu^+\mu^-H$ (right) processes at muon collider.}
	\label{fig2}
\end{figure}

It is well known that the main production channels of the SM Higgs $h$ at the lepton colliders are the Higgs-strahlung $\ell^+\ell^-\to Zh$ and $WW$-fusion $\ell^+\ell^-\to \bar{\nu}_\ell \nu_\ell h$  processes \cite{Kilian:1995tr}. Because the cross section of the Higgs-strahlung is suppressed by $1/s$, while the cross section of the $WW$-fusion increases logarithmically with $s$, the $WW$-fusion mechanism becomes the dominant channel when $\sqrt{s}\geq 400$ GeV. Under the new gauge symmetry $U(1)_{L_\mu-L_\tau}$, the multi-TeV muon collider is also a new gauge boson collider \cite{Asadi:2026kpt}. In this section, we first study the new gauge boson fusion processes $Z'Z'\to H\to NN$ at the muon collider.

\begin{figure}
	\begin{center}
		\includegraphics[width=0.45\linewidth]{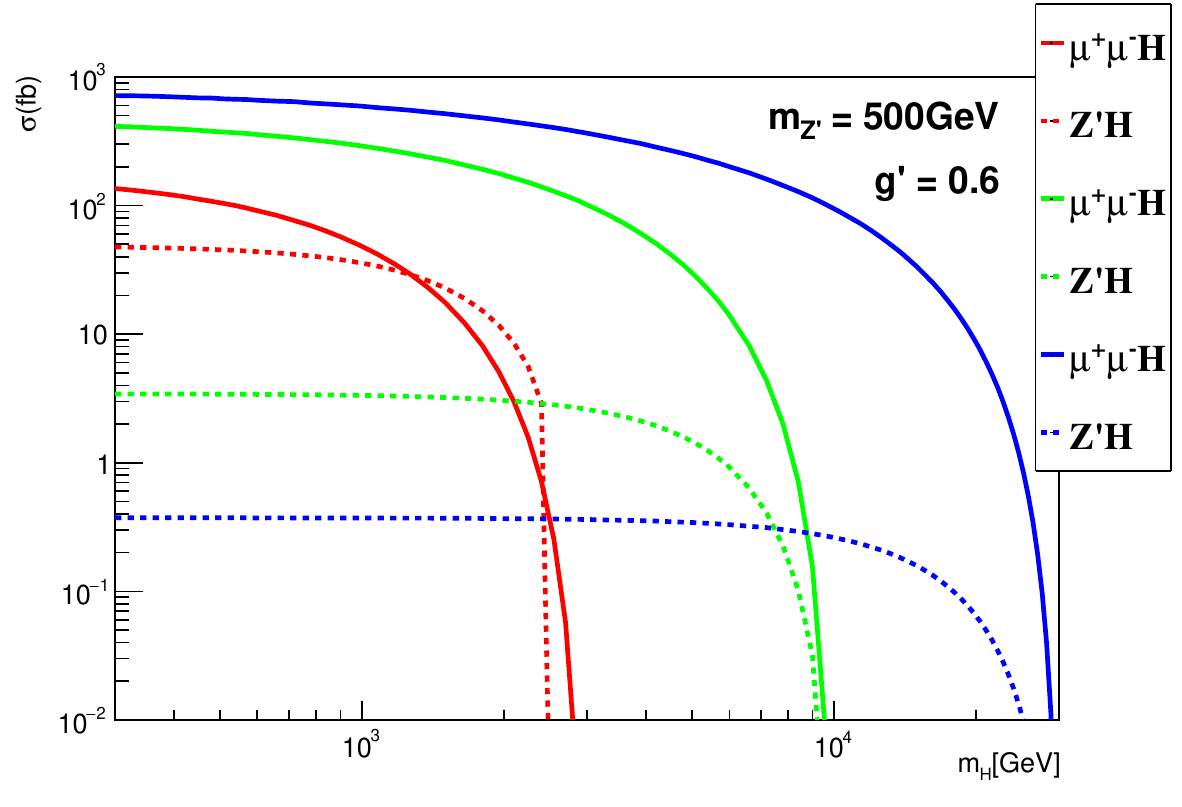}
		\includegraphics[width=0.45\linewidth]{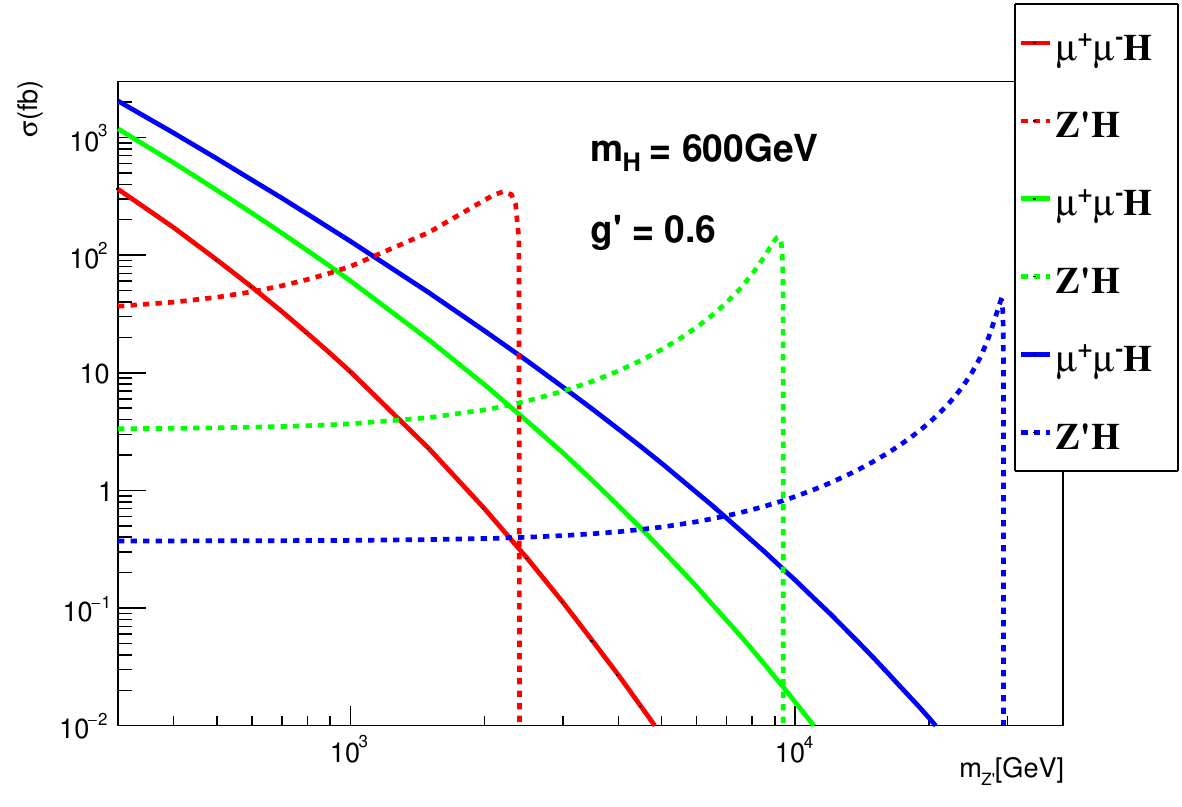}
		\includegraphics[width=0.45\linewidth]{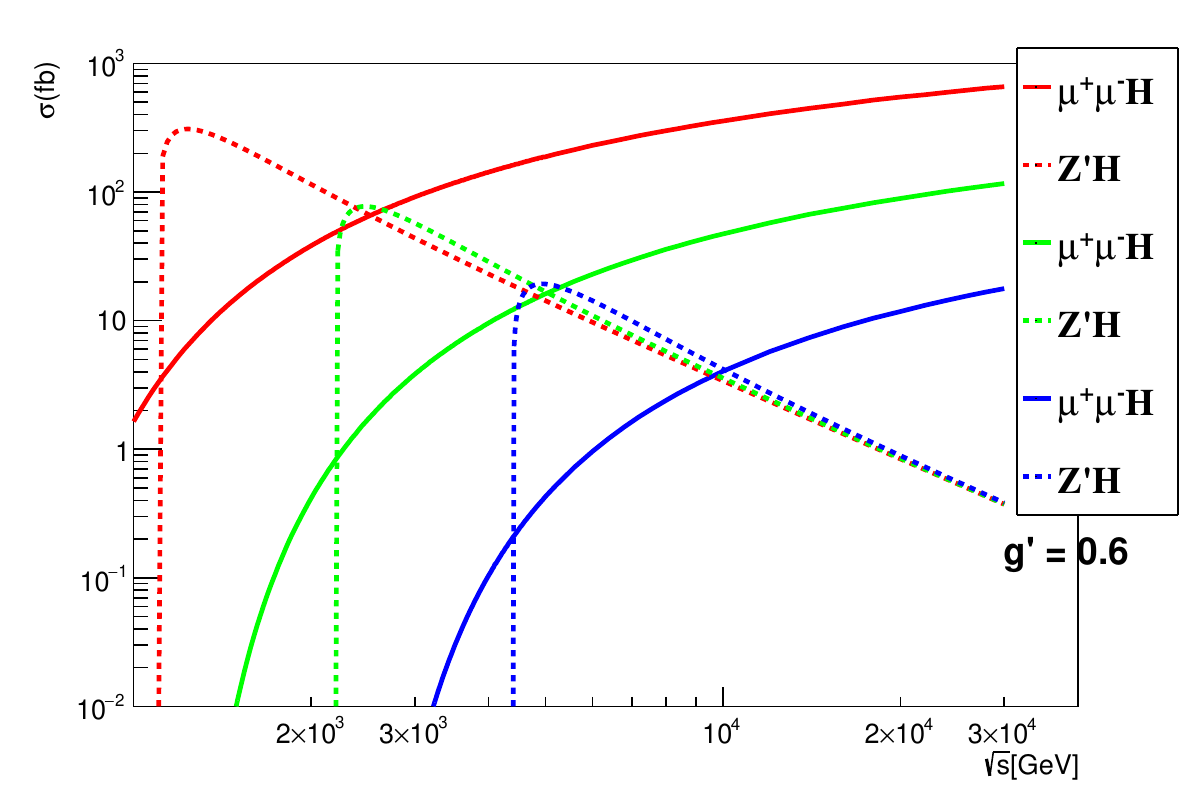}
	\end{center}
	\caption{Cross sections of the heavy Higgs-strahlung $\mu^+\mu^-\to Z'H$ (dashed lines)  and the $Z'Z'$-fusion $\mu^+\mu^-\to\mu^+\mu^-H$ (solid lines) process. Up-left panel: cross sections as a function of heavy Higgs mass $m_H$. Up-right panel: cross sections as a function of new gauge boson mass $m_{Z'}$. In these up panels, the red, green and blue lines are the results at the 3 TeV, 10 TeV and 30 TeV muon collider, respectively. In the down panel, the red, green and blue lines are the results for $(m_H,m_{Z'})=$(600 GeV, 500 GeV), (1200 GeV, 1000 GeV), and (2400 GeV, 2000 GeV).} 
	\label{fig3}
\end{figure}

Similar to the SM Higgs, there are also two production mechanisms of the heavy Higgs at the muon collider: the heavy Higgs-strahlung $\mu^+\mu^-\to Z'H$  and the $Z'Z'$-fusion $\mu^+\mu^-\to \mu^+\mu^-H$ processes. In Figure \ref{fig2}, we show the corresponding Feynman diagrams of these processes.  Cross sections of the $\mu^+\mu^-\rightarrow Z'H, \mu^+\mu^-H$ processes at the multi-TeV muon collider are shown in Figure \ref{fig3}, which are sensitive to the mass of heavy Higgs $m_H$ and the mass of new gauge boson $m_{Z'}$. In the up-left panel of Figure \ref{fig3}, we fix $m_{Z'}=500$ GeV, $g'=0.6$, and illustrate the impact of parameter $m_H$. We report that for both $\mu^+\mu^-\to Z'H$ and $\mu^+\mu^-\to \mu^+\mu^-H$ processes, increasing the heavy Higgs mass will lead to a smaller cross section. At the 3 TeV muon collider, the cross section of $\mu^+\mu^-\to \mu^+\mu^-H$ is larger when $m_H\lesssim1.3$ TeV, while in the mass range of 1.3 TeV$\sim$2.4 TeV, the cross section of $\mu^+\mu^-\to Z'H$ becomes larger. At the 10 TeV (30 TeV) muon collider, the cross section of $\mu^+\mu^-\to \mu^+\mu^-H$ is typically two (three) orders of magnitude  larger than that of $\mu^+\mu^-\to Z'H$ for the benchmark.

In the up-right panel of Figure \ref{fig3}, we then show the dependence of the cross sections on parameter $m_{Z'}$ by fixing $m_H=600$ GeV and $g'=0.6$. For the $Z'Z'$-fusion process $\mu^+\mu^-\to \mu^+\mu^-H$, the cross section quickly decreases as $m_{Z'}$ increases. However, for the $s$-channel heavy Higgs-strahlung $\mu^+\mu^-\to Z'H$ process, the cross section becomes larger when $Z'$ is heavier, which is mainly due to the  cancellation in the propagator. The maximum value of $\sigma(\mu^+\mu^-\to Z'H)$ is achieved slightly below the threshold $m_{Z'}+m_H<\sqrt{s}$. This indicates that for the considered  $Z'Z'$-fusion process $\mu^+\mu^-\to \mu^+\mu^-H$, a relatively light $Z'$ is favored. For example, the $Z'Z'$-fusion process $\mu^+\mu^-\to \mu^+\mu^-H$ is the dominant channel when $m_{Z'}\lesssim2$ TeV at the 10 TeV muon collider. It should be mentioned that above the threshold $m_{Z'}+m_H>\sqrt{s}$, the heavy Higgs-strahlung $\mu^+\mu^-\to Z'H$ process is kinematically forbidden, but the $\mu^+\mu^-\to \mu^+\mu^-H$ is still viable. In such a case, the $Z'Z'$-fusion process remains promising when the corresponding cross section is not too small.

In the down-panel of Figure \ref{fig3}, we show the cross section of the $\mu^+\mu^-\to Z'H, \mu^+\mu^-H$ process as a function of the center of collision energy $\sqrt{s}$. After sharply reaching the maximum value around the threshold $\sqrt{s}\gtrsim m_{Z'}+m_H$, the cross section of the heavy Higgs-strahlung $\mu^+\mu^-\to Z'H$ clearly decreases as $1/s$ at high energies. On the other hand, the cross section of the $Z'Z'$-fusion process $\mu^+\mu^-\to \mu^+\mu^-H$ increases logarithmically with $s$. For the benchmark with $m_H=600$ GeV and $m_{Z'}=500$ GeV, we find that the $Z'Z'$-fusion process is already the dominant contribution at the 3 TeV muon collider. When the masses of the new gauge boson and heavy Higgs become larger, the collision energy that the cross section of $Z'Z'$-fusion is larger than that of the heavy Higgs-strahlung  will also increase.

\begin{figure}
	\begin{center}
		\includegraphics[width=0.33\linewidth]{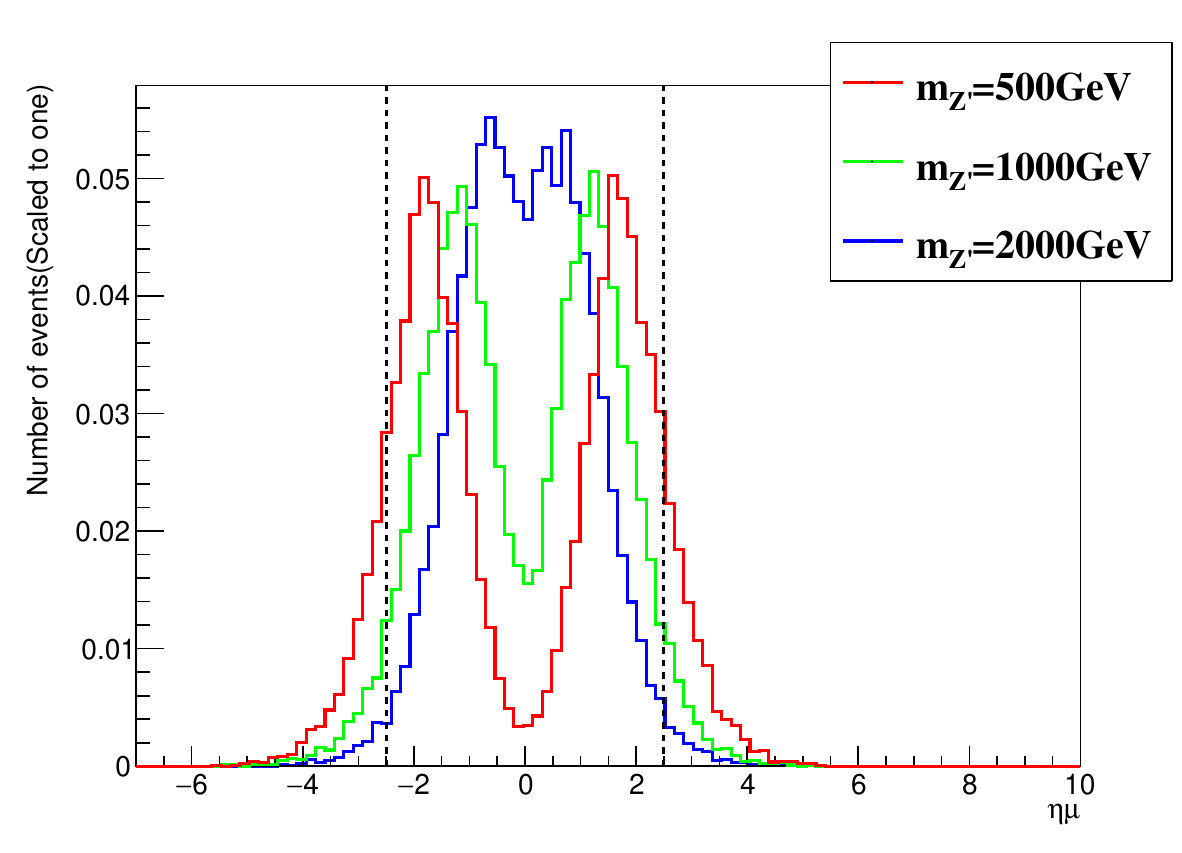}
		\includegraphics[width=0.33\linewidth]{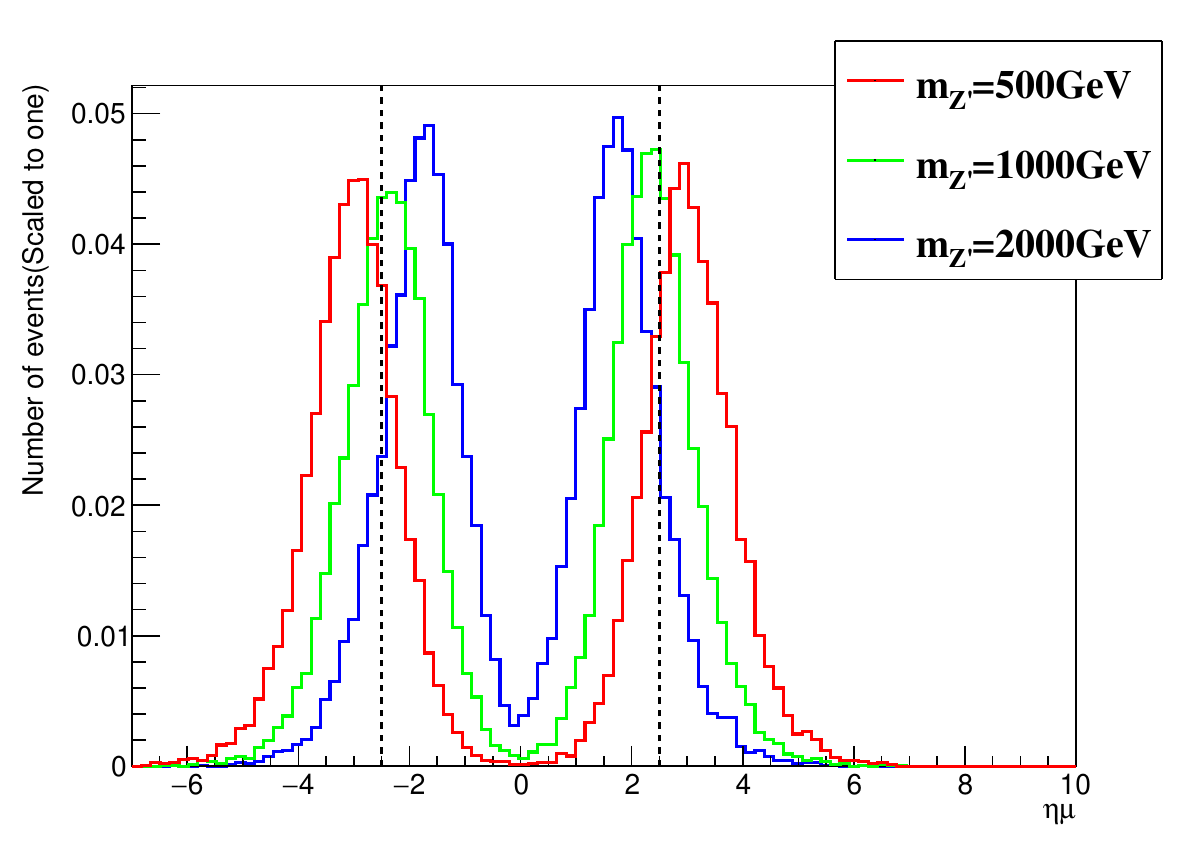}
		\includegraphics[width=0.33\linewidth]{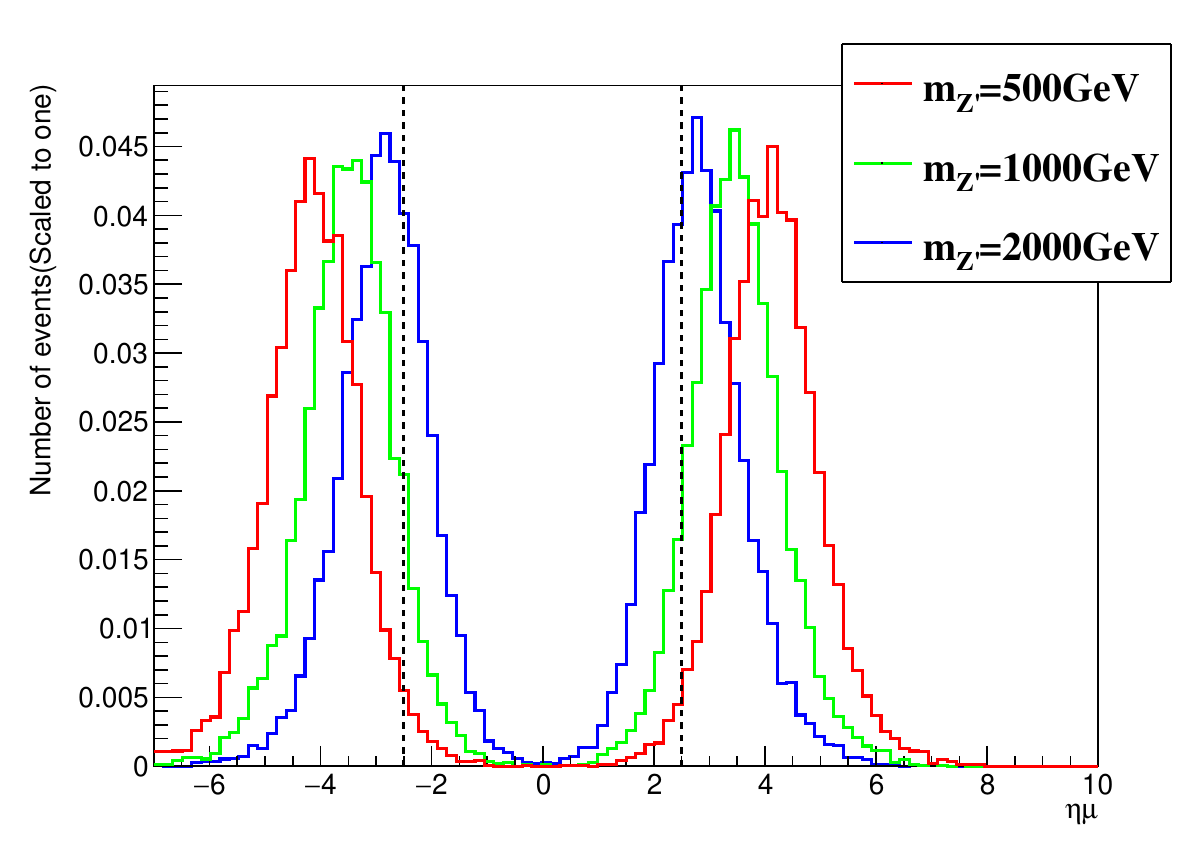}
	\end{center}
	\caption{Normalized distributions of pseudorapidity of final state muons $\eta(\mu)$ in the process $\mu^+\mu^-\to \mu^+\mu^- H$  at the 3 TeV (left), 10 TeV (middle) and 30 TeV (right) muon collider. The benchmark points are chosen as $m_{Z'}=$ 500~GeV, 1000 GeV and 2000 GeV  with $m_H=600$ GeV.}
	\label{fig4}
\end{figure}

It is well known that the final state muons in the vector boson fusion processes, such as the $ZZ$-fusion $\mu^+\mu^-\to \mu^+\mu^-h$, are typically in the forward region  ($|\eta(\mu)|>2.5$) \cite{Li:2024joa,Ruhdorfer:2024dgz}. Therefore, the forward muon detectors are strongly motivated during the construction of the muon collider \cite{Accettura:2023ked}.  For the new gauge boson $Z'Z'$-fusion process $\mu^+\mu^-\to \mu^+\mu^- H$, distributions of the final state muons might be different, as the mass of the new gauge boson $m_{Z'}$ is unknown. In Figure \ref{fig4}, we show the parton level normalized distributions of pseudorapidity of final state muons $\eta(\mu)$ in the process $\mu^+\mu^-\to \mu^+\mu^- H$ at the muon collider. We report that at the 3 TeV muon collider, most of the final state muons of the benchmark points are actually within the detection region of the main detector, i.e., $|\eta(\mu)|<2.5$. It is obvious that increasing the mass of the new gauge boson, the distribution of $|\eta(\mu)|$ tends to be smaller. Meanwhile, increasing the collision energy will lead to the growth of $|\eta(\mu)|$ for a fixed value of $m_{Z'}$. Roughly, the peak value of $|\eta(\mu)|$ is outside the main detector when $m_{Z'}\lesssim\sqrt{s}/10$.

Further decays of the new gauge boson and heavy Higgs could generate various interesting signatures. For instance, a novel same-sign tetralepton signature originates from the heavy Higgs-strahlung processes as $\mu^+\mu^-\to Z'H\to NN+NN\to 4\mu^\pm +4J$ \cite{Zhang:2026kwz}. In this paper, we focus on the $Z'Z'$-fusion process $\mu^+\mu^-\to \mu^+\mu^-H$ followed by $H\to NN$. For simplicity, we consider a muon flavor heavy neutral lepton, i.e., $|V_{\mu N}|\neq0$, and $|V_{e N}|=|V_{\tau N}|=0$. Due to the Majorana nature, the pair production of heavy neutral leptons $NN$ could generate the direct lepton number violation signature $ \mu^\pm jj+\mu^\pm jj$, with the two heavy neutral leptons both decay into a $\mu^+$ or $\mu^-$ at the same time.  The full process of the lepton number violation signature at the muon collider is given as 
\begin{equation}
	\mu^+ \mu^- \rightarrow \mu^+ \mu^- H \rightarrow \mu^+ \mu^- N N \rightarrow \mu^+ \mu^- + \mu^\pm jj +\mu^\pm jj.
\end{equation}
At the TeV scale muon collider, the dijets from the heavy neutral lepton decays could be highly boosted and could merge into one fat-jet $J$. Hence, the signature is further regarded as $\mu^+ \mu^-\mu^\pm\mu^\pm JJ$. The benchmark collision energies of the muon collider are regarded as $\sqrt{s}=$ 3 TeV, 10 TeV, 30 TeV with an integrated luminosity of 1 ab$^{-1}$, 10 ab$^{-1}$, 90 ab$^{-1}$ \cite{Delahaye:2019omf}, respectively.

The corresponding SM backgrounds are,
\begin{align}
	{\mu^+}{\mu^-}&\rightarrow {\mu^+}{\mu^-}WW,\\ 
	{\mu^+}{\mu^-}&\rightarrow{\mu^+}{\mu^-}WWZ,\\
	{\mu^+}{\mu^-}&\rightarrow \mu^{\pm}{\nu}WWW.
\end{align}
Contributions of the ${\mu^+}{\mu^-}\rightarrow {\mu^+}{\mu^-}WW$ and ${\mu^+}{\mu^-}\rightarrow \mu^{\pm}{\nu}WWW$ are from the leptonic decay of $W$ boson with misidentified lepton charge, which are significantly suppressed by the misidentification rate $0.1\%$ \cite{ATLAS:2019jvq}. Contributions of the ${\mu^+}{\mu^-}\rightarrow{\mu^+}{\mu^-}WWZ$ process are from the missing muon by the detector when the gauge boson $W$ or $Z$ decays leptonicly.
 
In the simulation, we use the {\bf FeynRules} package \cite{Arun:2022ecj} to implement the $U(1)_{L_\mu-L_\tau}$ gauged extension of the type-I seesaw model. We then use {\bf Madgraph5\_aMC@NLO} \cite{Alwall:2014hca} to generate the leading order signal and background events at the parton level. After this, {\bf Pythia8} \cite{Sjostrand:2014zea} is used to do parton showering and hadronization. The detector simulation is performed by {\bf Delphes3} \cite{deFavereau:2013fsa} with the corresponding muon collider card. The fat-jets are reconstructed by using the Valencia algorithm with $R = 1.2$.

Although a forward muon detector is implemented in {\bf Delphes3 } \cite{deFavereau:2013fsa}, the detailed performance is currently not so reliable. It should be mentioned that the final state muons from the backgrounds are also in the forward region. If forward muons are considered,  more sophisticated cuts on the forward muons should be applied \cite{Li:2024joa,Ruhdorfer:2024dgz}, which definitely decreases the selection efficiency of the signal. Since the final state muons from the signal might not be in the forward region, we only focus on the muons that can be detected by the main detector, i.e., $|\eta(\mu)|<2.5$. Pre-selection cuts on transverse momentum and pseudorapidity of the muons and the fat-jets are applied as following
\begin{equation}
	P_T(\mu)>10 ~{\rm{GeV}}, P_T(J)>50 ~\text{GeV}, |\eta(\mu)|<2.5, |\eta(J)|<2.5.
\end{equation}
which is correspond to the default set of {\bf Madgraph5\_aMC@NLO }\cite{Alwall:2014hca}  and {\bf Delphes3 } \cite{deFavereau:2013fsa}. 

\begin{figure}
	\begin{center}
		\includegraphics[width=0.33\linewidth]{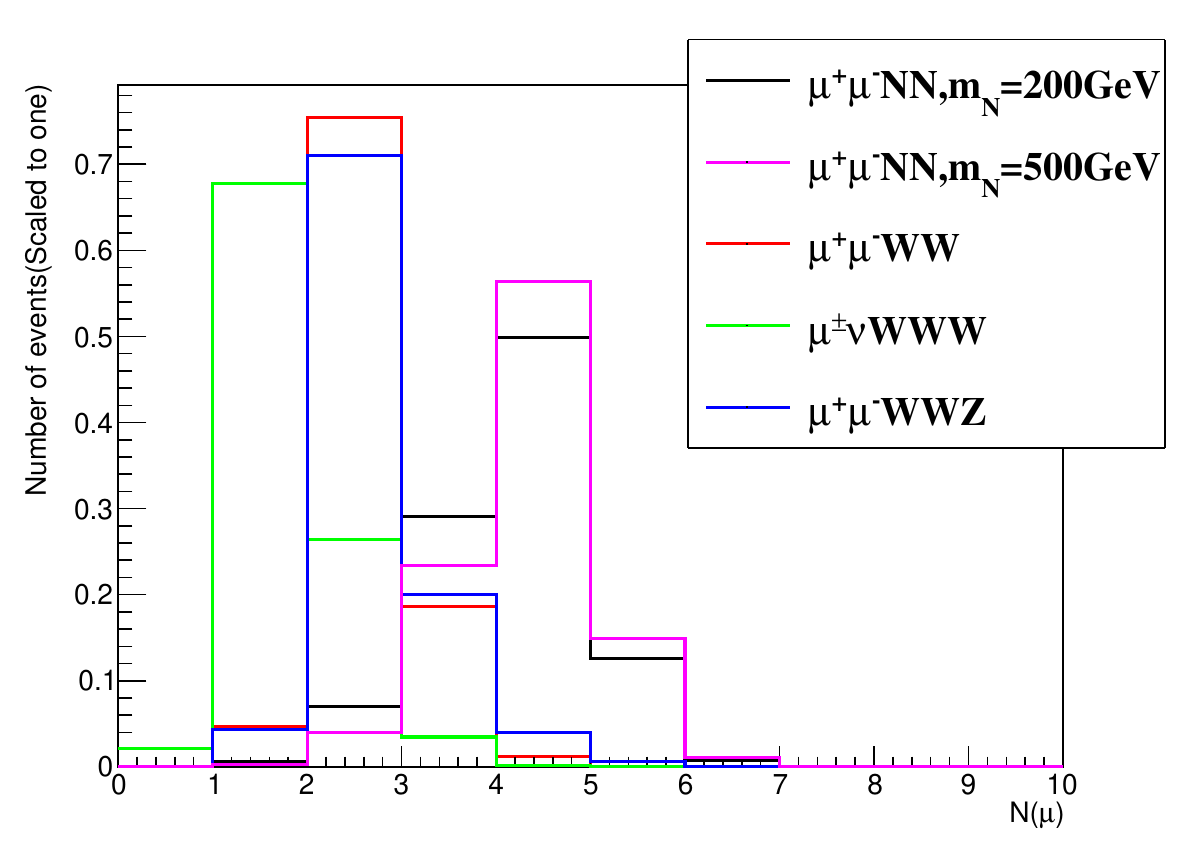}
		\includegraphics[width=0.33\linewidth]{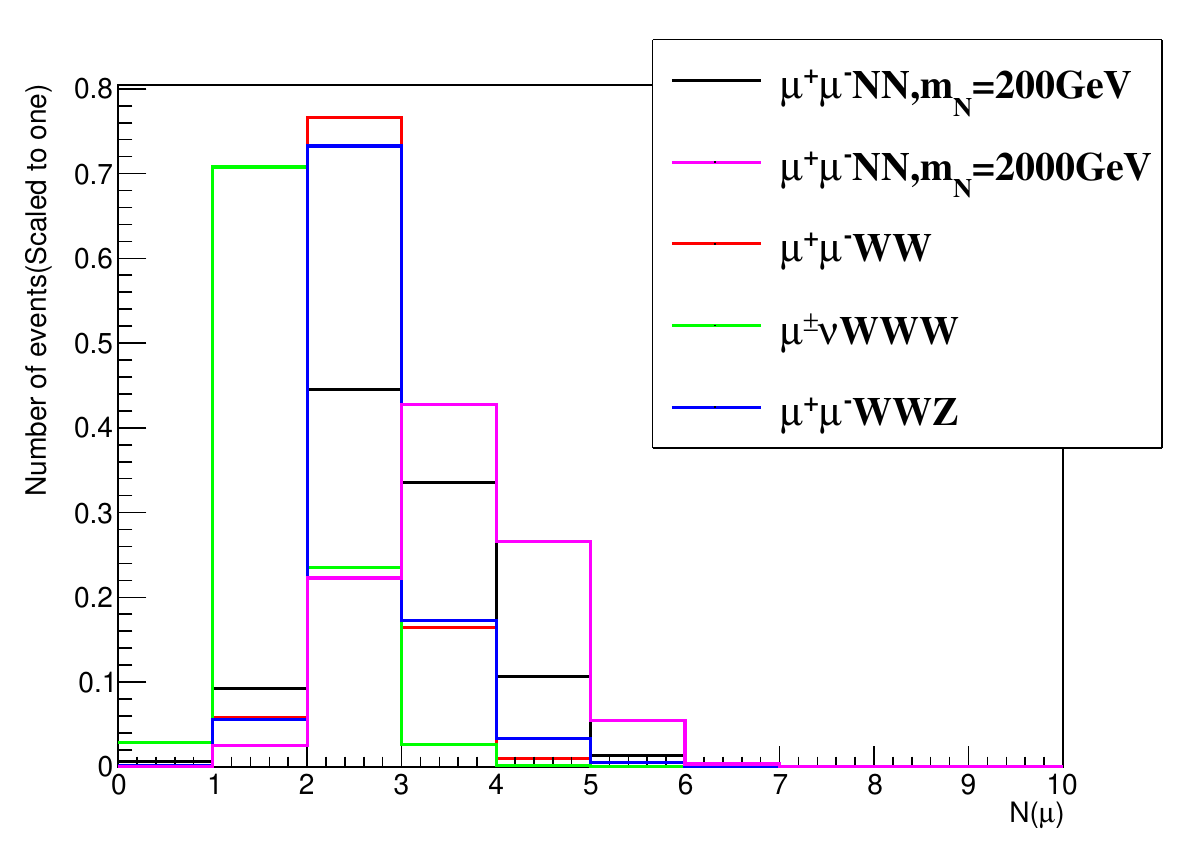}
		\includegraphics[width=0.33\linewidth]{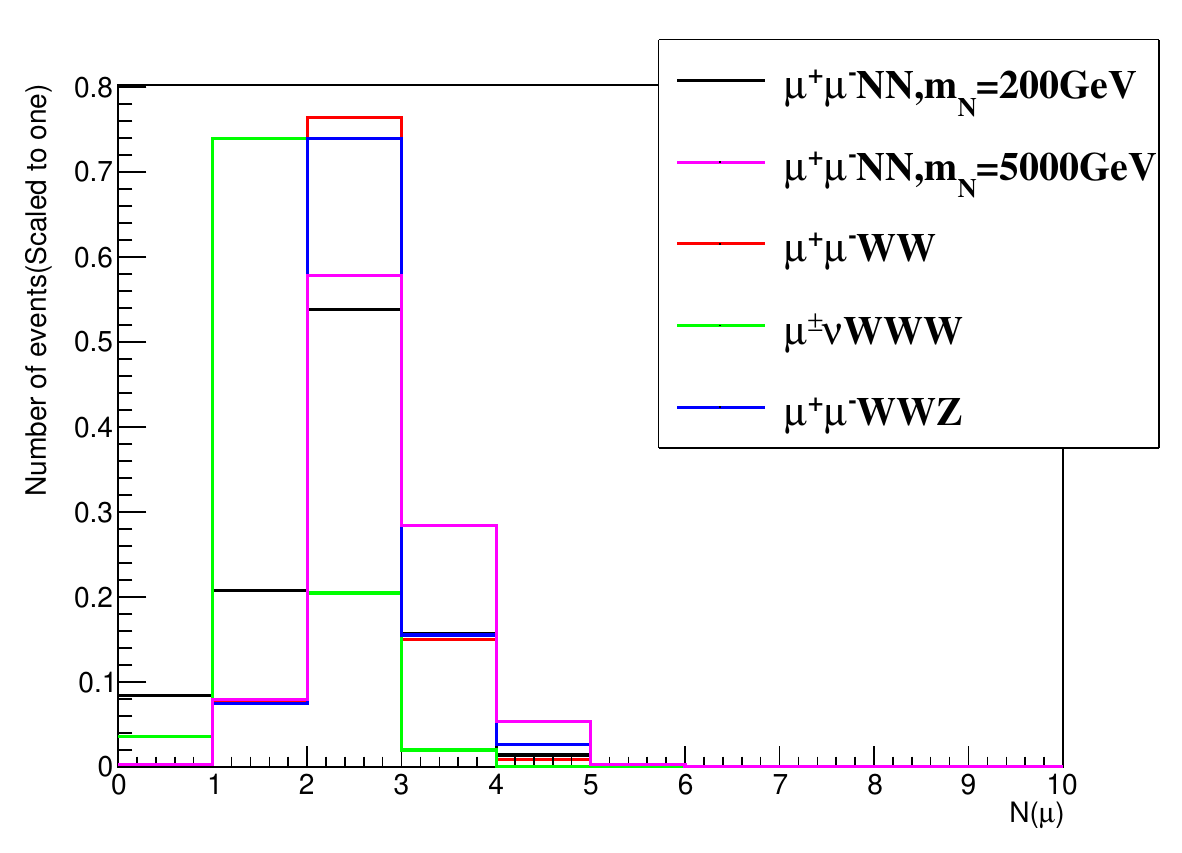}
		\includegraphics[width=0.33\linewidth]{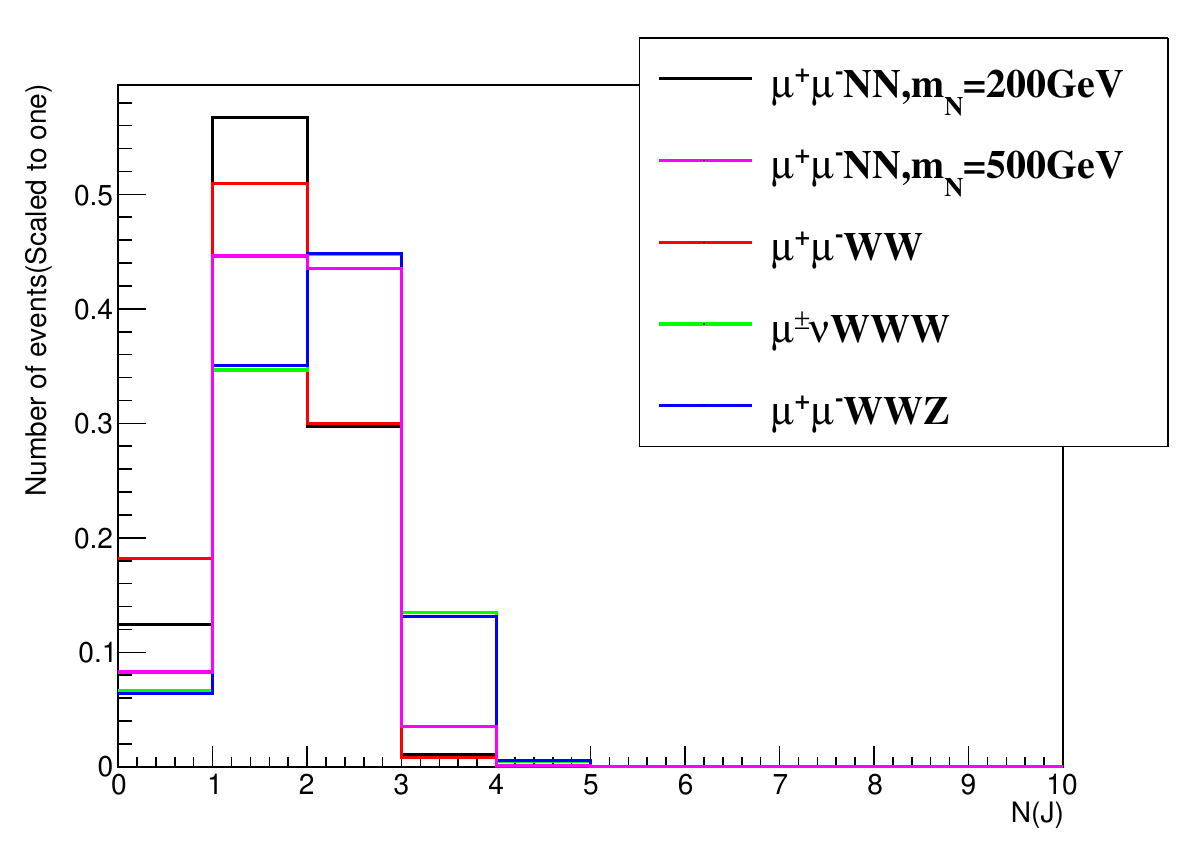}
		\includegraphics[width=0.33\linewidth]{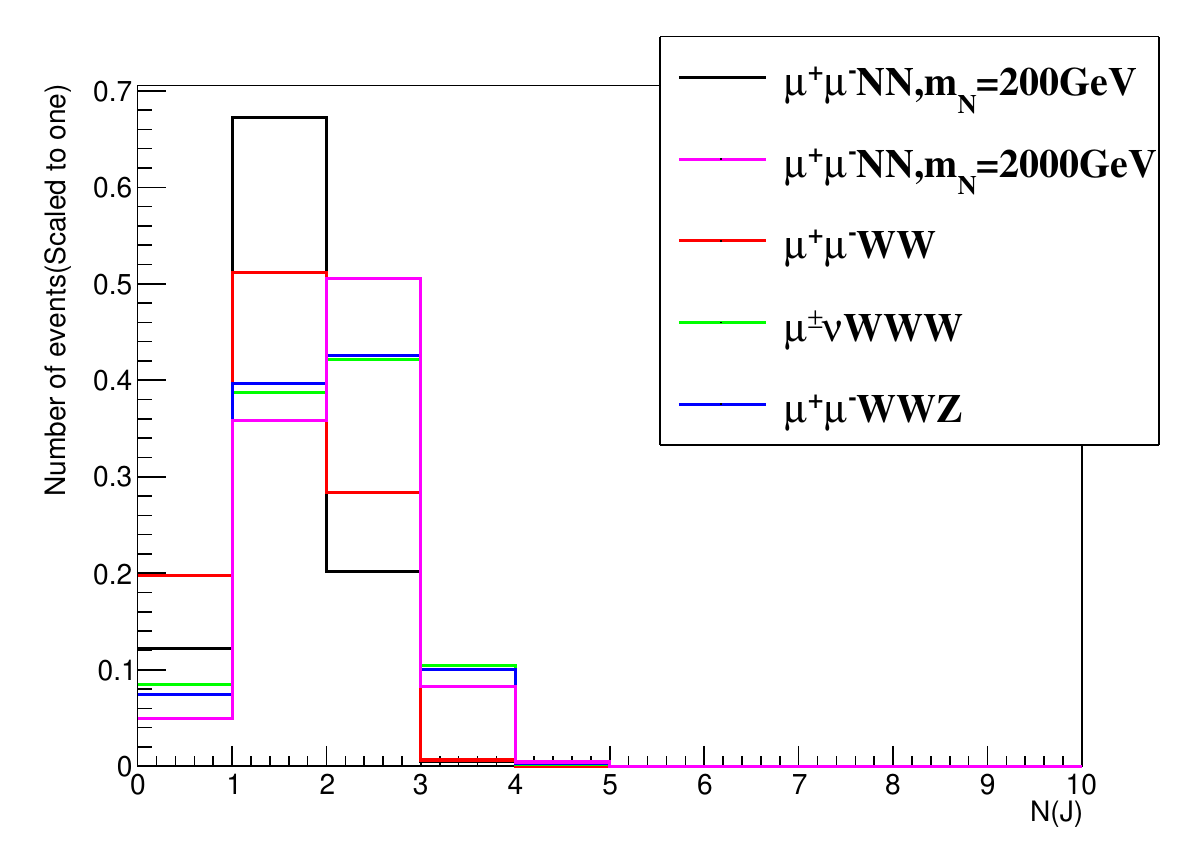}
		\includegraphics[width=0.33\linewidth]{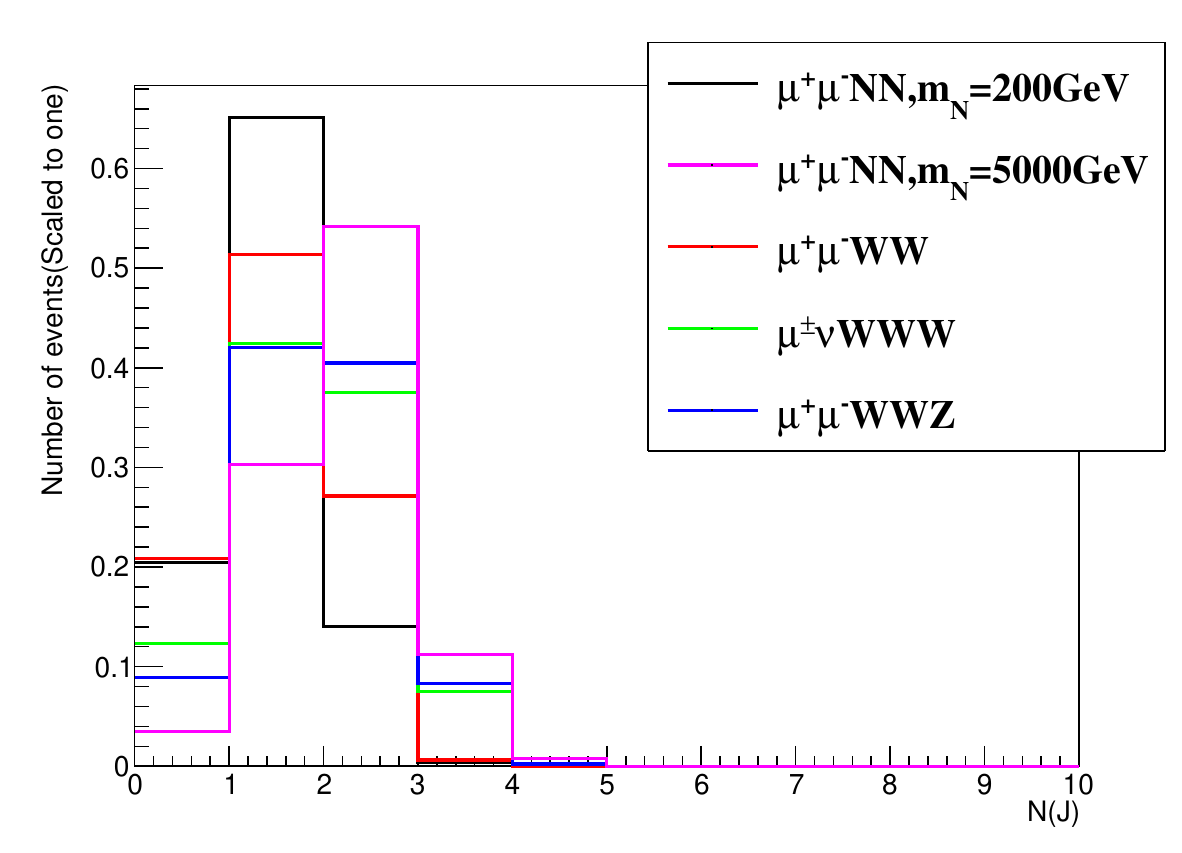}
		\includegraphics[width=0.33\linewidth]{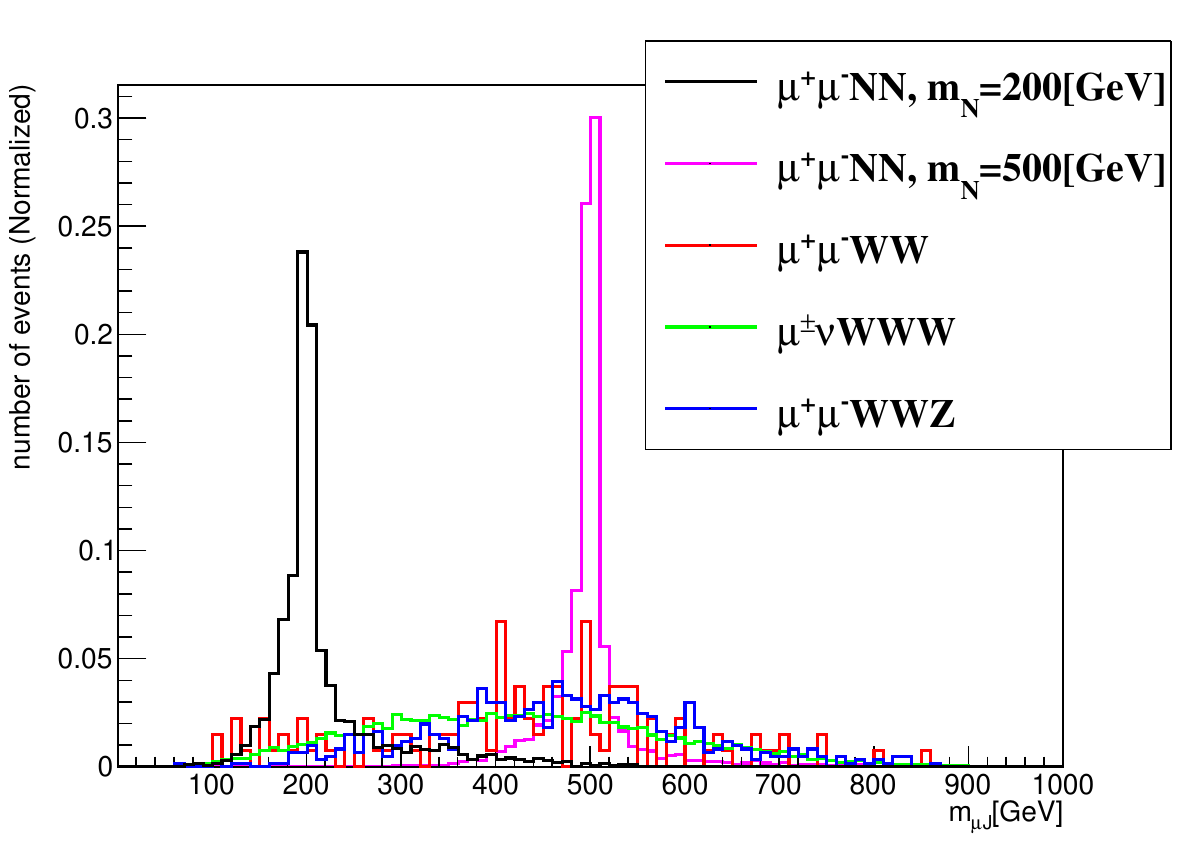}
		\includegraphics[width=0.33\linewidth]{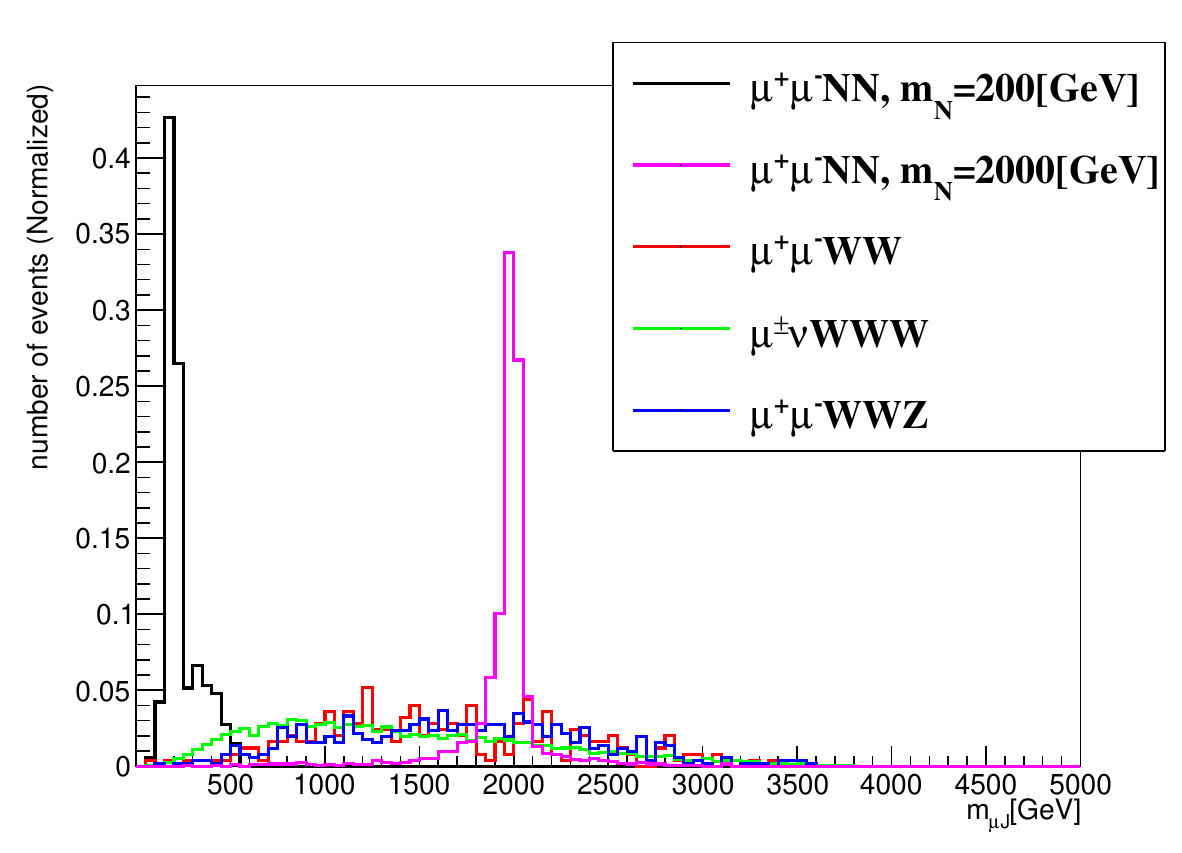}
		\includegraphics[width=0.33\linewidth]{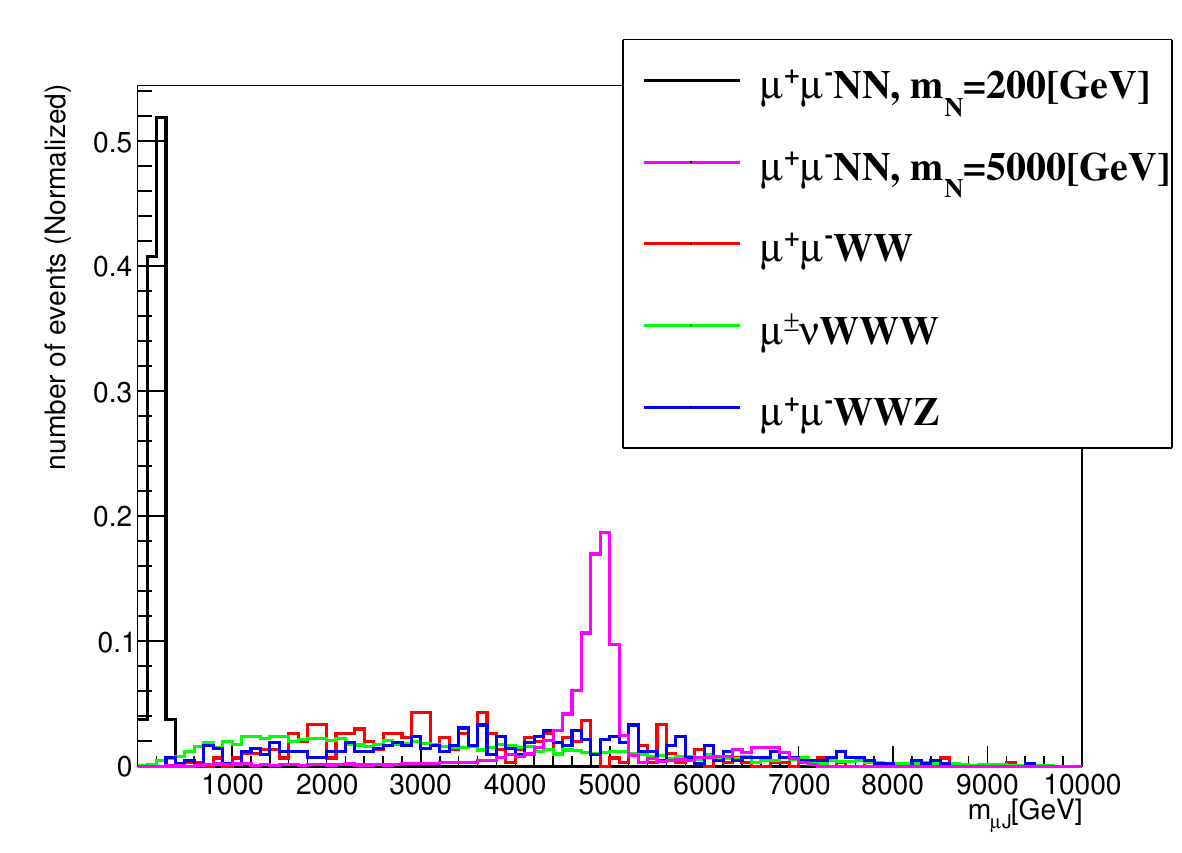}
	\end{center}
	\caption{Normalized distributions of number of muons $N(\mu)$ (up panels), number of fat-jets $N(J)$ (middle panels), and invariant mass of muon and fat-jet $m_{\mu J}$ (down panels) for the signal $\mu^+ \mu^- \mu^\pm\mu^\pm J J$ with heavy Higgs $H$ and corresponding backgrounds at the 3 TeV (left), 10 TeV (middle) and 30 TeV (right) muon collider. The benchmark points are chosen as $m_{Z'}=500$ GeV and $m_H=3\times m_N$.}
	\label{fig5}
\end{figure}

In Figure \ref{fig5}, we show the normalized distribution of variables both for the signal and background processes at the 3 TeV, 10 TeV and 30 TeV muon collider after the pre-selection cuts. Distributions of $N(\mu)$ from the backgrounds are independent of the collision energy. For instance, one muon final state is the dominant contribution from the $\mu^\pm\nu WWW$ process. Meanwhile, the $\mu^+\mu^-WW$ and $\mu^+\mu^- WWZ$ backgrounds mainly have two muons. 
According to the distributions of $\eta(\mu)$ in Figure  \ref{fig4}, we observe that at the 3 TeV muon collider, the signature is dominated by four muons. At the 10 TeV muon collider, the three muon signature becomes the dominant one for $m_N=2$ TeV, but it is two muons for $m_N=200$ GeV. Increasing the collision energy to 30 TeV, only the two muons from the heavy neutral leptons are within the main detector, so the two muons channel is always the dominant contribution. Since the number of detected muons from the signal depends on the relation between $m_{Z'}$ and $\sqrt{s}$, the dominant contribution of the signal varies when changing the related parameters. To facilitate the observation of the lepton number violation signature, all three kinds of channels are taken into account in our analysis. For the four muon channel, we require
\begin{equation} 
	N(\mu^\pm)=3, N(\mu^\mp)=1.
\end{equation}

For the three muon channel, we only consider the explicit  lepton number violation case with
\begin{equation} 
	N(\mu^\pm)=3,
\end{equation}
which can sufficiently suppress the backgrounds. The less exotic three muons channel as $\mu^\pm \mu^\pm \mu^\mp$ has the same cross section, but the corresponding backgrounds are much larger.

For the two muon channel, two same-sign muons are selected
\begin{equation} 
	N(\mu^\pm)=2.
\end{equation}
According to the distributions in Figure \ref{fig5}, there could be relatively large backgrounds for the two muon  channel. Therefore, we further apply the cut $P_T(\mu)<m_N$ to suppress the backgrounds.

Then, to make sure the $\mu^+ \mu^- \mu^\pm\mu^\pm J J$ signature comes from the heavy neutral lepton, cuts on fat-jets are required. In the middle panels of Figure \ref{fig5}, we observe that the one fat-jet event is dominant for the electroweak scale $m_N$. Increasing the heavy neutral lepton mass, the fraction of two fat-jets event becomes larger.  In the following analysis, we require at least one fat-jet in the final states,
\begin{align}
	N (J)\geq1.
\end{align} 

The mass of the heavy neutral lepton $m_N$ can be reconstructed by the invariant mass of $m_{\mu J}$. As shown in the down panels of Figure \ref{fig5}, the distributions of $m_{\mu J}$ have a sharp peak at $m_N$ for the signal, while the distributions of $m_{\mu J}$ are mild for the backgrounds.  Therefore,  $m_{\mu J}$ is further used to select events in the mass range of
\begin{align}\label{cut03}
	0.8m_N<m_{\mu{J}}<1.2m_N.
\end{align}

In summary, the selection cuts on muons can effectively suppress the backgrounds much smaller than the signal. After the muon cuts, the dominant background is from the $\mu^+\mu^-WWZ$ process. Cuts on fat-jet and on $m_{\mu J}$ are a supplement of the muon cuts. We use it in a conservative way, aiming to reduce the backgrounds while preserving the signal as much as possible.

\begin{table}
	\begin{center}
		\begin{tabular}{c | c | c | c | c | c} 
			\hline
			\hline
			\multicolumn{2}{c|}{$\mu^+\mu^- H\to \mu^+ \mu^- \mu^\pm\mu^\pm J J$} & After Selection (fb)& Backgrounds (fb) & Significance & $5\sigma$ Luminosity (fb$^{-1}$) \\
			\hline
			\multirow{3}{*}{3 TeV}	& $3\mu^\pm + \mu^\mp$   &6.15   &$5.97\times10^{-4}$    & $3.08\times10^2$   &1.35   \\			
			\cline{2-6}
			& $3 \mu^\pm$   &1.57   &$2.71\times10^{-4}$   & $1.42\times10^2$    &5.29  \\			
			\cline{2-6}
			& $2 \mu^\pm$   & 0.42    &$3.10\times10^{-2}$   & $3.94\times10^1$    &76.5  \\			
			\hline
			\hline
			\multirow{3}{*}{10 TeV}	& $3\mu^\pm + \mu^\mp$   &2.75  &$5.53\times10^{-4}$   & $6.42\times10^2$   &5.62   \\			
			\cline{2-6}
			& $3 \mu^\pm$   &3.91     &$1.92\times10^{-4}$    &$8.35\times10^2$    &2.66  \\			
			\cline{2-6}
			& $2 \mu^\pm$   &7.97   &$1.80\times10^{-3}$    &$1.09\times10^3$  &3.15  \\			
			\hline
			\hline
			\multirow{3}{*}{30 TeV}	& $3\mu^\pm + \mu^\mp$   & 0.28  &$1.53\times10^{-4}$   &$5.72\times10^2$   &82.5  \\			
			\cline{2-6}
			& $3 \mu^\pm$   & 0.61   &$5.13\times10^{-5}$   &$9.58\times10^2$   &23.7 \\			
			\cline{2-6}
			& $2 \mu^\pm$  &12.6    &$1.93\times10^{-4}$   &$4.28\times10^3$   &2.26   \\			
			\hline
			\hline
		\end{tabular}
	\end{center}
	\caption{ Results for the $\mu^+ \mu^- \mu^\pm\mu^\pm J J$ signature with heavy Higgs $H$ and backgrounds at the muon collider. The benchmark point is selected as $m_N=200~\rm{GeV}$, $m_H=3\times m_N$, $m_{Z'}=500~\rm{GeV}$ and $g'=0.6$.  }
	\label{Tab01}
\end{table}

In Table \ref{Tab01}, we show the results for the $\mu^+ \mu^- \mu^\pm\mu^\pm J J$ signature with heavy Higgs $H$. The benchmark point is fixed as $m_N=200~\rm{GeV}$, $m_H=3\times m_N$, $m_{Z'}=500~\rm{GeV}$ and $g'=0.6$. Here, the significance is calculated by using the algorithm proposed in Ref.~\cite{Cowan:2010js} 
\begin{align}
	S=\sqrt{2\left[(N_S+N_B)\log\left(1+\frac{N_S}{N_B}\right)-N_S\right]},
\end{align}
where $N_S$ and $N_B$ are the event numbers of signal and backgrounds, respectively. The precision of the signature is then calculated through $\sqrt{N_S+N_B}/N_S$ \cite{Forslund:2022xjq}. We notice that the lepton number violation signatures from the SM background processes at the muon collider are rare events. When the derived event number of backgrounds $N_B$ after all selection cuts is less than one, we will take $N_B$ as one for a conservative estimation.

As expected from the distributions of $N(\mu)$ in Figure \ref{fig5}, the four muon  channel $3\mu^\pm+\mu^\mp$ is the dominant contribution at the 3 TeV muon collider. After all the selection cuts, the cross section of the $3\mu^\pm+\mu^\mp$ channel is 6.15 fb, leading to a significance of $3.08\times10^2$ with 1 ab$^{-1}$ data. Because the backgrounds are tiny, a luminosity of 1.35 fb$^{-1}$ is enough to discover this benchmark point. For the three muons channel $3\mu^\pm$, the cross section is smaller than that of the $3\mu^\pm+\mu^\mp$ channel, but it is at the same order of magnitude. So 5.29 fb$^{-1}$ data is required to discover the benchmark through the three muon  channel $3\mu^\pm$. On the other hand, the cross section of the two muon  channel $2\mu^\pm$ is much smaller at the 3 TeV muon collider. However, the corresponding backgrounds become much larger. The two muon  channel $2\mu^\pm$ could reach a significance of $3.94\times10^1$, which is approximately an order of magnitude smaller than that of the four muon  channel $3\mu^\pm+\mu^\mp$. With an integrated luminosity of 1 ab$^{-1}$ at the 3 TeV muon collider, the precision of the $3\mu^\pm+\mu^\mp$, $3\mu^\pm$, and  $2\mu^\pm$ channels are 1.3\%, 2.5\%, and 5.1\%, respectively.

We find that the cross sections of all these three channels are around a few fb after the selection cuts at the 10 TeV muon collider. So all these three channels could reach the $5\sigma$ discovering limit with less than 10~fb$^{-1}$ data. Since the corresponding backgrounds are the cleanest, the three muon  channel $3\mu^\pm$ is the most promising one for discovering the benchmark. The two muon  channel has the largest signal cross section, which leads to the largest significance of $1.09\times10^3$ with full luminosity. The precision of the $3\mu^\pm+\mu^\mp$, $3\mu^\pm$, and  $2\mu^\pm$ channels are 0.60\%, 0.51\%, and 0.35\% for an integrated luminosity of 10 ab$^{-1}$ at the 10 TeV muon collider.

The results become totally different at the 30 TeV muon collider. The two muon  channel $2\mu^\pm$ has a cross section of 12.6 fb after the selection cuts, while the cross sections of the other two channels are less than one fb. Hence, the two muon channel $2\mu^\pm$ is the most promising one, which could be discovered with 2.26 fb$^{-1}$ luminosity for the benchmark. The four muon  $3\mu^\pm+\mu^\mp$ and three muon  $3\mu^\pm$ channels are less promising, but these two channels still can reach the 5$\sigma$ discovery limit with less than 100 fb$^{-1}$ data. With an integrated luminosity of 90 ab$^{-1}$ at the 30 TeV muon collider, we obtain a precision of 0.63\%, 0.43\%, and 0.094\% for the $3\mu^\pm+\mu^\mp$, $3\mu^\pm$, and  $2\mu^\pm$ channels, respectively. The great improvement in the precision for the two muon  channel is mainly due to the increasing luminosity at the 30 TeV stage.

\begin{figure}
	\begin{center}
		\includegraphics[width=0.33\linewidth]{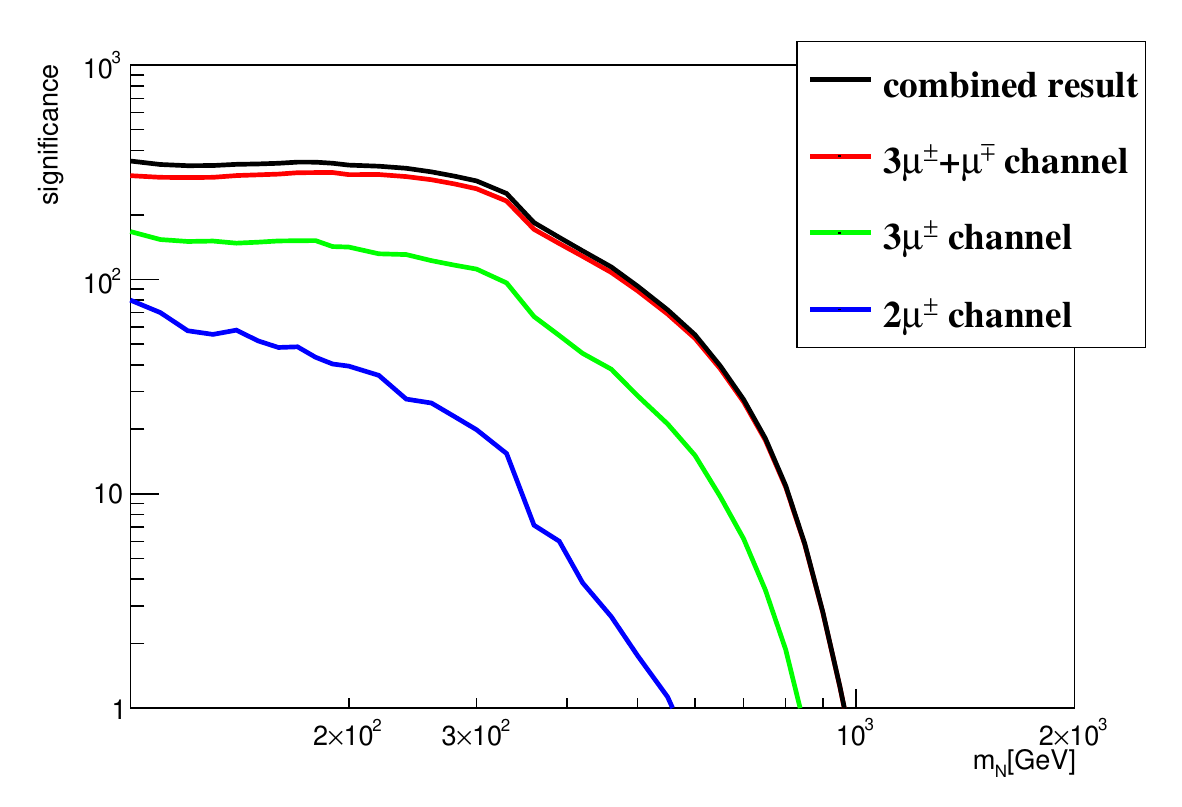}		
		\includegraphics[width=0.33\linewidth]{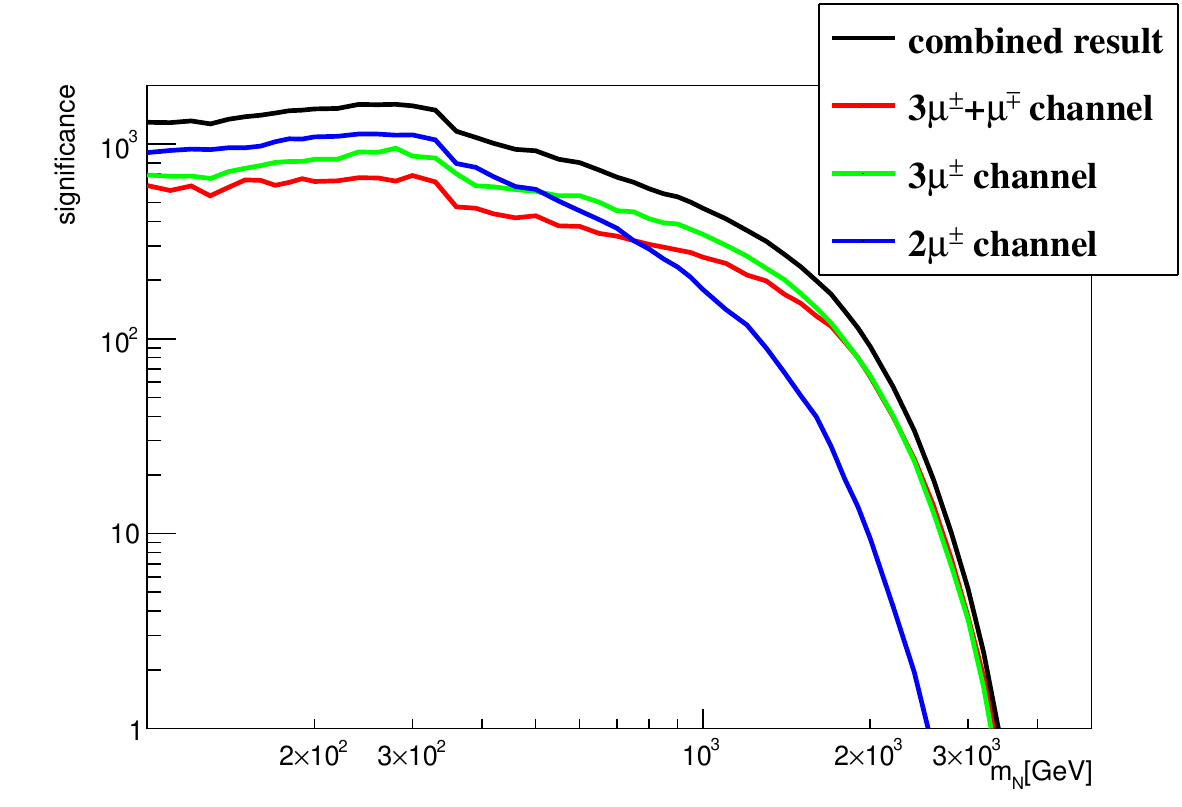}
		\includegraphics[width=0.33\linewidth]{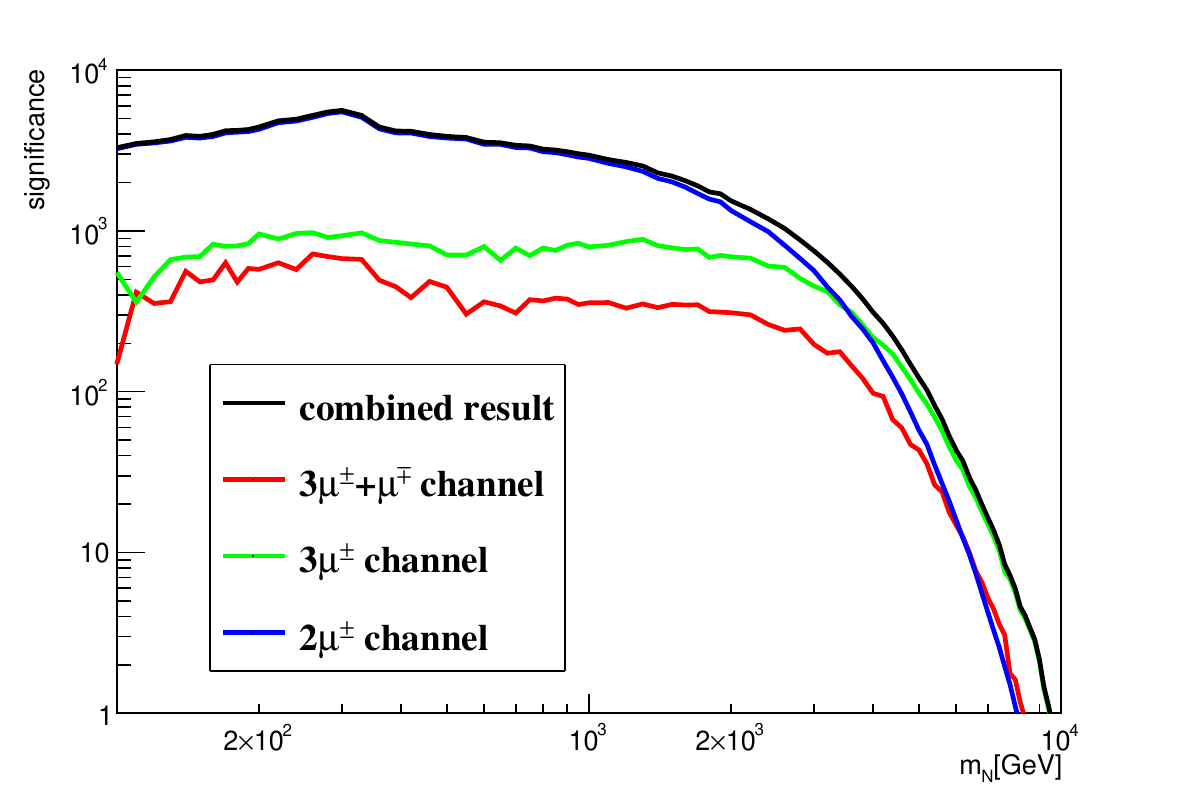}
		\includegraphics[width=0.33\linewidth]{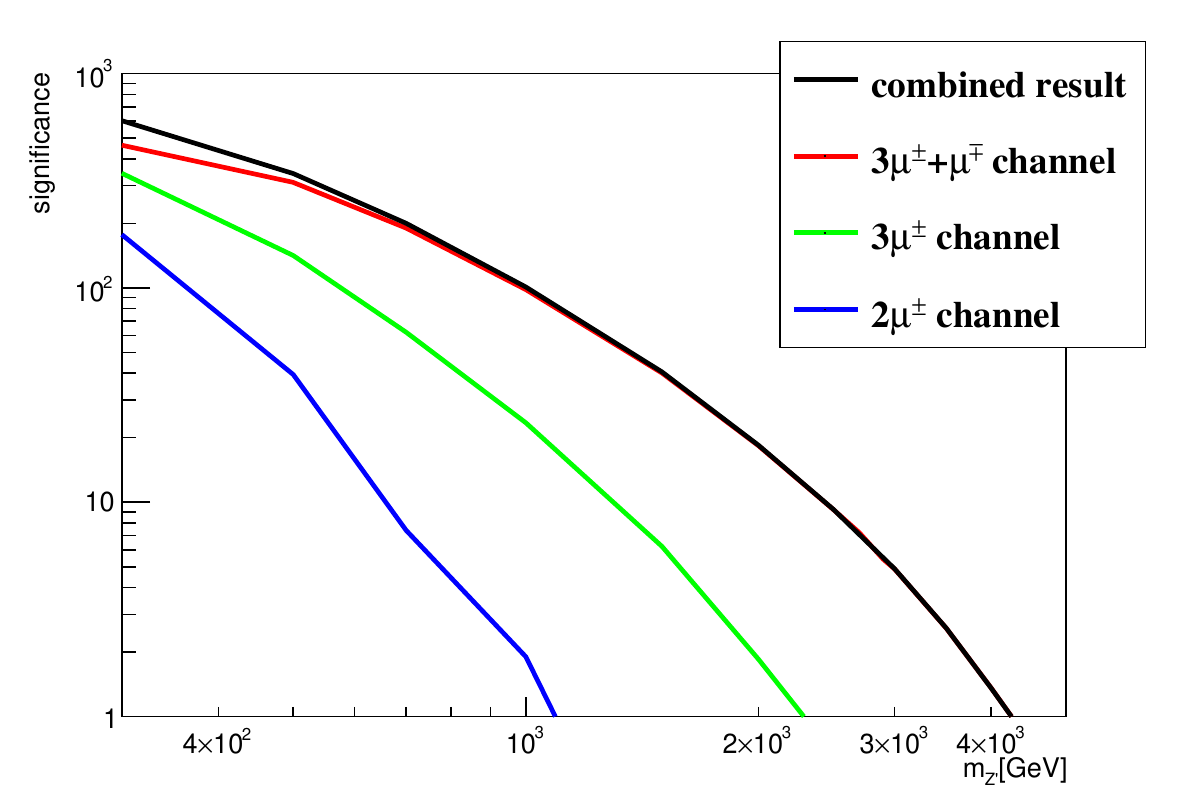}
		\includegraphics[width=0.33\linewidth]{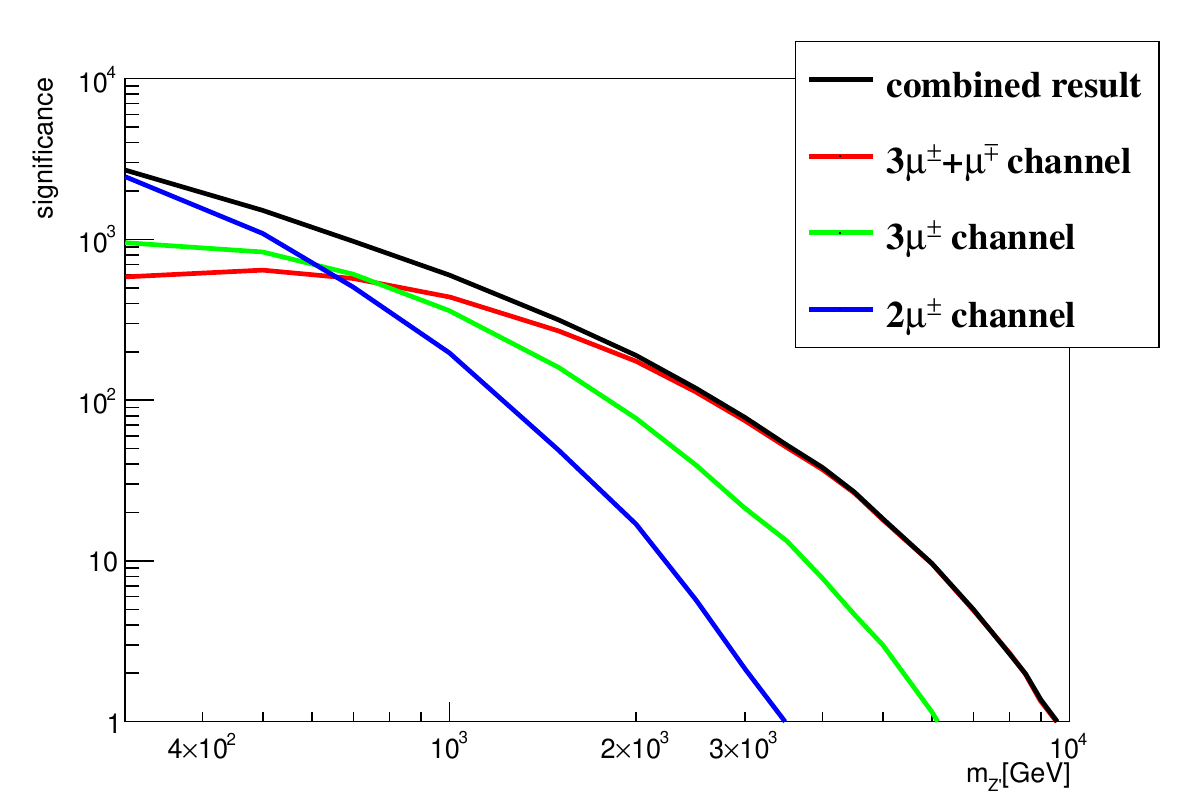}	
		\includegraphics[width=0.33\linewidth]{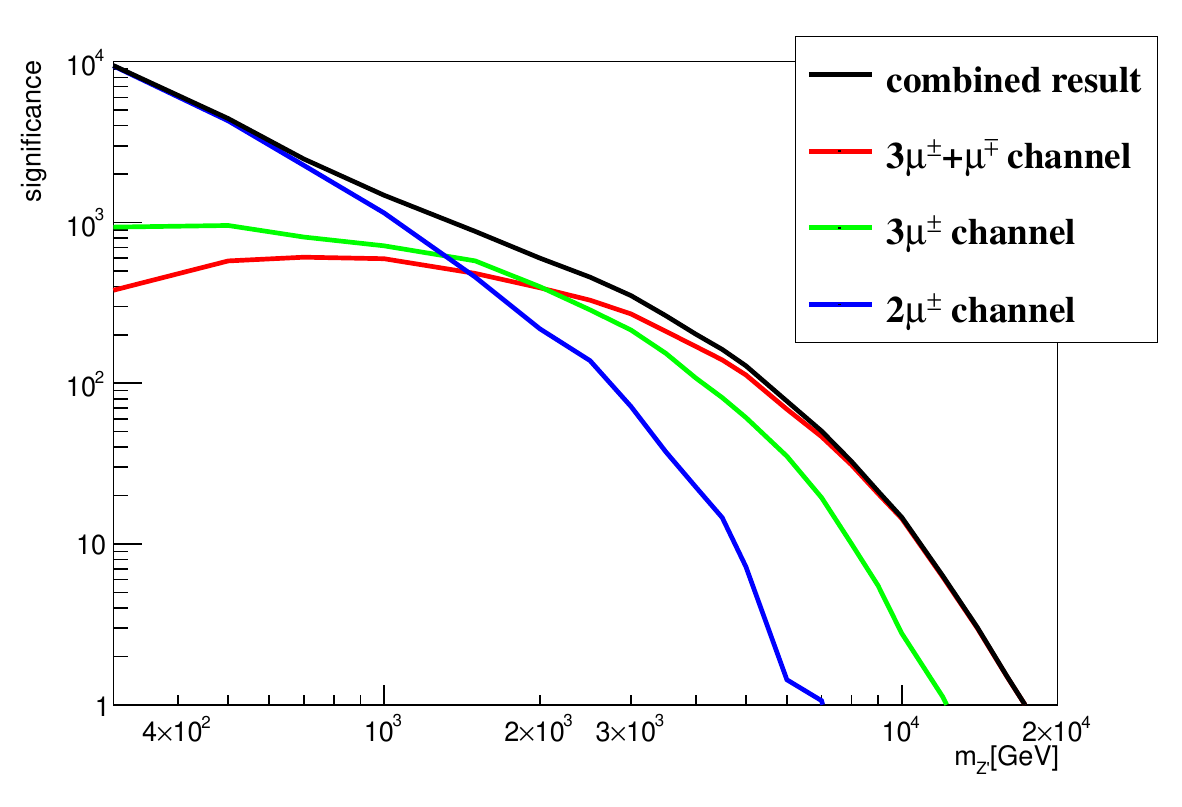}
	\end{center}
	\caption{The significance of the $\mu^+ \mu^- \mu^\pm\mu^\pm J J$ signature with heavy Higgs $H$ at the 3 TeV (left panel), 10 TeV (middle panel) and 30 TeV (right panel) muon collider  with the fixed mass relation of $m_H=3\times m_N$ and $g'=0.6$. In the up panels we have set $m_{Z'}=500$ GeV and in the down panels we have set $m_{N}=200$ GeV. The red, green, and blue lines are the results for the $3\mu^\pm+\mu^\mp$, $3\mu^\pm$, and  $2\mu^\pm$ channels, respectively. The black lines are the combined results of these three channels.}
	\label{fig6}
\end{figure}

Based on the analysis of the benchmark point in Table \ref{Tab01}, we then explore the impact of relevant parameters. Results are shown in Figure \ref{fig6}, where the combined significance is calculated as
\begin{equation}
	S_\text{comb}=\sqrt{S_{3\mu^\pm+\mu^\mp}^2+S_{3\mu^\pm}^2+S_{2\mu^\pm}^2}.
\end{equation}
Qualitatively speaking, increasing the masses of heavy neutral lepton $m_N$ or new gauge boson $m_{Z'}$ will lead to the decreasing of the significance. It is obvious that the four muon  channel $3\mu^\pm+\mu^\mp$ is always the most promising one at the 3 TeV muon collider. The two muon  channel $2\mu^\pm$ can only be discovered with $m_N$ below about 400 GeV or $m_{Z'}$ below about 800 GeV at the 3~TeV stage. The two muon  channel $2\mu^\pm$ is the dominant one under the circumstance of relatively small $m_{N}$ or $m_{Z'}$ at the 10 TeV and 30 TeV muon collider.

For the benchmark scenario with $m_{Z'}=$500~GeV, we find that the three muon  channel  $3\mu^\pm$ becomes the dominant contribution when $m_{N}\gtrsim$500 GeV at the 10~TeV muon collider. Above 2 TeV of $m_N$, the $3\mu^\pm$ and  $3\mu^\pm+\mu^\mp$ channels would have similar significance. At the 30 TeV muon collider, the $2\mu^\pm$ channel dominant region could extend to $m_N\lesssim4$ TeV. Meanwhile, the significance of the $3\mu^\pm+\mu^\mp$ channels is always smaller than that of the $3\mu^\pm$ channel.

In the benchmark scenarios with varying $m_{Z'}$ at the 10 TeV and 30 TeV muon collider, we report that the $2\mu^\pm$ channel is dominant in the light $Z'$ region, while the $3\mu^\pm+\mu^\mp$ channel is dominant in the heavy $Z'$ region. The $3\mu^\pm$ channel could become the dominant one only in a quite narrow region, e.g., $m_{Z'}\sim700$~GeV at the 10 TeV  stage. 

\begin{figure}
	\begin{center}
		\includegraphics[width=0.45\linewidth]{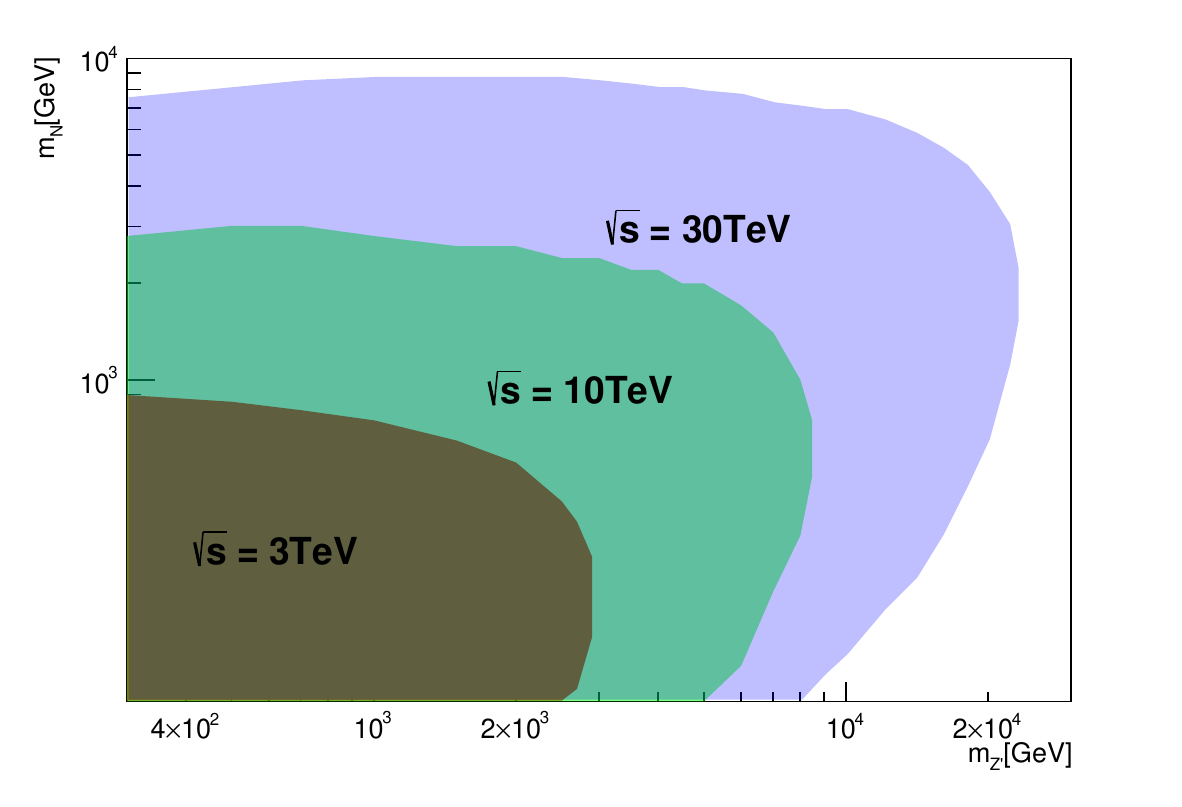}
		\includegraphics[width=0.45\linewidth]{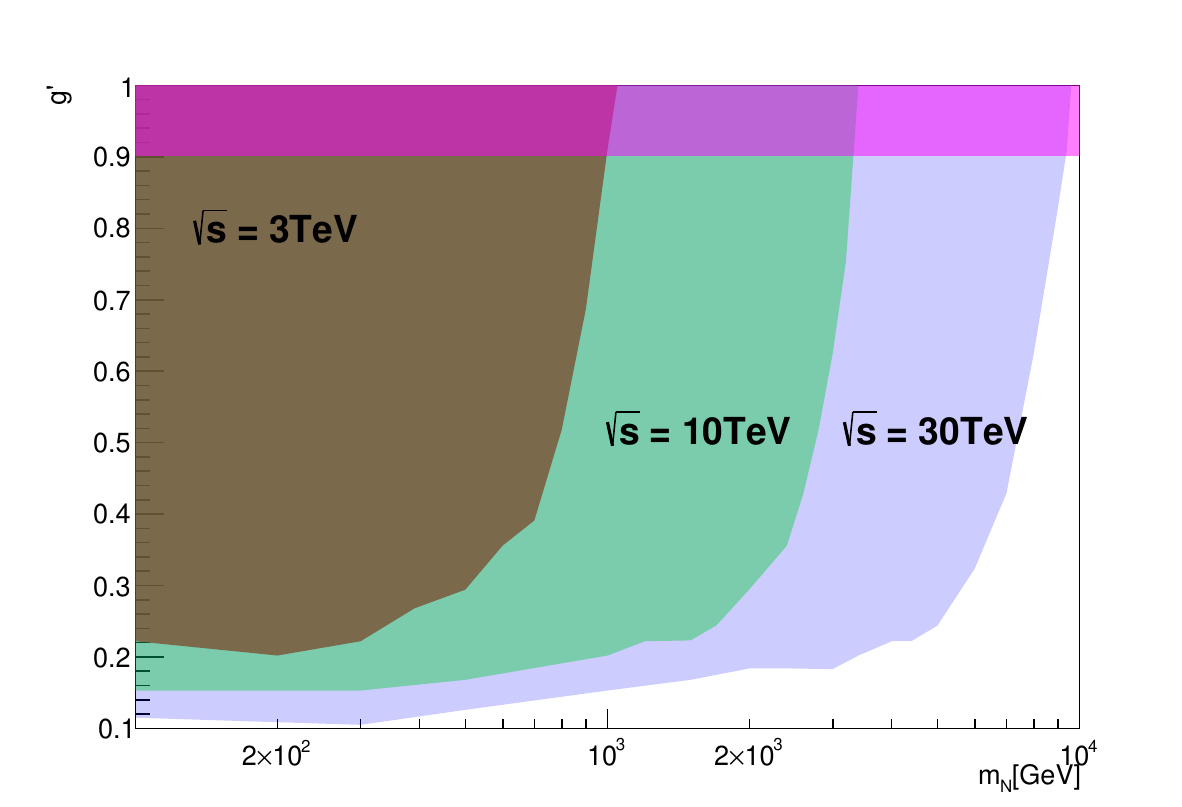}
	\end{center}
	\caption{The combined $5\sigma$ discovery reach of the $\mu^+ \mu^- \mu^\pm\mu^\pm J J$ signature with heavy Higgs $H$ at muon collider. We fix the mass ratio as $m_H=3\times m_N$.  The brown, green, and purple areas are the results of the 3 TeV, 10 TeV, and 30~TeV muon collider, respectively. The pink region is excluded by the neutrino trident production \cite{Altmannshofer:2014pba}. In the left panel, the new gauge coupling $g'=0.6$ is assumed. In the right panel, the new gauge boson mass $m_{Z'}=500$ GeV is considered. }
	\label{fig7}
\end{figure}

The combined $5\sigma$ discovery reaches of the  $\mu^+ \mu^- \mu^\pm\mu^\pm J J$ signature with heavy Higgs $H$ at muon collider are shown in Figure \ref{fig7}. Since we fix the mass relation $3m_N=m_H$, the upper limits on $m_N$ are close to the kinematic threshold $3m_N=m_H<\sqrt{s}$. In the left panel of Figure \ref{fig7}, we investigate the promising area in the $m_N-m_{Z'}$ plane by assuming $g'=0.6$. At the 3 TeV muon collider, the parameter space can be detected for $m_{Z'}<2.9$ TeV and $m_N<900$ GeV. A large parameter space can be detected at the 10 TeV muon collider as $m_{Z'}<8.5$ TeV and $m_N<2.8$ TeV. The region within $m_{Z'}<23$ TeV and $m_N<8.4$ TeV can be discovered at the 30 TeV muon collider. We notice that when the mass of the heavy neutral lepton $m_N$ is much smaller than the collision energy, the muons from the highly boosted $N$ decay are hardly isolated from the jets. Therefore, the small $m_N$ region is less promising than the large $m_N$ one, especially at the 30 TeV muon collider. In the right panel of Figure \ref{fig7}, we show the discovery reach in the $g'-m_N$ plane by fixing $m_{Z'}=500$ GeV. We report that the 3 TeV, 10 TeV, and 30 TeV muon collider could probe $g'\gtrsim0.2$, $g'\gtrsim0.16$, and $g'\gtrsim0.12$, respectively.


\section{Signature without Heavy Higgs }\label{SEC:Sig2}

\begin{figure}
	\begin{center}
		\includegraphics[width=0.45\linewidth,,height=0.36\linewidth]{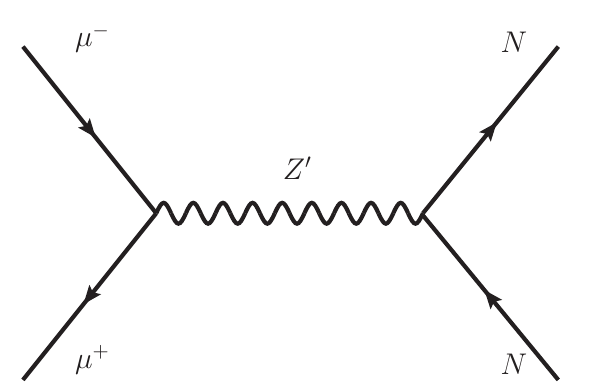}
		\includegraphics[width=0.45\linewidth]{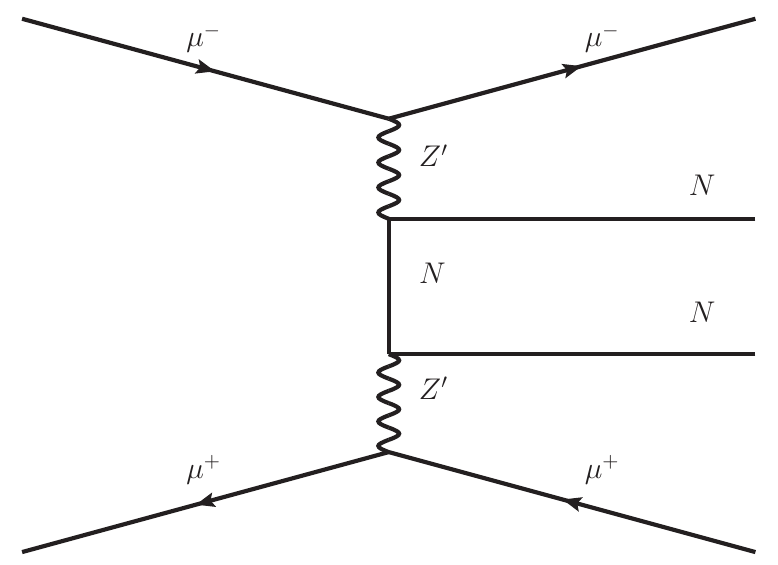}
	\end{center}
	\caption{ Feynman diagrams of the direct heavy neutral lepton pair $\mu^+\mu^-\to Z^{\prime *}\to NN $  and $Z'Z'$-fusion $\mu^+\mu^-\to\mu^+\mu^- NN$ processes at muon collider.}
	\label{fig8}
\end{figure}

In the previous Section \ref{SEC:Sig1}, we investigate the new vector boson fusion channel with heavy Higgs $Z'Z'\to H\to  NN$. However, when the mass of the heavy Higgs is larger than the collision energy $m_H > \sqrt{s}$, such a resonance heavy Higgs process is kinematically forbidden. In this scenario, the heavy neutral leptons still can be pair produced via the new vector boson fusion process $Z'Z'\to NN$ through the exchange of $N$. For comparison, we also consider the direct pair production process $\mu^+\mu^-\to Z^{\prime *}\to NN$. The relevant Feynman diagrams are shown in Figure \ref{fig8}.

\begin{figure}
	\begin{center}
		\includegraphics[width=0.45\linewidth]{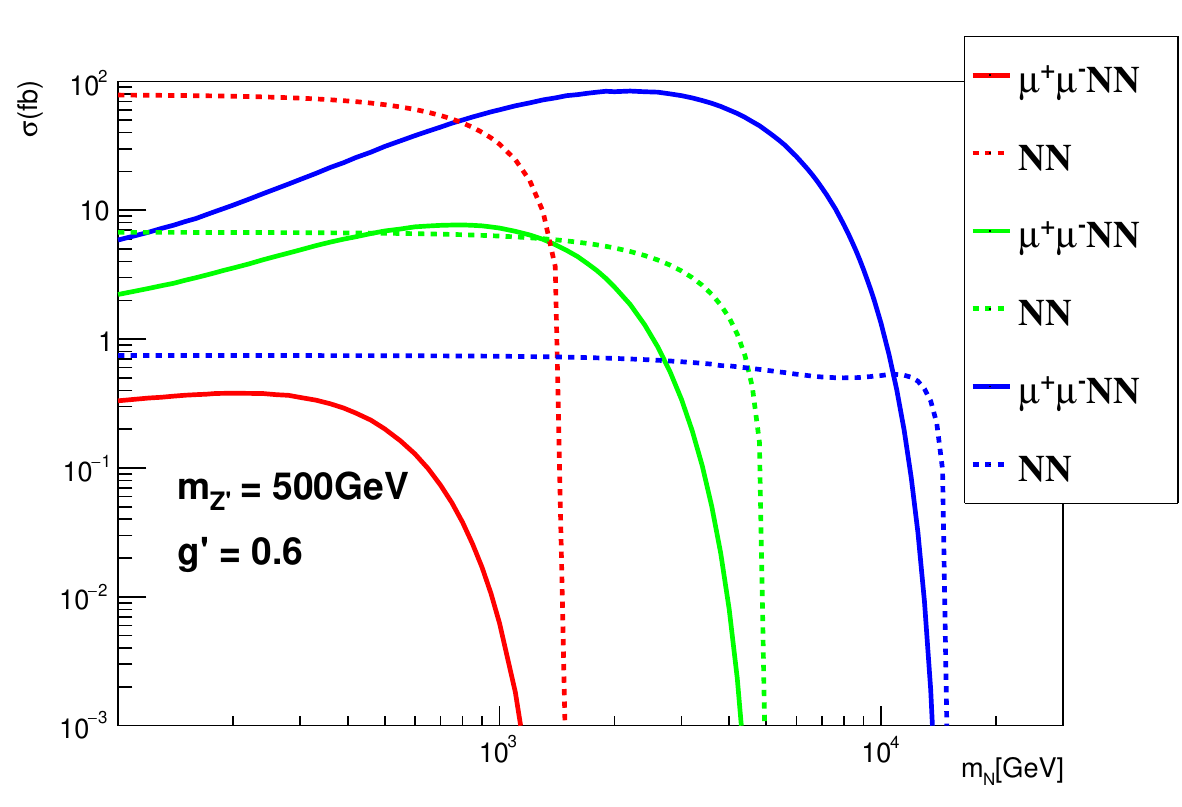}
		\includegraphics[width=0.45\linewidth]{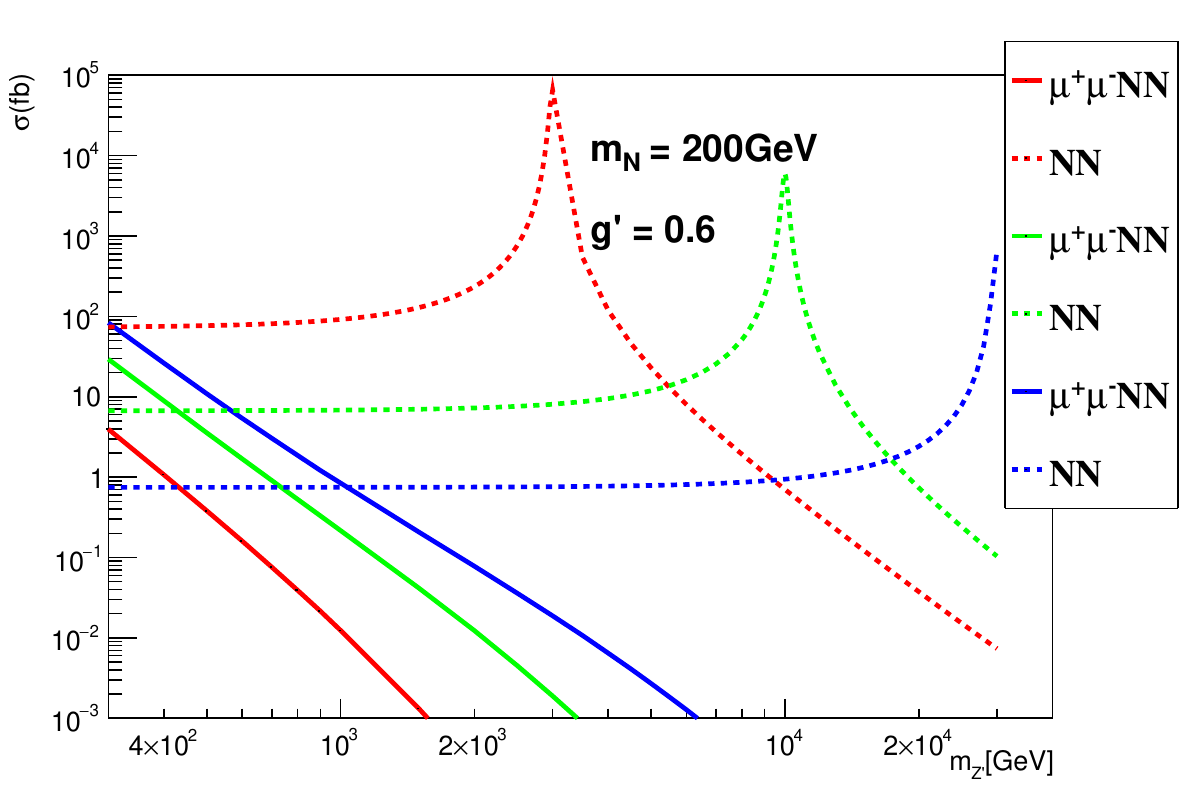}
		\includegraphics[width=0.45\linewidth]{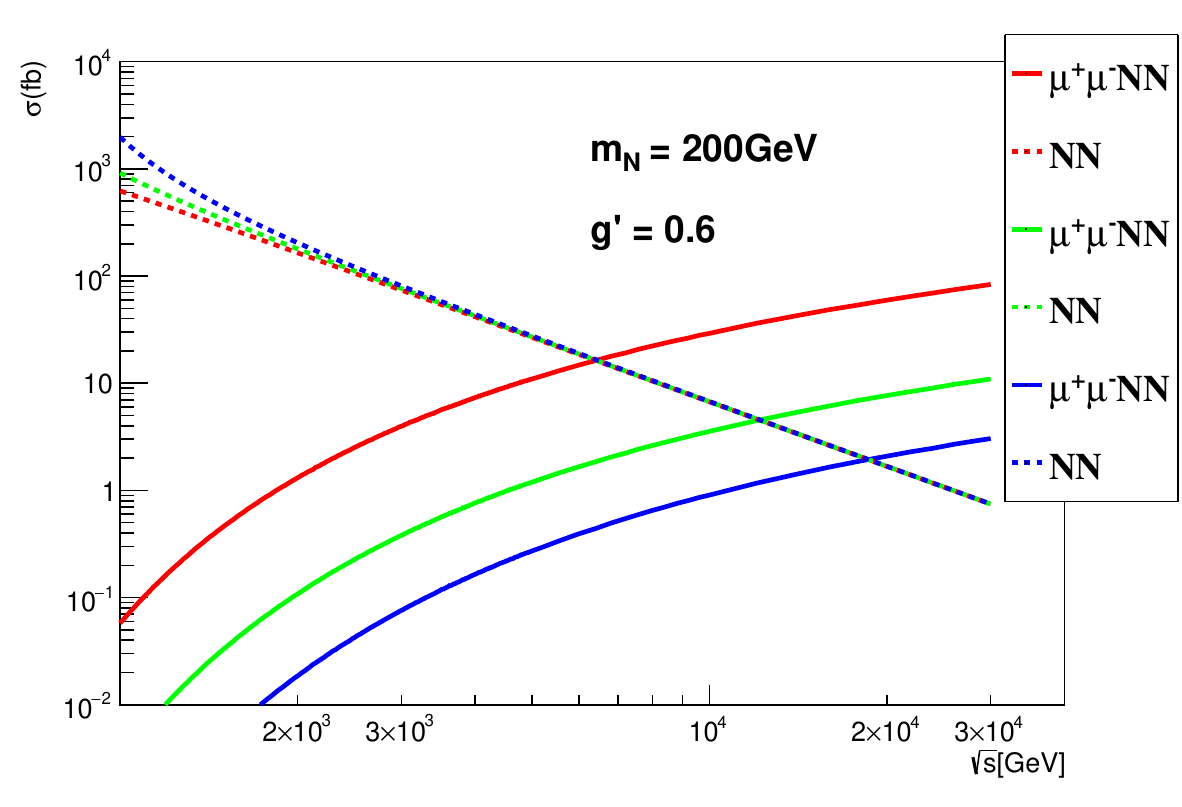}
	\end{center}
	\caption{Cross sections of the the direct heavy neutral lepton pair $\mu^+\mu^-\to Z^{\prime *}\to NN $ (dashed lines)  and $Z'Z'$-fusion $\mu^+\mu^-\to\mu^+\mu^- NN$ (solid lines) process. Up-left panel: cross sections as a function of heavy neutral lepton mass $m_N$. Up-right panel: cross sections as a function of new gauge boson mass $m_{Z'}$. In these up panels, the red, green and blue lines are the results at the 3 TeV, 10 TeV and 30 TeV muon collider, respectively. In the down panel, the red, green and blue lines are the results for $m_{Z'}=$300 GeV, 500 GeV and 700 GeV.}
	\label{fig9}
\end{figure}

Cross sections of the $\mu^+\mu^-\to Z^{\prime *}\to NN $ and  $\mu^+\mu^-\to\mu^+\mu^- NN$ processes are shown in Figure~\ref{fig9}, which are free from the heavy Higgs mass $m_H$ but sensitive to the heavy neutral lepton mass $m_N$. In the up-left panel of Figure \ref{fig9}, we consider the impact of parameter $m_N$ by fixing $m_{Z'}=500$ GeV and $g'=0.6$. The dependence of the direct pair production process $\mu^+\mu^-\to Z^{\prime *}\to NN$ on the parameter $m_N$ is different from the new vector boson fusion process $\mu^+\mu^-\to\mu^+\mu^- NN$. For the $s$-channel process $\mu^+\mu^-\to Z^{\prime *}\to NN$, the cross section gradually decreases as $m_N$ increases. Provided the phase-space is sufficiently large, the cross section of the $t$- and $u$-channel process $\mu^+\mu^-\to\mu^+\mu^- NN$ increases with larger heavy neutral lepton mass. At the 3 TeV muon collider, the cross section of $\mu^+\mu^-\to\mu^+\mu^- NN$ is over two orders of magnitudes smaller than the cross section of $\mu^+\mu^-\to Z^{\prime *}\to NN$ for the benchmark scenario. Cross sections of these two processes are comparable for TeV scale $m_N$ at the 10 TeV muon collider. Increasing the collision energy to 30 TeV, the cross section of $\mu^+\mu^-\to\mu^+\mu^- NN$ could be much larger than the cross section of $\mu^+\mu^-\to Z^{\prime *}\to NN$.

In the up-right panel of Figure \ref{fig9}, we show the cross sections as a function of new gauge boson mass $m_{Z'}$, where $m_N=200$ GeV and $g'=0.6$ are fixed. The cross section of the $Z'Z'$-fusion process $\mu^+\mu^-\to \mu^+\mu^- NN$ clearly decreases as $m_{Z'}$ becomes larger. For the direct pair production process $\mu^+\mu^-\to Z^{\prime *} \to NN$, the cross section is approximately a constant in the light $m_{Z'}$ limit. On the $Z'$-pole, i.e., $m_{Z'}=\sqrt{s}$, the cross section of $\mu^+\mu^-\to Z^{\prime *} \to NN$ can be enhanced by about three orders of magnitudes. Therefore, the  $Z'Z'$-fusion process $\mu^+\mu^-\to \mu^+\mu^- NN$ might become the dominant channel when $m_{Z'}\ll \sqrt{s}$. Specifically speaking, $m_{Z'}\lesssim420$ GeV and $m_{Z'}\lesssim1000$ GeV should be satisfied for the benchmark point at the 10 TeV and 30 TeV muon collider, respectively.

In the down panel of Figure \ref{fig9}, we depict the cross sections as a function of the collision energy $\sqrt{s}$. We fix $m_{N}=200$ GeV and $g'=0.6$, while vary $m_{Z'}=300$ GeV, 500 GeV and 700 GeV for illustration. It is obvious that the cross section of the $s$-channel process $\mu^+\mu^-\to Z^{\prime *} \to NN$ decreases as $1/s$ at high energies. In contrast, the cross section of the $Z'Z'$-fusion process $\mu^+\mu^-\to \mu^+\mu^- NN$ increases when the collision energy is larger. We report that the $Z'Z'$-fusion process becomes the dominant contribution when the collision energy is larger than about 10 TeV.

\begin{figure}
	\begin{center}
		\includegraphics[width=0.33\linewidth]{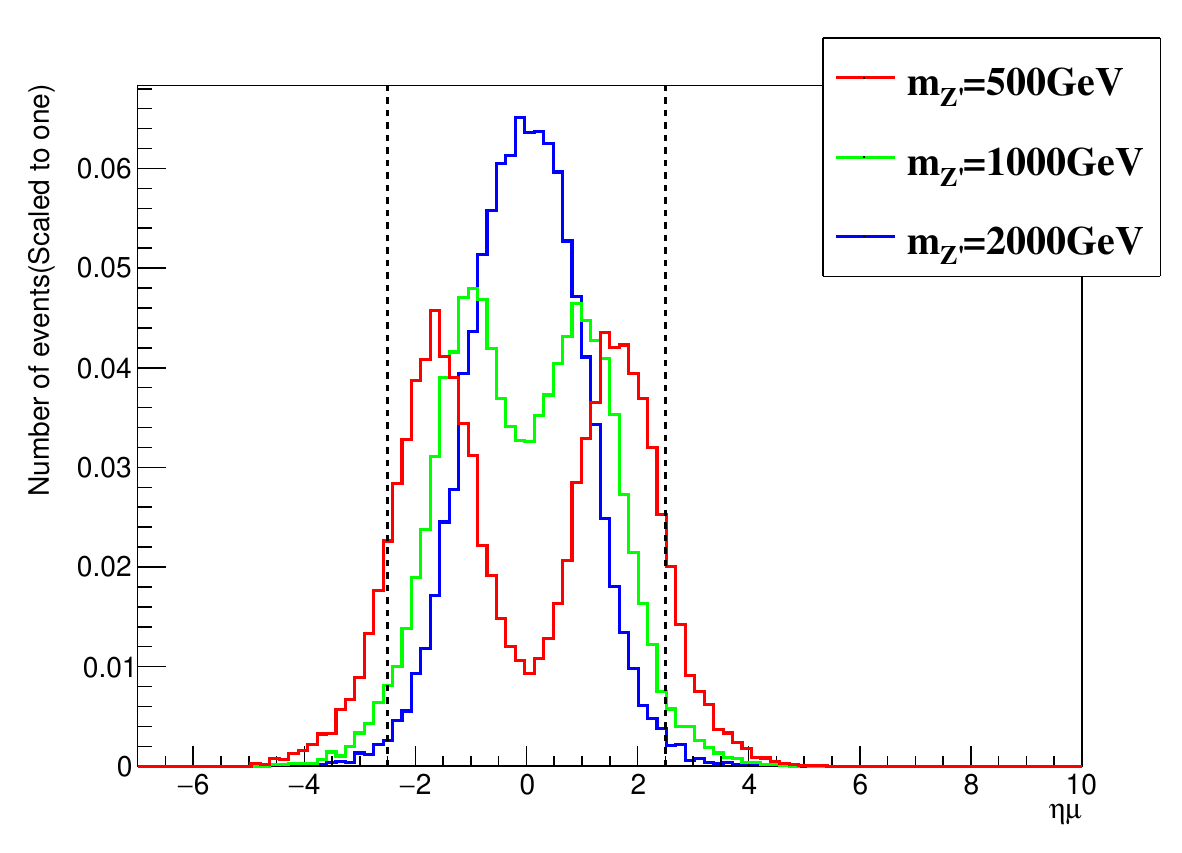}
		\includegraphics[width=0.33\linewidth]{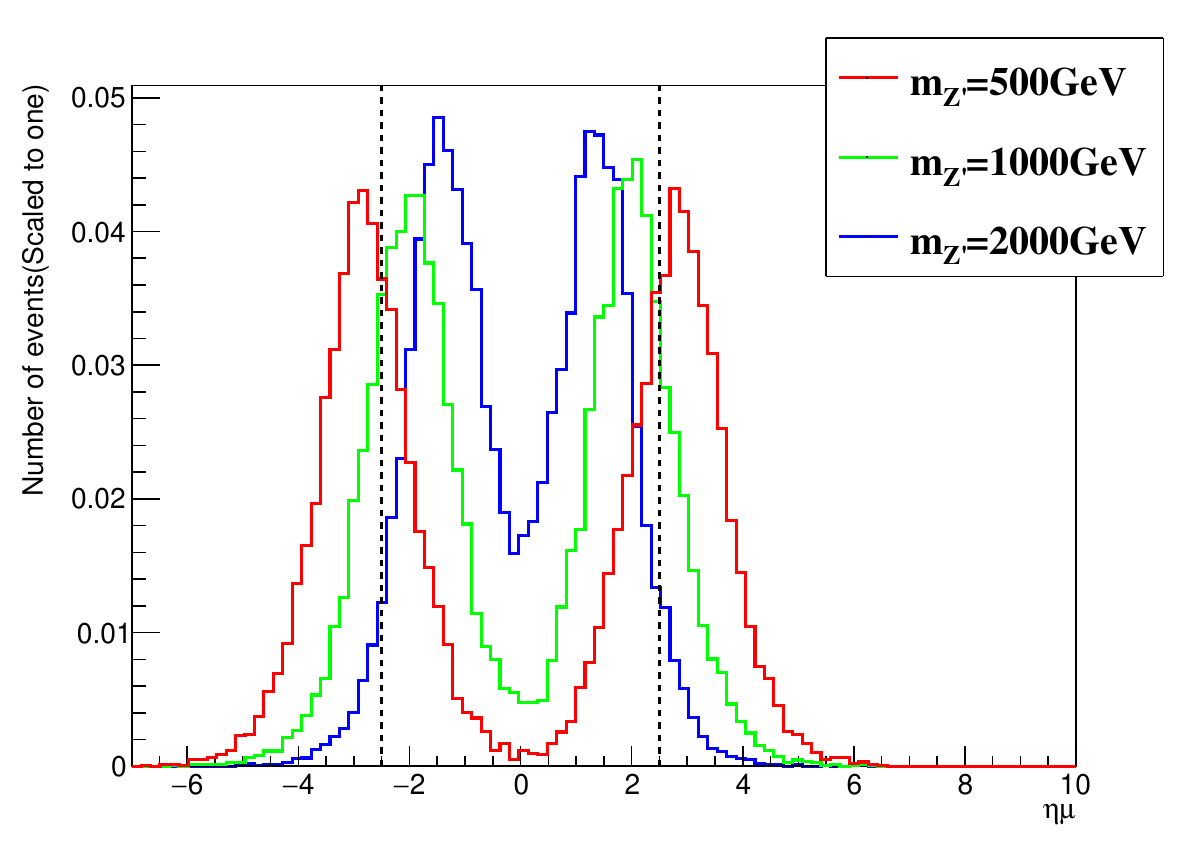}
		\includegraphics[width=0.33\linewidth]{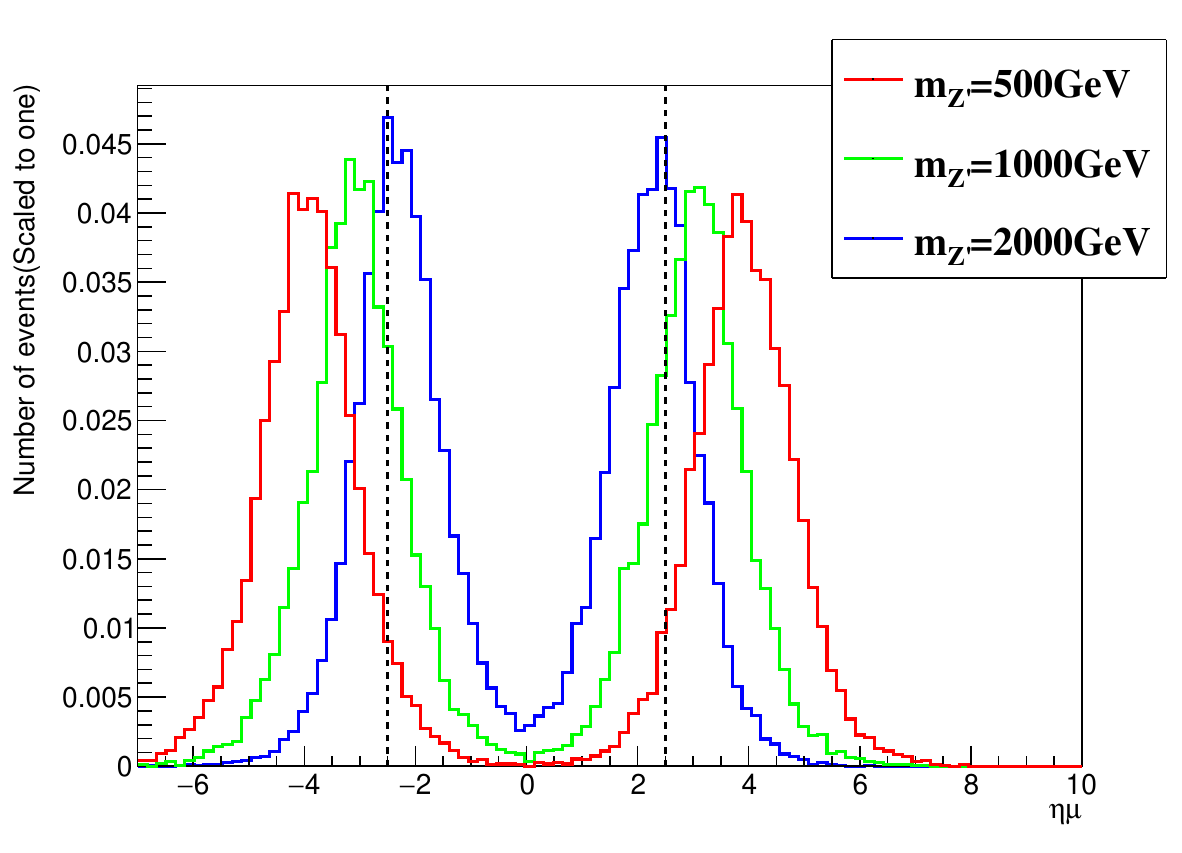}
	\end{center}
	\caption{Normalized distributions of pseudorapidity of final state muons $\eta(\mu)$ in the process $\mu^+\mu^-\to \mu^+\mu^- N N$ without heavy Higgs $H$ at the 3 TeV (left), 10 TeV (middle) and 30 TeV (right) muon collider. The benchmark points are chosen as $m_{Z'}=$ 500~GeV, 1000 GeV and 2000 GeV  with $m_N=200$ GeV.}
	\label{fig10}
\end{figure}

In Figure \ref{fig10}, we also show the parton level normalized distributions of pseudorapidity of final state muons $\eta(\mu)$ in the process $\mu^+\mu^-\to \mu^+\mu^- NN$ without heavy Higgs $H$ at the muon collider. The results are similar to those with heavy Higgs $H$. Generally speaking, the final state muons are outside the main detector when $m_{Z'}\ll \sqrt{s}$. Otherwise, they will come into the main detector. For instance, when $m_{Z'}=1000$ GeV, the two final state muons are mainly within the main detector at the 3 TeV and 10 TeV stage, but they are mainly outside at the 30 TeV stage.

The cascade decays of heavy neutral lepton in the process  $\mu^+\mu^-\rightarrow \mu^+\mu^-NN$
also generate the lepton number violation signature
\begin{equation}
	\mu^+ \mu^- \rightarrow \mu^+ \mu^- N N \rightarrow \mu^+ \mu^- + \mu^\pm jj +\mu^\pm jj,
\end{equation}
where the two jets from the same $N$ can merge into one fat-jet $J$. Hence, we study the same signature $\mu^+\mu^-\mu^\pm\mu^\pm JJ$ as in Section \ref{SEC:Sig1}. The considered SM backgrounds are also $\mu^+\mu^-\to \mu^+\mu^- WW$, $\mu^+\mu^-WWZ$, and $\mu^\pm \nu WWW$.

\begin{figure}
	\begin{center}
		\includegraphics[width=0.33\linewidth]{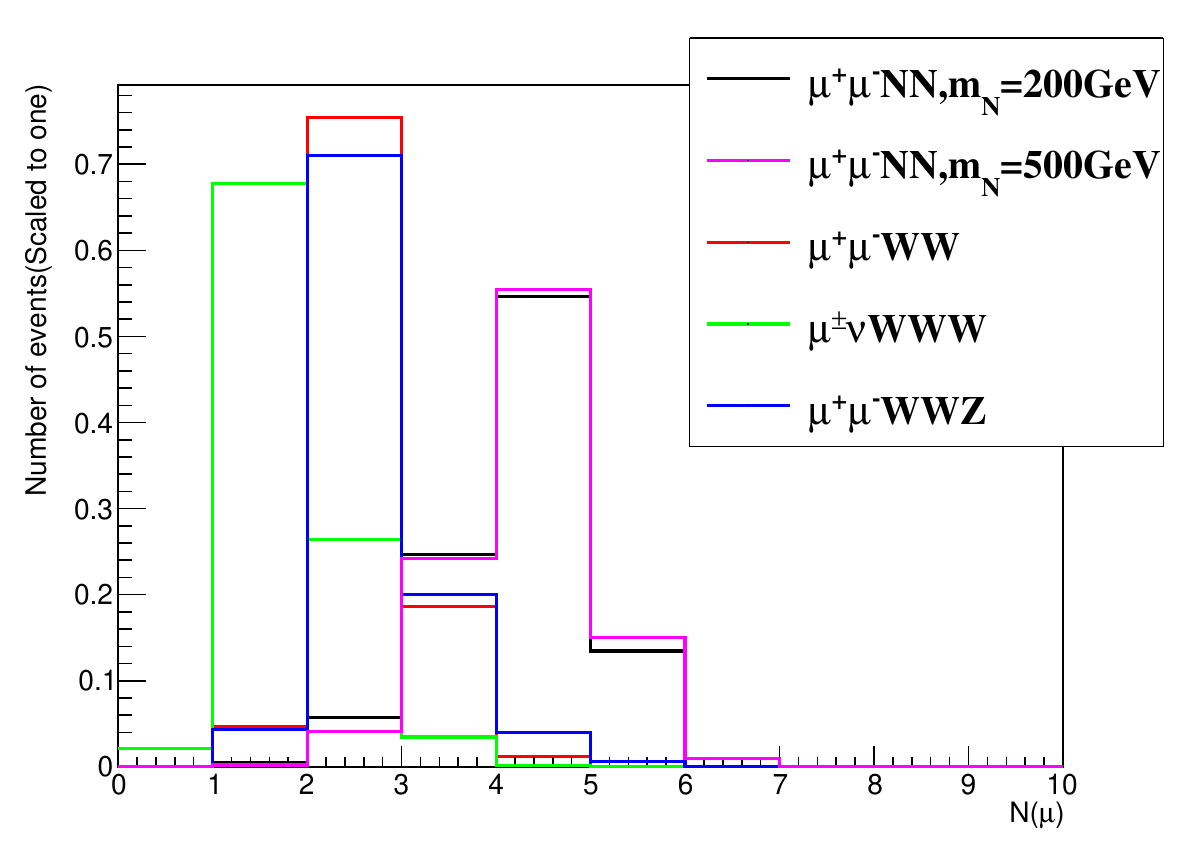}
		\includegraphics[width=0.33\linewidth]{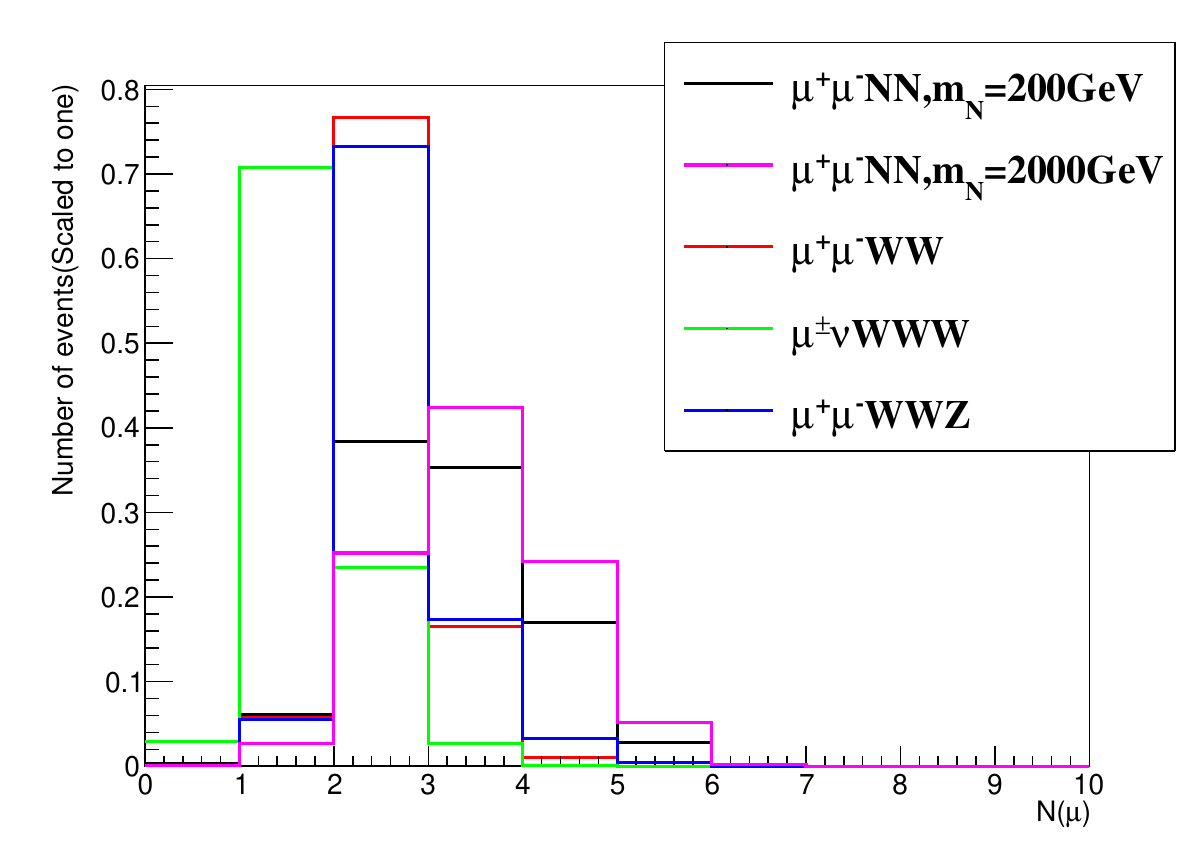}
		\includegraphics[width=0.33\linewidth]{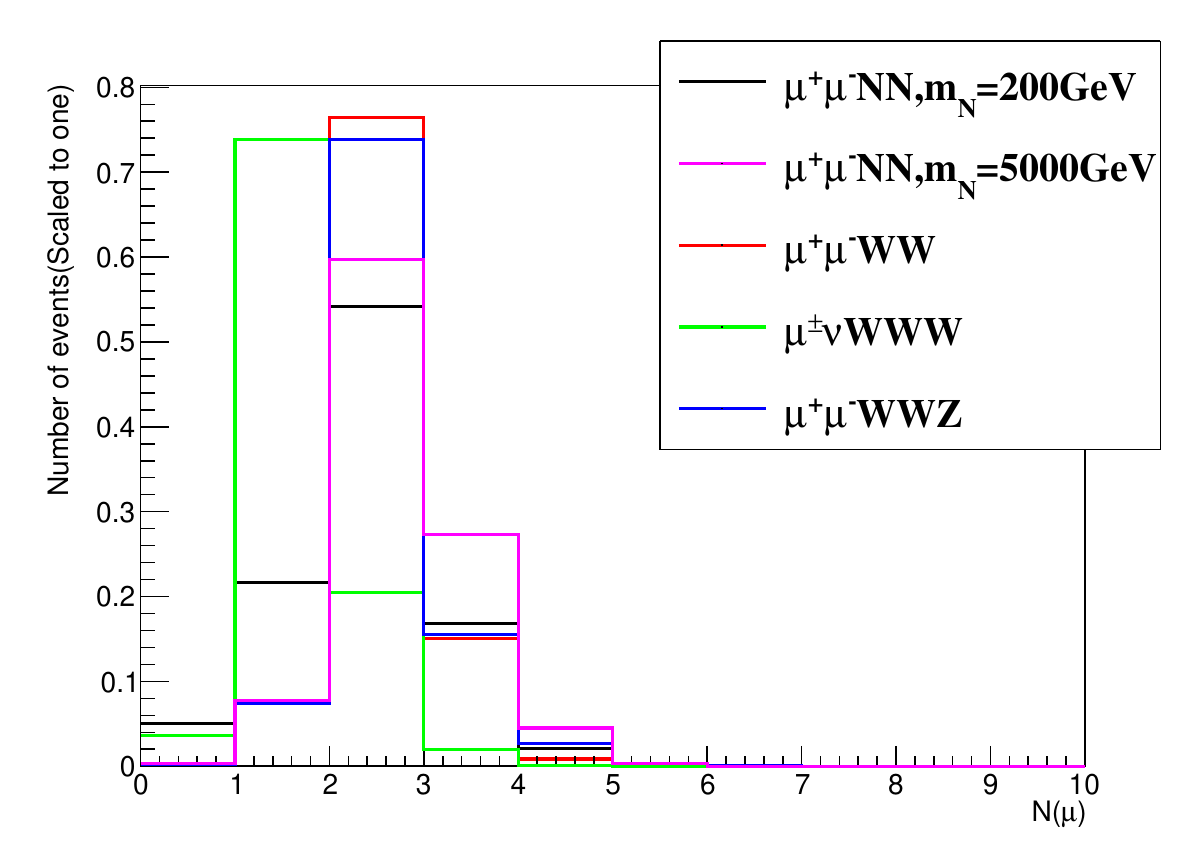}
		\includegraphics[width=0.33\linewidth]{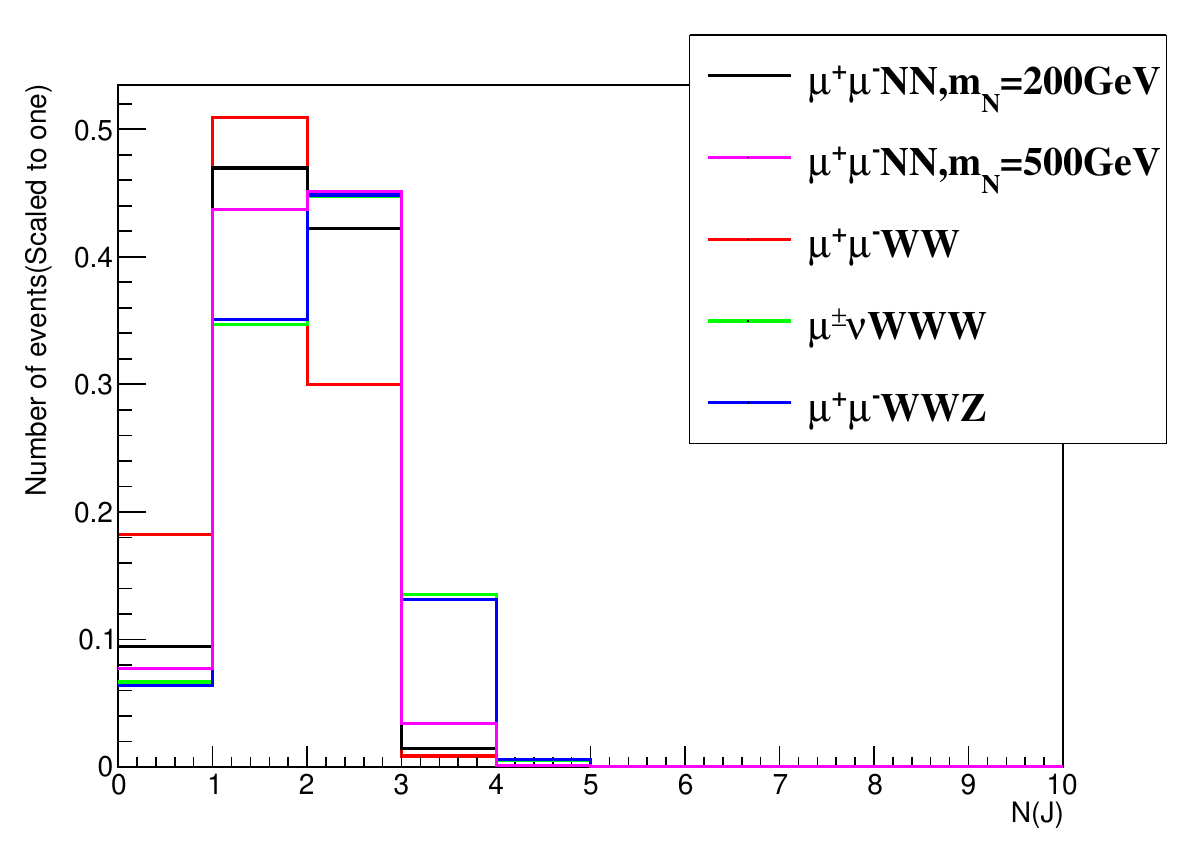}
		\includegraphics[width=0.33\linewidth]{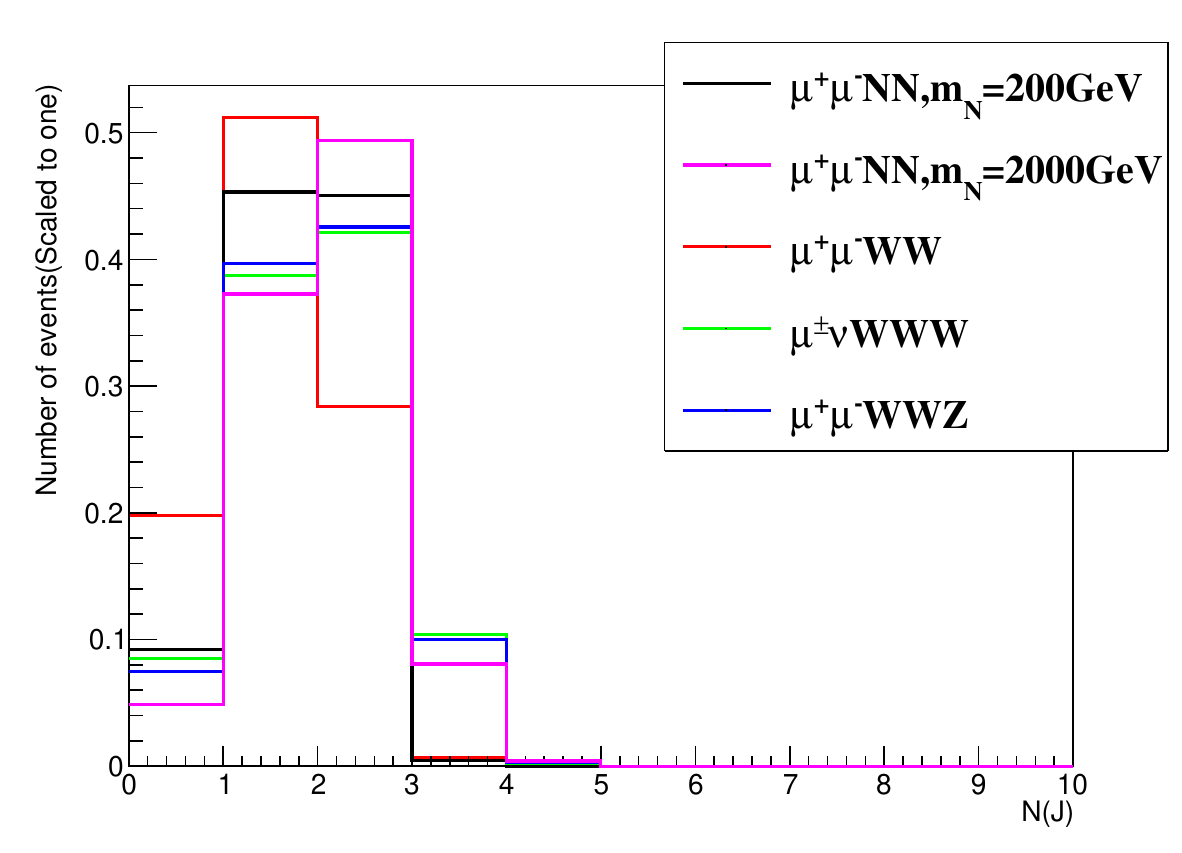}
		\includegraphics[width=0.33\linewidth]{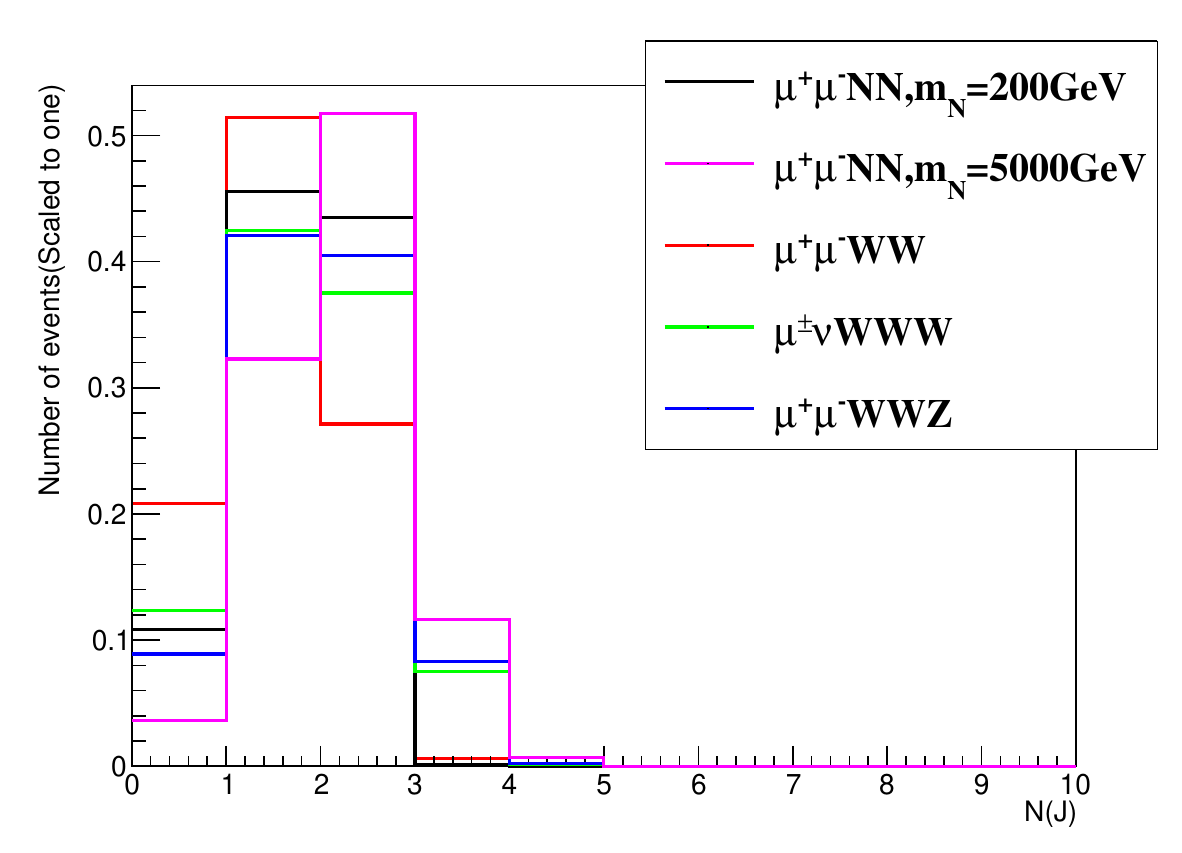}		
		\includegraphics[width=0.33\linewidth]{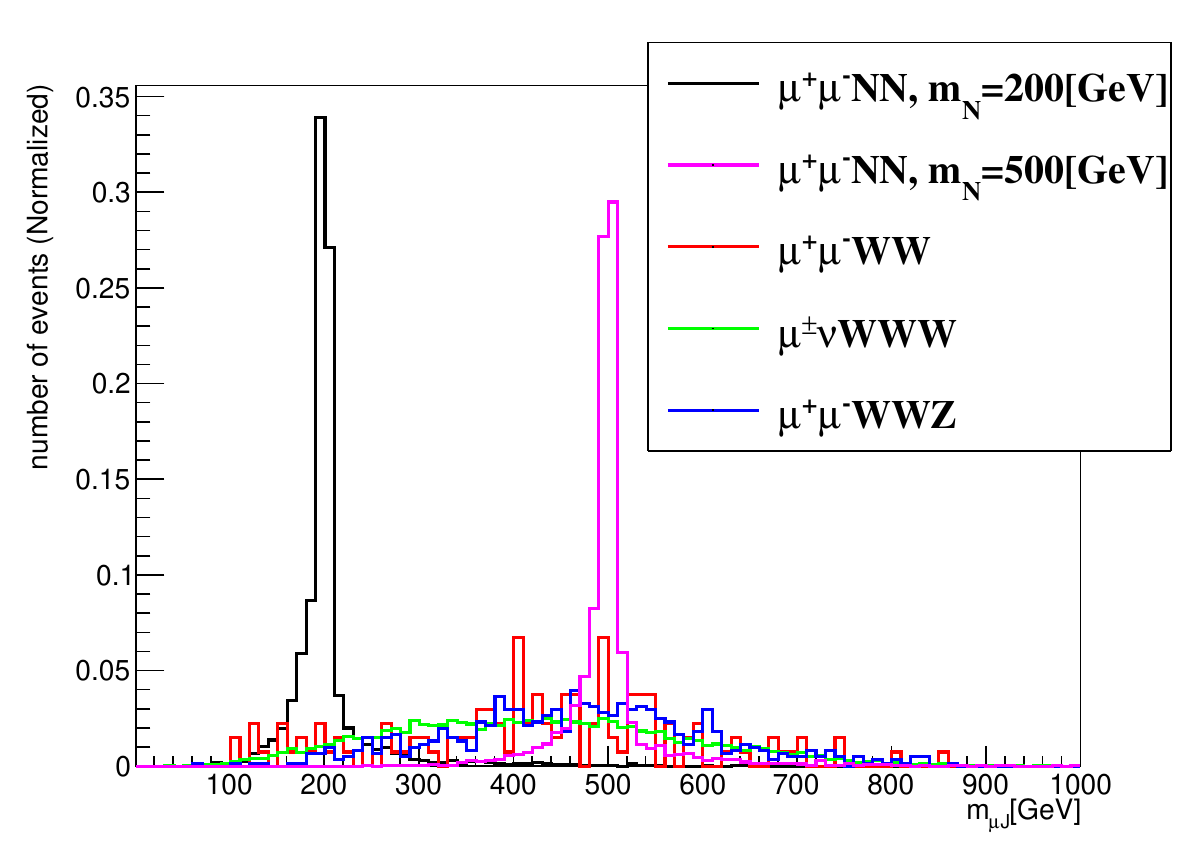}
		\includegraphics[width=0.33\linewidth]{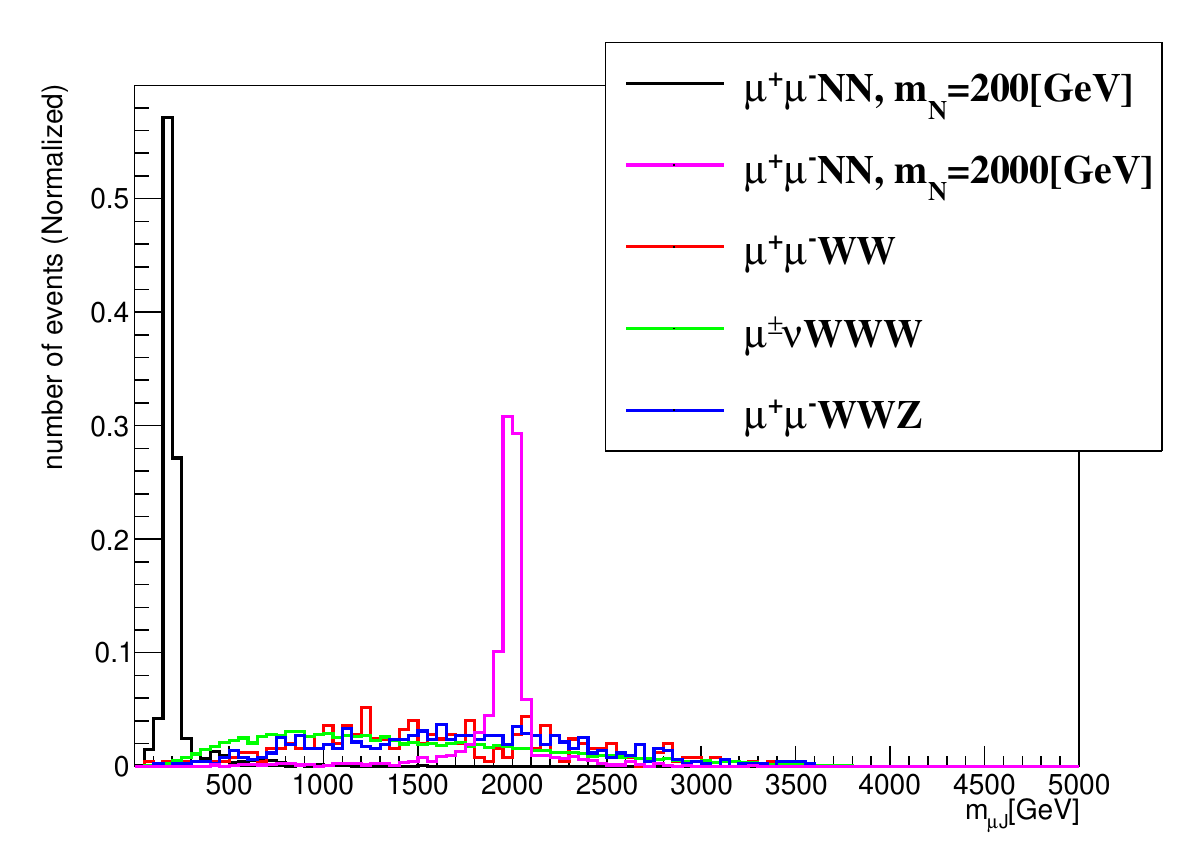}
		\includegraphics[width=0.33\linewidth]{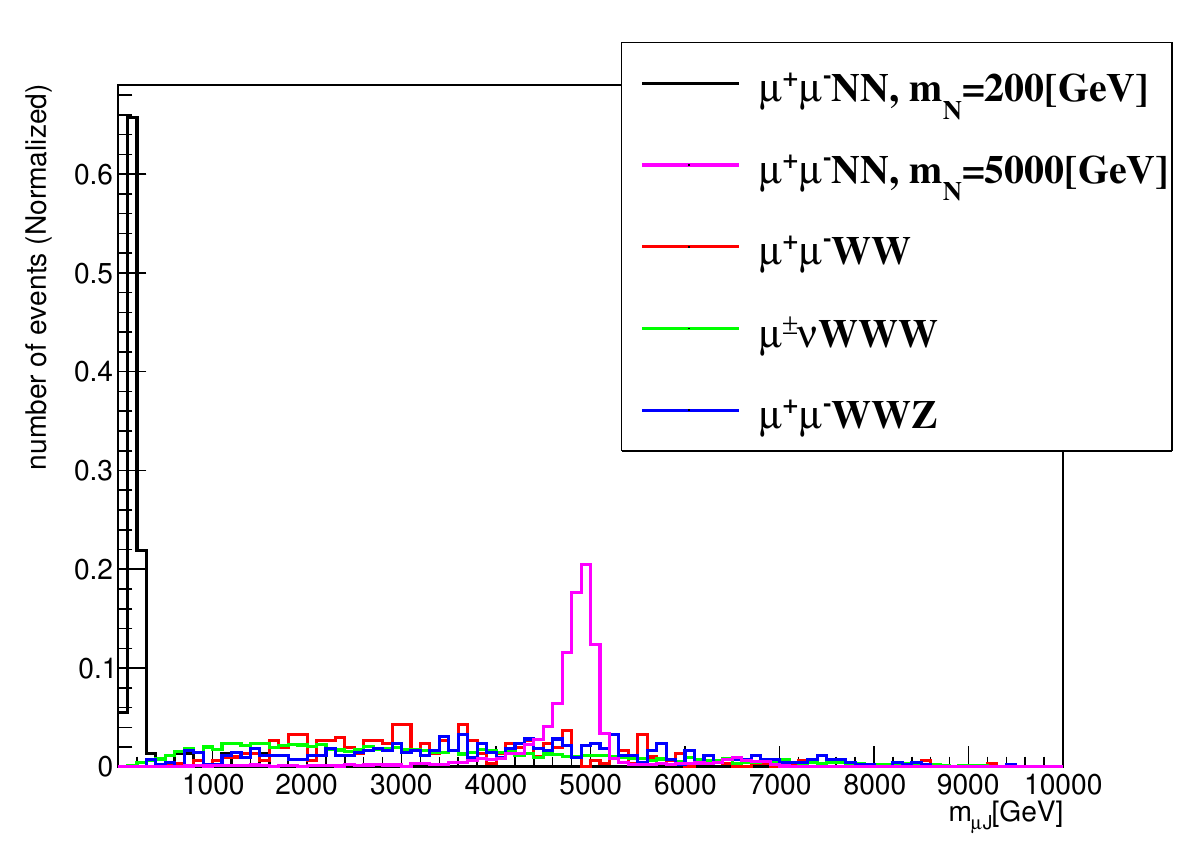}
	\end{center}
	\caption{Normalized distributions of number of muons $N(\mu)$ (up panels), number of fat-jets $N(J)$ (middle panels), and invariant mass of muon and fat-jet $m_{\mu J}$ (down panels) for the signal $\mu^+ \mu^- \mu^\pm\mu^\pm J J$ without heavy Higgs $H$ and corresponding backgrounds at the 3 TeV (left), 10 TeV (middle) and 30 TeV (right) muon collider. The benchmark points are chosen as $m_{Z'}=500$ GeV.}
	\label{fig11}
\end{figure}

In Figure \ref{fig11}, we show the normalized distribution of variables both for the signal without heavy Higgs and backgrounds process at the 3 TeV, 10 TeV and 30 TeV muon collider after the pre-selection cuts. Despite the cross section of $\mu^+\mu^-NN\to \mu^+\mu^- \mu^\pm \mu^\pm JJ$ being much smaller than that of $\mu^+\mu^-H\to \mu^+\mu^- \mu^\pm \mu^\pm JJ$, the distributions of the two signal processes are quite similar for the same benchmark points. That is to say, the four muon  channel is dominant at the 3 TeV muon collider, while the two muon  channel becomes dominant at the 30 TeV stage. On the other hand, the three/two muon  channel is dominant for relatively heavy/light $m_N$ at the 10 TeV muon collider. Therefore, all  three kinds of channels are considered in this section. And we apply the same selection cuts as in Section \ref{SEC:Sig1}.

\begin{table}
	\begin{center}
		\begin{tabular}{c | c | c | c | c | c} 
			\hline
			\hline
			\multicolumn{2}{c|}{$ \mu^+\mu^- NN\to \mu^+ \mu^- \mu^\pm\mu^\pm J J$} & After Selection (fb)& Backgrounds (fb) & Significance & $5\sigma$ Luminosity (fb$^{-1}$) \\
			\hline
			\multirow{3}{*}{3 TeV}	& $3\mu^\pm + \mu^\mp$   &$3.46\times10^{-2}$    &$5.97\times10^{-4}$    &  13.6   & $2.41\times10^2$    \\			
			\cline{2-6}
			& $3 \mu^\pm$   &$5.69\times10^{-3}$   &$2.71\times10^{-4}$   &3.75   &   $1.46\times10^3$   \\			
			\cline{2-6}
			& $2 \mu^\pm$   &$7.84\times10^{-4}$    &$3.10\times10^{-2}$   &0.14   & $4.07\times10^4$  \\			
			\hline
			\hline
			\multirow{3}{*}{10 TeV}	& $3\mu^\pm + \mu^\mp$   &$7.31\times10^{-2}$   &$5.53\times10^{-4}$   &75.8   & $2.11\times10^2$   \\			
			\cline{2-6}
			& $3 \mu^\pm$   &$7.58\times10^{-2}$     &$1.92\times10^{-4}$    &87.1   & $1.37\times10^2$  \\			
			\cline{2-6}
			& $2 \mu^\pm$  &$1.89\times10^{-2}$    &$1.80\times10^{-3}$    &25.2   & $1.32\times10^3$  \\			
			\hline
			\hline
			\multirow{3}{*}{30 TeV}	& $3\mu^\pm + \mu^\mp$   &$1.25\times10^{-2}$   &$1.53\times10^{-4}$   &88.3   & $1.79\times10^3$  \\			
			\cline{2-6}
			& $3 \mu^\pm$  &$3.03\times10^{-2}$   &$5.13\times10^{-5}$   &171   & $4.75\times10^2$ \\			
			\cline{2-6}
			& $2 \mu^\pm$  &$3.36\times10^{-2}$   &$1.93\times10^{-4}$   &159   & $7.34\times10^2$   \\			
			\hline
			\hline
		\end{tabular}
	\end{center}
	\caption{ Results for the $\mu^+ \mu^- \mu^\pm\mu^\pm J J$ signature without heavy Higgs $H$ and backgrounds at the muon collider. The benchmark point is selected as $m_N=200~\rm{GeV}$, $m_{Z'}=500~\rm{GeV}$ and $g'=0.6$. }
	\label{Tab02}
\end{table}

The results for the $\mu^+ \mu^- \mu^\pm\mu^\pm J J$ signature without heavy Higgs $H$ and corresponding backgrounds are shown in Table \ref{Tab02}. For comparison, the benchmark point is also selected as $m_N=200$ GeV, $m_{Z'}=500$~GeV, and $g'=0.6$. With the same cuts as in  Section \ref{SEC:Sig1}, the results for the backgrounds after all selection cuts are the same as in Table \ref{Tab01}. The backgrounds for the $3\mu^\pm + \mu^\mp$ and $3\mu^\pm$ channels are relatively small, which are typically at the order of $\mathcal{O}(10^{-4})$ fb. Hence, there are only a few background events even with 10 ab$^{-1}$ luminosity at the 10 TeV muon collider. The $2\mu^\pm$ channel has a relatively large background. Increasing the collision energy clearly leads to the decreasing of backgrounds for the $2\mu^\pm$ channel as shown in Table \ref{Tab02}.

Without the $s$-channel heavy Higgs $H$ enhancement, the cross sections of the lepton number violation signatures are several orders of magnitude  smaller. For an integrated luminosity of 1 ab$^{-1}$, the $3\mu^\pm$ and $2\mu^\pm$ channels are not promising at the 3 TeV muon collider. The benchmark point can only be discovered through the  $3\mu^\pm + \mu^\mp$ channel with at least $2.41\times10^{2}$ fb$^{-1}$ data. The significance of the $3\mu^\pm + \mu^\mp$ channel could reach 13.6, with a precision of 17\% at the 3 TeV stage.

At the 10 TeV muon collider, the cross sections of these three channels are all around $\mathcal{O}(10^{-2})$ fb after selection cuts. With the largest signal cross section and smallest backgrounds, the three muons channel is the most promising one. The benchmark point can be discovered with $2.11\times10^2$ fb$^{-1}$, $1.37\times10^2$ fb$^{-1}$ data, and $1.32\times10^3$ fb$^{-1}$ data for the $3\mu^\pm + \mu^\mp$, $3\mu^\pm$ and $2\mu^\pm$ channels, respectively. For an integrated luminosity of 10 ab$^{-1}$, the significance of the $3\mu^\pm + \mu^\mp$,  $3\mu^\pm$, and $2\mu^\pm$ channels are 75.8, 87.1, and 25.2, with the precision of 3.7\%, 3.6\%, and 7.6\% accordingly.

At the 30 TeV stage, the cross sections of the $3\mu^\pm + \mu^\mp$ and $3\mu^\pm$ channels decrease, while the cross section of the $2\mu^\pm$ channel increases. With the smallest backgrounds, the $3\mu^\pm$ channel is still the most promising one. The $2\mu^\pm$ channel has the largest signal cross section, but the corresponding background is also the largest.
The $3\mu^\pm$ and $2\mu^\pm$ channels can be discovered with less than 1 ab$^{-1}$ luminosity. The  $3\mu^\pm + \mu^\mp$ channel is the most unpromising one, but it still could be probed with over 1.79 ab$^{-1}$ data. The final significance of the $3\mu^\pm + \mu^\mp$,  $3\mu^\pm$, and $2\mu^\pm$ channels could reach 88.3, 171, and 159 with a luminosity of 90 ab$^{-1}$. The corresponding precisions are 3.0\%, 1.9\%, and 1.8\%, respectively.

\begin{figure}
	\begin{center}
		\includegraphics[width=0.33\linewidth]{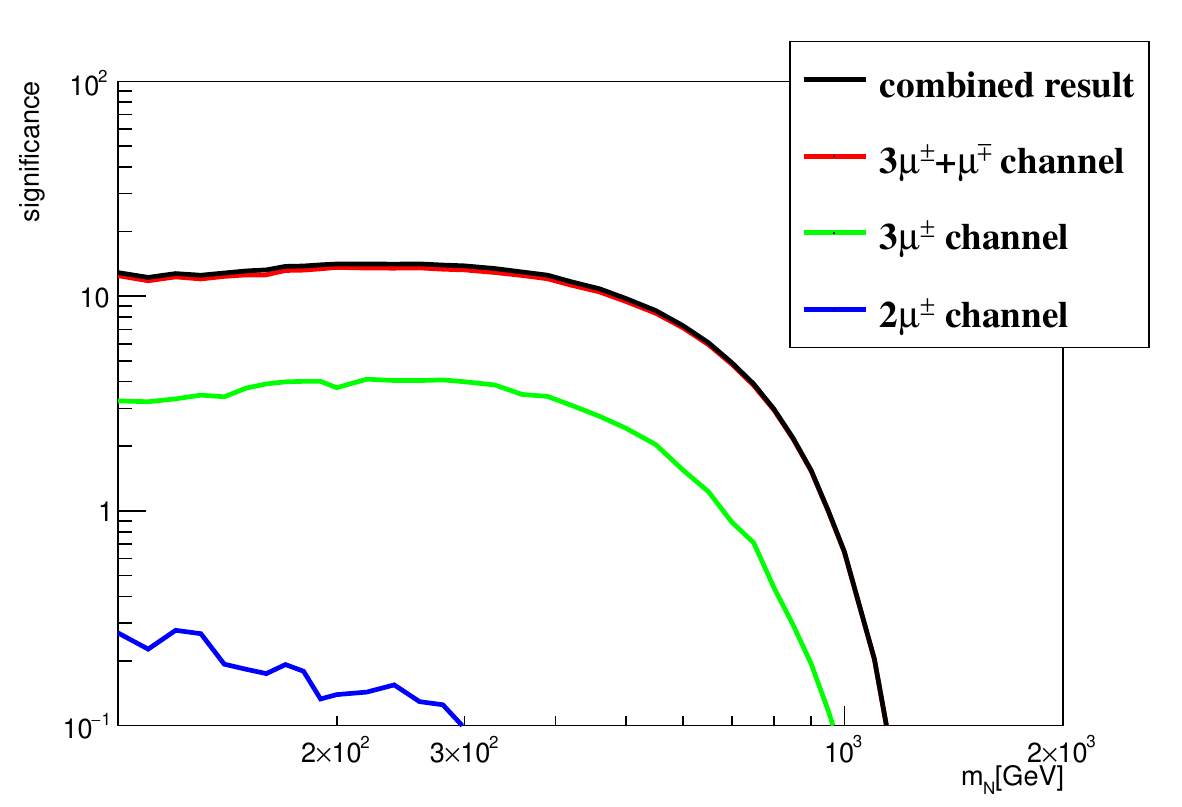}
		\includegraphics[width=0.33\linewidth]{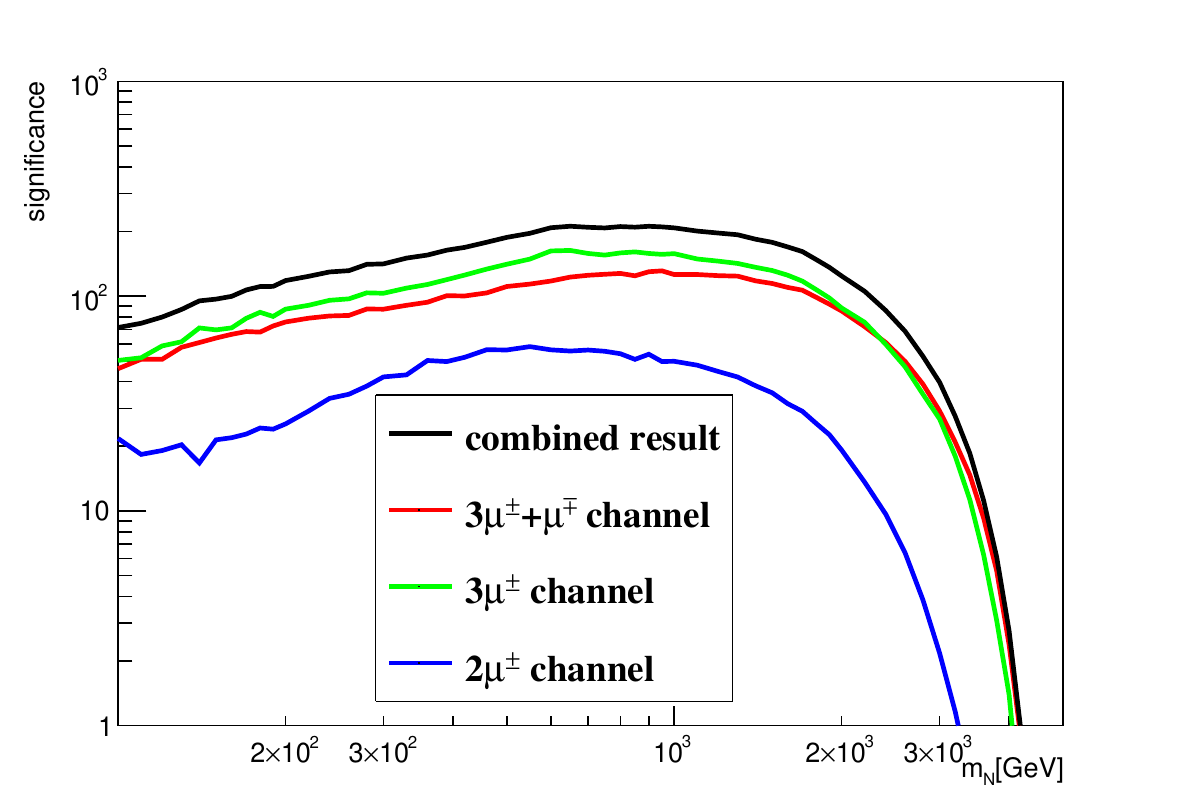}
		\includegraphics[width=0.33\linewidth]{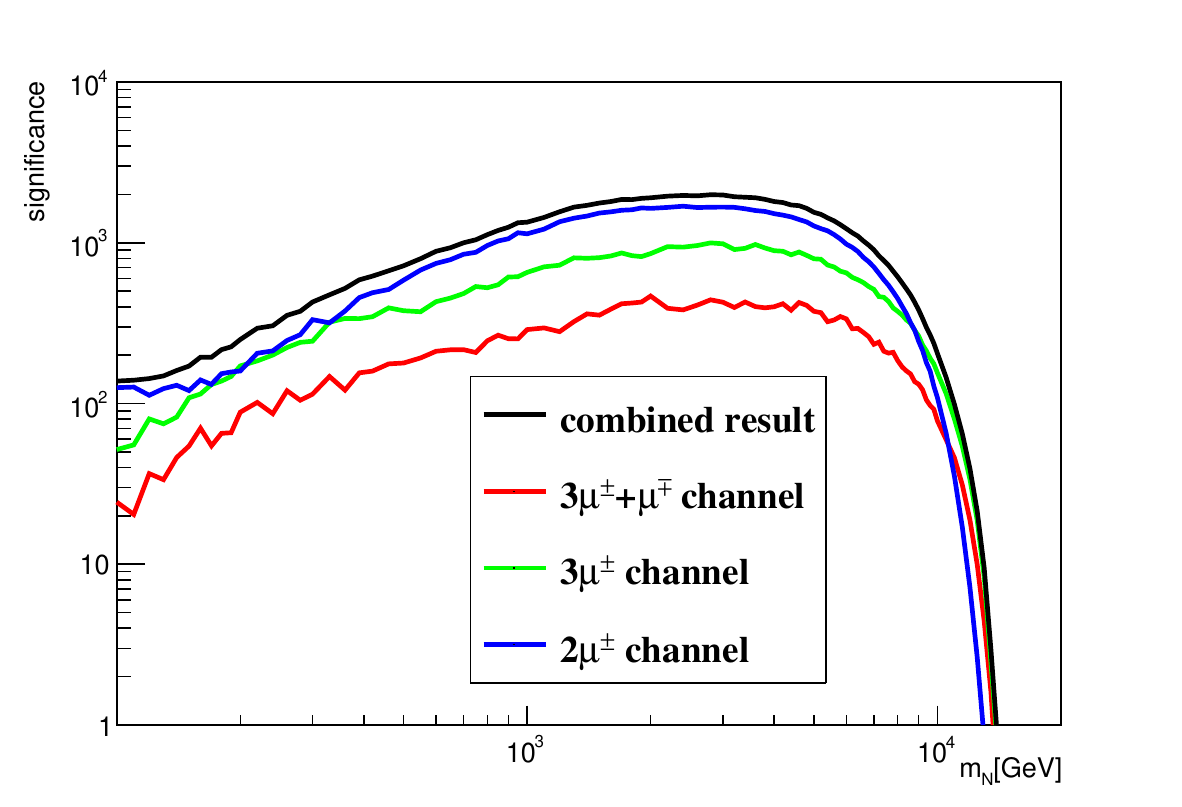}
		\includegraphics[width=0.33\linewidth]{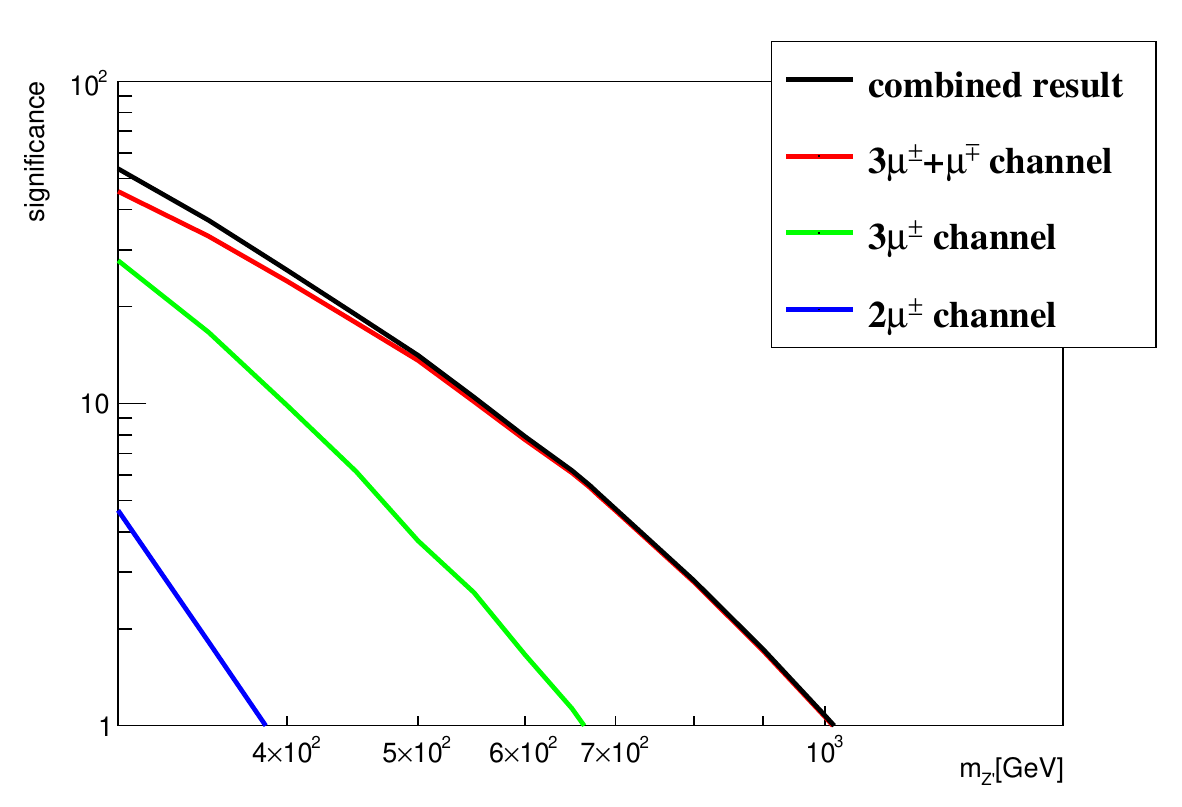}
		\includegraphics[width=0.33\linewidth]{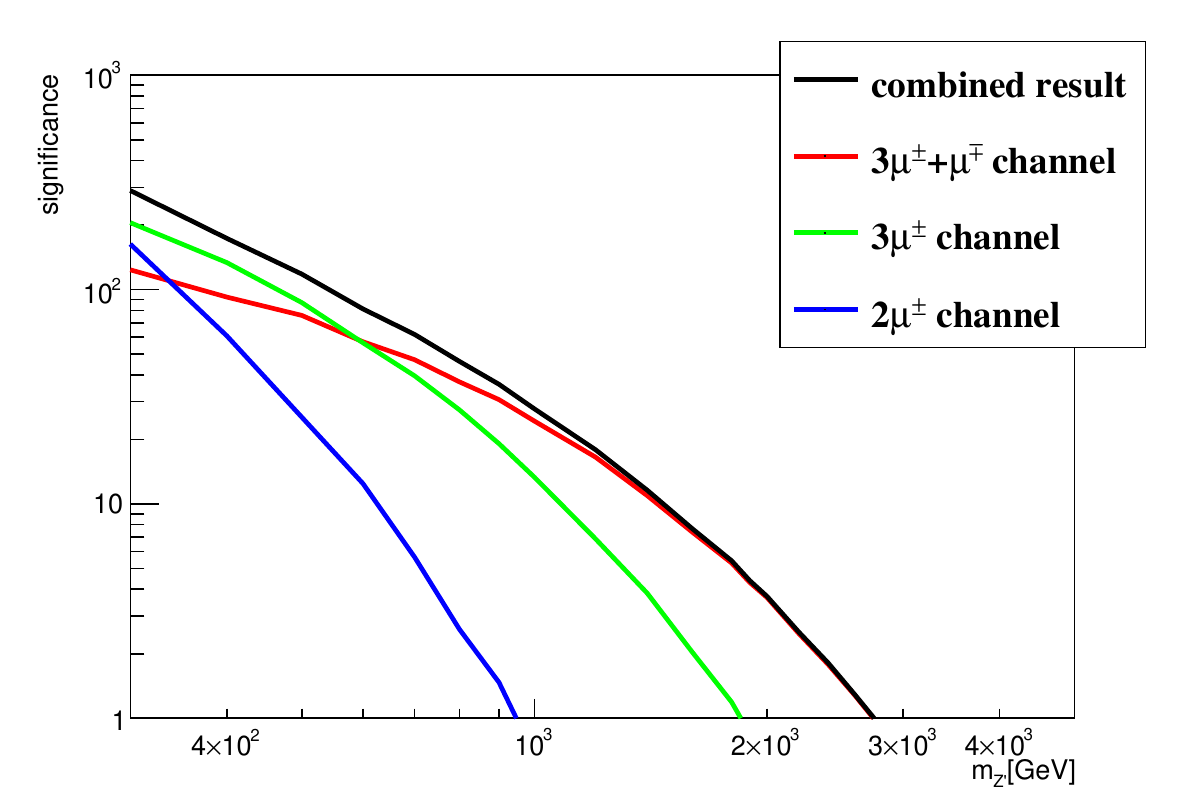}		
		\includegraphics[width=0.33\linewidth]{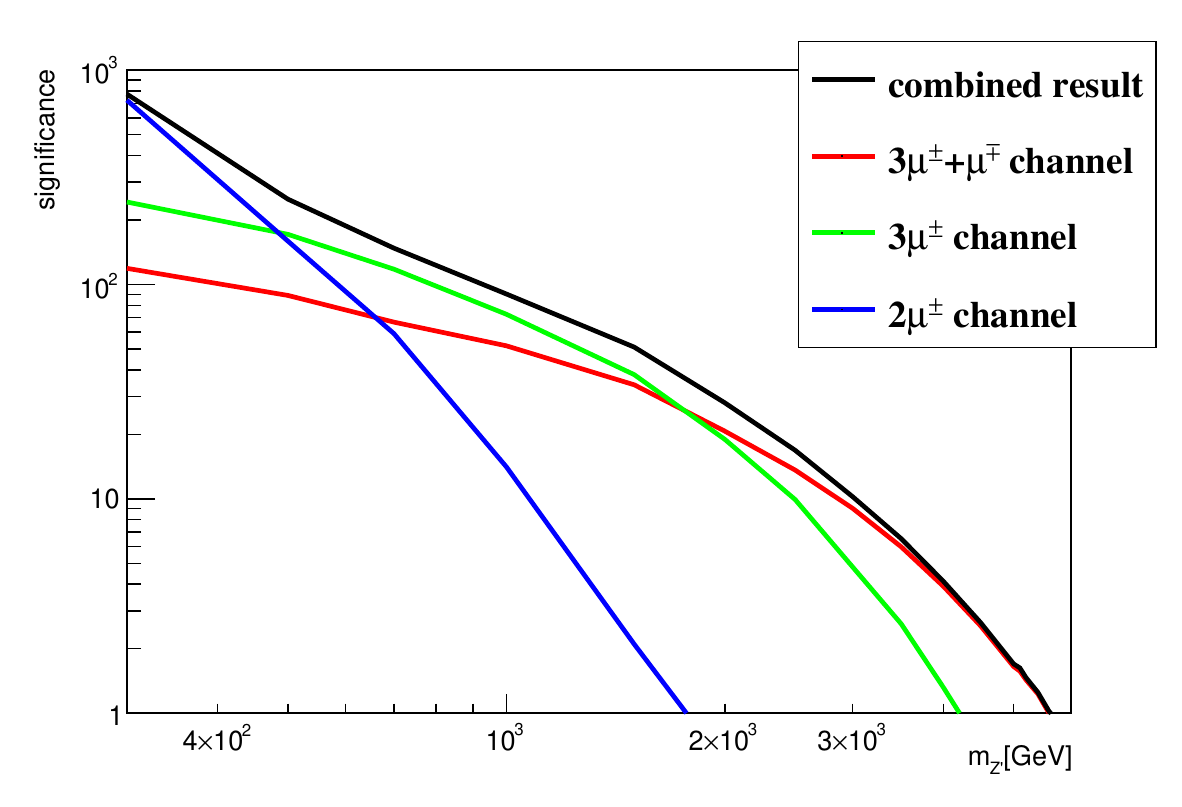}
	\end{center}
	\caption{The significance of the $\mu^+ \mu^- \mu^\pm\mu^\pm J J$ signature without heavy Higgs $H$ at the 3 TeV (left panel), 10 TeV (middle panel) and 30 TeV (right panel) muon collider  with  $g'=0.6$. In the up panels we have set $m_{Z'}=500$ GeV and in the down panels we have set $m_{N}=200$ GeV. The red, green, and blue lines are the results for the $3\mu^\pm+\mu^\mp$, $3\mu^\pm$, and  $2\mu^\pm$ channels, respectively. The black lines are the combined results of these three channels.}
	\label{fig12}
\end{figure}

In Figure \ref{fig12}, we show the significance of the $\mu^+ \mu^- \mu^\pm\mu^\pm J J$ signature without heavy Higgs $H$ at muon collider as a function of $m_N$($m_{Z'}$) in the up(down) panels. Different from the results in Figure \ref{fig6}, an increasing in $m_N$ also leads to the increasing of the significance without heavy Higgs before the phase-space suppression becomes important. Meanwhile, a heavier new gauge boson leads to smaller significance. It is obvious that the $3\mu^\pm+\mu^\mp$ channel is the absolute dominant contribution at the 3 TeV stage. The significance of the $3\mu^\pm$ channel at the 3 TeV muon collider at best could reach about four.

For the benchmark scenario with $m_{Z'}=500$ GeV, the significance of the $3\mu^\pm$ channel approximately equals that of the $3\mu^\pm+\mu^\mp$ channel at the 10 TeV stage. One may also discover the heavy neutral lepton through the $2\mu^\pm$ channel at the 10 TeV muon collider when $m_N\lesssim2.6$ TeV. The $2\mu^\pm$ channel becomes the most promising one at the 30 TeV muon collider when $m_N\lesssim9$ TeV. Above $m_N>9$ TeV, the $3\mu^\pm$ channel is the dominant contribution.

For the benchmark scenario with $m_{N}=200$ GeV, we notice that the $2\mu^\pm$ channel can hardly become the dominant one when varying $m_{Z'}$ at the 10 TeV stage. Before $m_{Z'}\lesssim600$ GeV, the $3\mu^\pm$ channel is the most promising one at the 10 TeV muon collider. Afterwards, the $3\mu^\pm+\mu^\mp$ channel becomes the most promising one. At the 30 TeV muon collider, the $2\mu^\pm$ channel could be dominant when $m_{Z'}\lesssim500$ GeV, and the $3\mu^\pm+\mu^\mp$ channel is the dominant one when $m_{Z'}\gtrsim2$ TeV. In the mass interval of (0.5,2) TeV, the dominant contribution is from the $3\mu^\pm$ channel

\begin{figure}
	\begin{center}
		\includegraphics[width=0.45\linewidth]{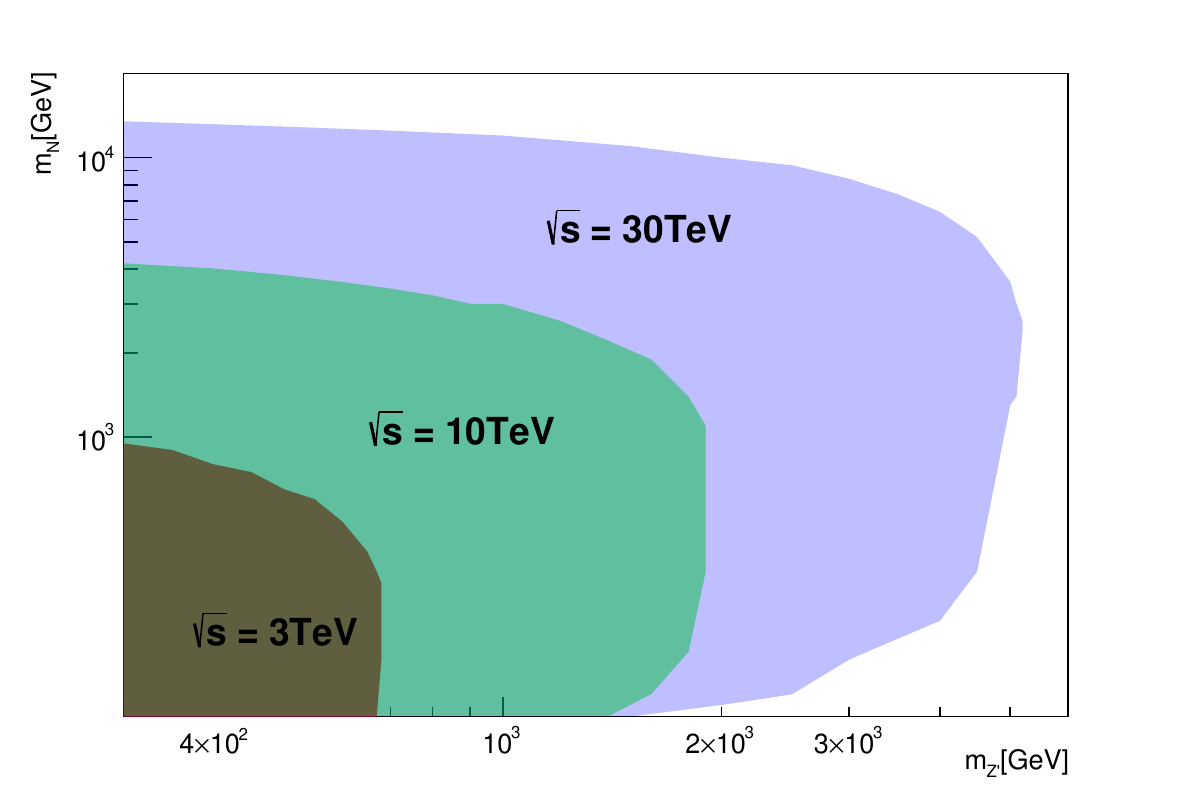}
		\includegraphics[width=0.45\linewidth]{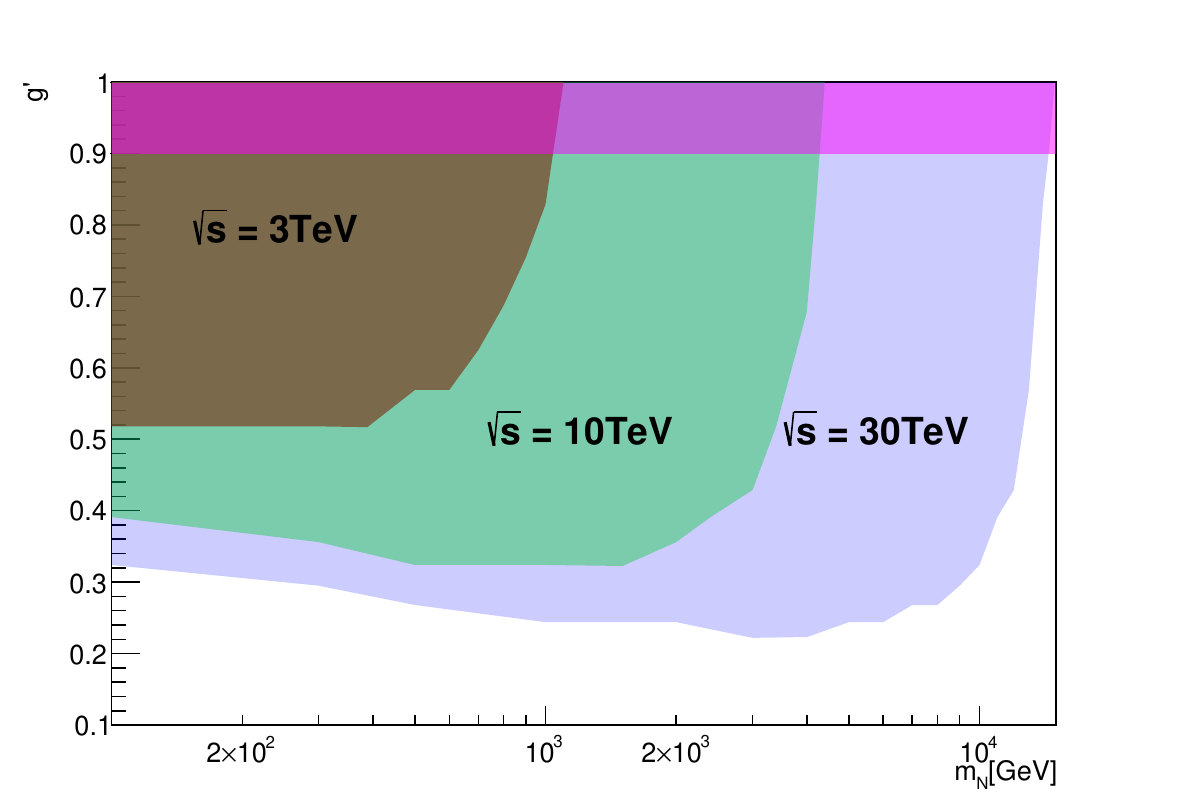}
	\end{center}
	\caption{The combined $5\sigma$ discovery reach of the $\mu^+ \mu^- \mu^\pm\mu^\pm J J$ signature without heavy Higgs $H$ at muon collider.  The brown, green, and purple areas are the results of the 3 TeV, 10 TeV, and 30~TeV muon collider, respectively. The pink region is excluded by the neutrino trident production \cite{Altmannshofer:2014pba}. In the left panel, the new gauge coupling $g'=0.6$ is assumed. In the right panel, the new gauge boson mass $m_{Z'}=500$ GeV is considered. }
	\label{fig13}
\end{figure}

In Figure \ref{fig13}, we show the combined $5\sigma$ discovery reach of the $\mu^+ \mu^- \mu^\pm\mu^\pm J J$ signature without heavy Higgs $H$ at the muon collider. Qualitatively speaking, the discovery regions of this signature are much smaller than those with heavy Higgs as shown in Figure \ref{fig7}. Provided $g'=0.6$, the promising region in the $m_N-m_{Z'}$ plane is shown in the left panel of  Figure \ref{fig13}. The region within $m_{Z'}<680~\text{GeV}$ and $m_N<950~\text{GeV}$ can be discovered at the 3 TeV muon collider. The discovered region is extended to $m_{Z'}<1.9~\text{TeV}$ and $m_N<4~\text{TeV}$ at the 10~TeV stage. The 30 TeV muon collider could detect the parameter space within $m_{Z'}<5.1~\text{TeV}$ and $m_N<13~\text{TeV}$, but it is not sensitive to the electroweak scale $m_N$ with TeV scale $m_{Z'}$. In the right panel of Figure \ref{fig7},  the discovery reach in the $g'-m_N$ plane by fixing $m_{Z'}=500$~GeV are shown. The 3 TeV, 10 TeV, and 30 TeV muon collider could discover $g'\gtrsim0.52$, $g'\gtrsim0.32$, and $g'\gtrsim0.22$, respectively. The 10(30)  TeV stage is mostly sensitive to $m_N\sim1(4)$ TeV.

\section{Conclusion}\label{SEC:CL}

The gauged $U(1)_{L\mu-L\tau}$ extension of the type-I seesaw mechanism is well motivated by various evidence of new physics, but it is still less constrained by current limits. This model introduces three generations of heavy neutral leptons $(N_e,N_\mu,N_\tau)$ to explain the origin of tiny neutrino mass. The scalar singlet $S$ with $U(1)_{L_\mu-L_\tau}$ charge $+1$ breaks the $U(1)_{L_\mu-L_\tau}$ symmetry spontaneously, which then results in the new gauge boson $Z'$ and heavy Higgs boson $H$. 

Since the multi-TeV muon collider is effectively a gauge boson collider, we investigate the new vector boson fusion processes $Z'Z'\to H\to NN$ with and $Z'Z'\to NN$ without the heavy Higgs at the multi-TeV muon collider. Focus on the visible decay channel of heavy neutral lepton $N\to \mu^\pm jj$ with the dijets regarded as one fat-jet $J$, both processes could generate the lepton number violation signature $\mu^+\mu^- \mu^\pm\mu^\pm JJ$. We report that the number of detected final state muons in the main detector is highly dependent on the parameters, such as $m_{Z'}$ and $m_N$. For instance, the $2\mu^\pm$ channel is the dominant one for relatively light $m_N$ or $m_{Z'}$. On the other hand, the $3\mu^\pm+\mu^\mp$ channel becomes the dominant one for relatively heavy $m_N$ or $m_{Z'}$. In this paper, we perform a comprehensive analysis of three different signals, i.e.,  $3\mu^\pm+\mu^\mp$, $3\mu^\pm$, and $2\mu^\pm$ channels.

For the signature $\mu^+ \mu^- \mu^\pm\mu^\pm J J$ with heavy Higgs $H$, we find that all three channels are promising at the muon collider. For the benchmark point with $m_N=200~\rm{GeV}$, $m_H=3\times m_N$, $m_{Z'}=500~\rm{GeV}$ and $g'=0.6$, the $3\mu^\pm+\mu^\mp$ channel is the dominant one at the 3 TeV muon collider. At the 10 TeV stage, all three channels have similar cross section after the selection cuts. While the $2\mu^\pm$ channel becomes the dominant one at the 30 TeV stage. With an integrated luminosity of $90~\text{ab}^{-1}$, the 30~TeV muon collider could discover the largest parameter space. For example, the area of $m_N<8.4 ~\text{TeV}$ and $m_{Z'}<23~\text{TeV}$ with $g'=0.6$ can be discovered. When fixing $m_{Z'}=500$ GeV and the relation $m_{H}=3m_N$, the 30 TeV stage could probe $g'\gtrsim0.12$ and $m_{N}<9$ TeV.

Without the $s$-channel enhancement, the signature $\mu^+ \mu^- \mu^\pm\mu^\pm J J$ without heavy Higgs $H$ is less promising at the muon collider. For the benchmark point with $m_N=200~\rm{GeV}$, $m_{Z'}=500~\rm{GeV}$ and $g'=0.6$, only the $3\mu^\pm+\mu^\mp$ channel can be discovered at the 3 TeV muon collider. At the 10 TeV muon collider, the $3\mu^\pm+\mu^\mp$ and $3\mu^\pm$ channel could be tested with $\mathcal{O}(10^2)$ fb data, while the $2\mu^\pm$ channel requires $\mathcal{O}(10^3)$ fb data. At the 30 TeV stage, the $2\mu^\pm$ channel has the largest signal cross section, but the corresponding background is also the largest. Therefore, the $3\mu^\pm$ channel is the most promising one at the 30 TeV muon collider. With $90~\text{ab}^{-1}$ data, the 30 TeV muon collider could discover the parameter space of $m_{Z'}<5.1 ~\text{TeV}$ and $m_N<13~\text{TeV}$ with $g'=0.6$. Provided $m_{Z'}=500$ GeV, the parameter space of $g'>0.22$ and $m_N<13$ TeV could be probed at the 30 TeV muon collider.

\section{Acknowledgments}

This work is supported by the National Natural Science Foundation of China under Grant No. 12375074, Natural Science Foundation of Shandong Province under Grant No. ZR2022MA056, and University of Jinan Disciplinary Cross-Convergence Construction Project 2024 (XKJC-202404).


\end{document}